\documentclass[11pt]{amsart}
\usepackage{graphicx} 
\usepackage{geometry}
\usepackage{amsmath}
\usepackage{enumerate}
\usepackage{natbib}
\usepackage{verbatim}
\usepackage{amsmath, amssymb, amsthm}
\usepackage{mathtools}
\usepackage{bbm}
\usepackage{tcolorbox}
\usepackage{hyperref}
\usepackage{makeidx}
\usepackage{todonotes}
\usepackage{color}
\usepackage{xcolor}
\usepackage[T1]{fontenc}
\usepackage[utf8]{inputenc}
\usepackage{pdfpages}
\usepackage{dsfont}
\usepackage{booktabs}
\usepackage{float}
\usepackage[shortlabels]{enumitem}
\usepackage{array}
\usepackage[normalem]{ulem}
\usepackage{enumitem}
\usepackage{algorithm} 
\usepackage{algpseudocode}
\usepackage{parskip}
\newcommand{\fix}{\textbf{Fix  }}

\definecolor{violet}{RGB}{148,0,211}

\usepackage[capitalize]{cleveref}
\usepackage{url} 
\usepackage{graphicx}
\graphicspath{{./images/}}

\newcommand{\test}{\Phi}
\newcommand{\indifunc}[1]{\mathds{1}\{#1\}}
\newcommand{\R}{\mathbb{R}}
\newcommand{\N}{\mathbb{N}}

\newcommand{\PP}{\mathbb{P}}

\DeclareMathOperator*{\argmax}{\text{argmax}}

\usepackage{amsthm}

\definecolor{darkred}{rgb}{.8, .1,.1}
\definecolor{darkgreen}{rgb}{0,0.8,0.2}

\newtheoremstyle{spacedtheorem}
  {20pt} 
  {20pt} 
  {} 
  {} 
  {\bfseries} 
  {.} 
  { } 
  {} 

\theoremstyle{spacedtheorem}

\newtheorem{theorem}{Theorem}
\newtheorem{remark}{Remark}
\newtheorem{lemma}{Lemma}

\newtheorem{definition}{Definition}
\setlist[itemize]{topsep=5pt, partopsep=0pt, itemsep=2pt, parsep=0pt}

\title{Fatigue detection via sequential testing of biomechanical  data using martingale statistic}
\author{Rupsa Basu$^+$}
\author{Katharina Proksch$^{*}$}

\thanks{$^*$University of Twente, The Netherlands, $^+$University of Cologne, Germany}

\begin{document}

\begin{abstract}
Injuries to the knee joint are very common for long-distance and frequent runners, an issue which is often attributed to fatigue. We address the problem of fatigue detection from biomechanical data from different sources, consisting of lower extremity joint angles and ground reaction forces from running athletes with the goal of better understanding the impact of fatigue on the biomechanics of runners in general and on an individual level. This is done by sequentially monitoring for change in a datastream using a simple martingale test statistic under minimal assumptions. Problem specific time-uniform probabilistic martingale bounds are provided, which are used as thresholds for the test statistic. Sharp bounds can be developed by a hybrid of a piece-wise linear- and a law of the iterated logarithm- bound over all time regimes, where the probability of an early detection is controlled in a uniform way. If the underlying distribution of the data gradually changes over the course of a run, then a timely upcrossing of the martingale over these bounds is expected. The methods are developed for a setting when change sets in gradually in an incoming stream of data. Parameter selection for the bounds are based on simulations and methodological comparison is done with respect to existing advances.  The algorithms presented here can be easily adapted to an online change-detection setting. Finally, we provide a detailed data analysis based on extensive measurements of several athletes. 
 Qualitative conclusions on the biomechanical profiles of the athletes can be made based on the shape of the martingale trajectories even in the absence of an upcrossing of the threshold. 
\end{abstract}
\maketitle


\pagestyle{plain}
\section{Introduction}
\label{sec:intro}

Recent decades have seen an immense increase in data collection, processing and analysis, with modern data sets being too large or too complex to be analyzed by traditional approaches, in particular if the goal is to use as much of the information provided by the data as quickly as possible in an online setting. More precisely, some applications require for the information processing to be real-time, which means, sequential methods have to be devised to relay results `on the fly’. Advent of smart technology in the form of phones or wearable watches requires that data analysis is \textit{low cost} in terms of computational time and memory usage. Added to these constraints is the fact that scientific teams are becoming more multidisciplinary requiring that theoretical statistical methods and techniques are easily comprehensible by such cross teams. And with this, we introduce the main goal of this paper, which is to statistically analyse biomechanics of human movement, in particular in running athletes. The focus of the methods presented here is to ensure easy adaptability to online or real-time settings.

\vspace{0.5cm}
Running is an immensely popular recreational sport worldwide, with a large number of people suffering from running related injuries; see, \cite{running_popularity},  \cite{runner_injury} and \cite{ZANDBERGEN202360}. The main research question  pursued in this work pertains to addressing fatigue detection in biomechanical sports data obtained from running athletes. In particular, fatigue during the physical exertion of an activity like running is expected to change movement patterns of the athletes. This comes with increased risk of getting injured due to overexertion. Our study is based  on a  dataset which is collected following a fatiguing protocol (details in \cref{application_section}). Similar monitoring of biomechanical movements have seen other widespread application areas including but not restricted to; engineering (\cite{postema1997energy}), clinical (\cite{lu2012biomechanics}), rehabilitation (\cite{yoshioka2009biomechanical}) and sports applications (\cite{zandbergen2022drift}).

\begin{figure}[h]

\centering
  \includegraphics[width=0.65\textwidth]{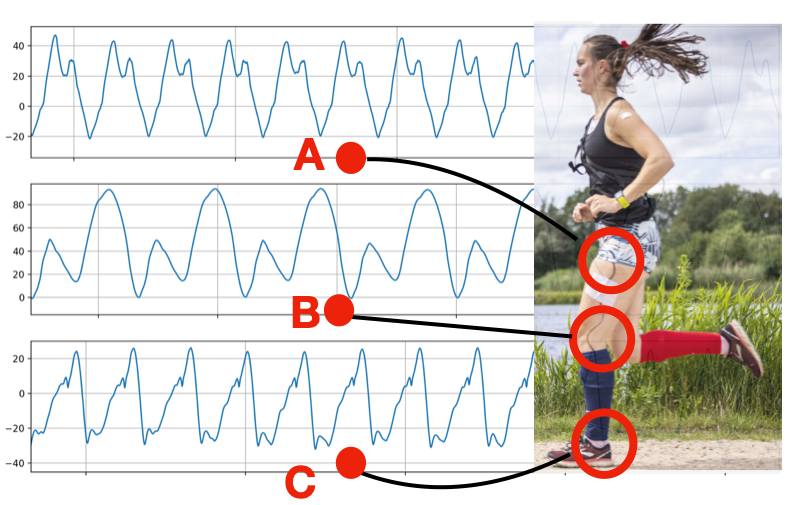}
  \caption{Joint angle patterns for  runner.(A) Hip (B) Knee (C) Ankle angles, over time. Photo courtesy: \href{https://rikkertharink.nl/}{Rikkert Harink}}
  \label{fig:sketch}
\end{figure}

\begin{figure}
     \centering
     \includegraphics[width=\textwidth]{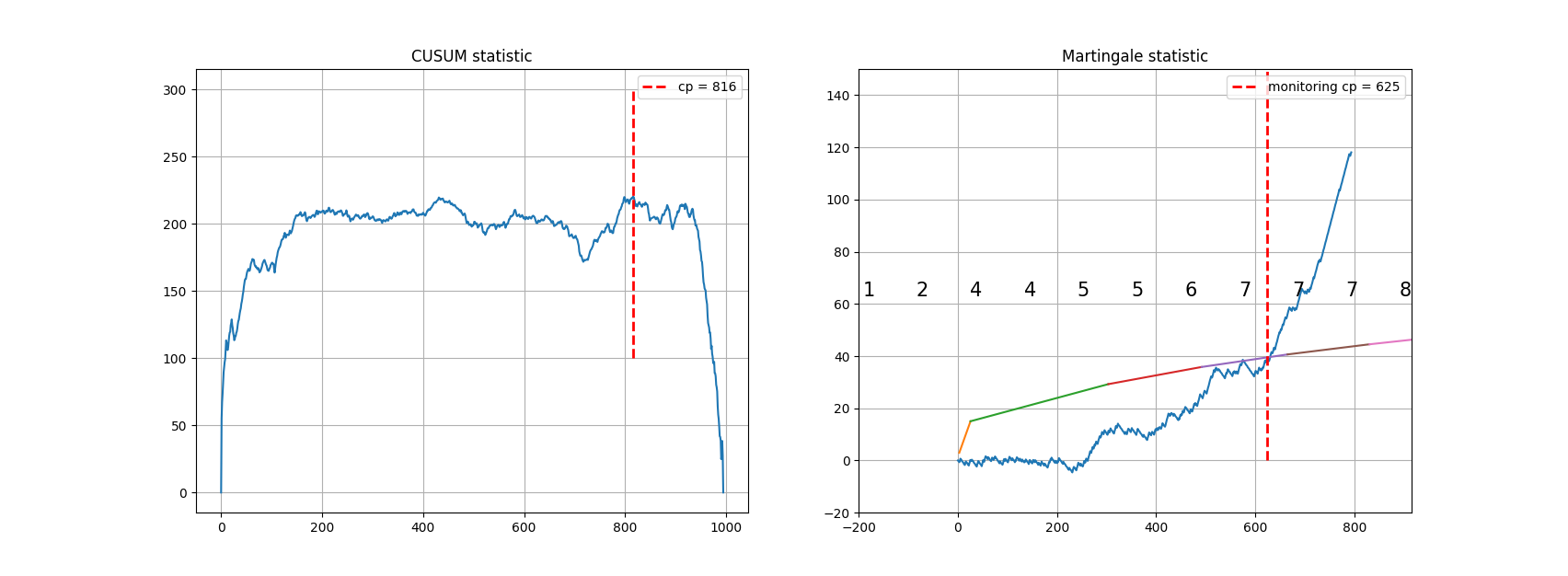}
     \caption{Comparison of our monitoring statistic (right plot) showing initial stable phase (below the bound) with steep rise over the bound for higher fatigue levels with respect to a standard CUSUM statistic for which no such conclusions can be made. Magnitude of the martingale at the point 625 of upcrossing is $40.8$.}
     \label{fig:introfigMonitoring}
 \end{figure}

\vspace{0.5cm}
A commonality in these aforementioned datasets (in particular for activities like walking and running) is that the lower extremity joint angle data show cyclic patterns as seen in the curves shown in \cref{fig:sketch}. Changes caused in such datasets (due to fatigue or otherwise) are not expected to alter the curves completely but more subtly in terms of regional changes within the curves. Biomechanically, this corresponds to a change in a specific movement (like extent of bending the knee) when the athlete is tired. Mathematically, it is possible to study such data via functional time series methods, in particular via change point analysis. And while excellent works exist in these areas (see \cref{CP_literature review} for a review), it is hard for the practitioner to make distributional assumptions or select an exact model for specific datasets. Challenges for data analysis may be in the form of estimating parameters consistently for the underlying model (like coefficients in AR- models) as well as hard interpretability in functional spaces, for example basis expansion and dimension reduction when working with $L^2 [0,1]$.  Further, our question is slightly different than looking for change points for functional data. While detecting a location of change is certainly advantageous, our requirement lies in a monitoring framework which can progressively track our curve data  during the course of the run and monitor deviation from an initial part of the run (rest phase of athlete). In the right plot of \cref{fig:introfigMonitoring}, we visualise the monitoring statistic of our method and track the onset of fatigue. Mathematically, this is modelled as the loss of a certain property $\mathcal{P}$ (see test problem in \cref{local_null}) in later parts of the dataset. The monitoring statistic rises with increasing tiredness (level of fatigue given by the numbers at the top of the plot, with 1 being lowest and 10 being highest). The varicolored line is our monitoring threshold which is crossed by the monitoring statistic with a slight delay, leaving no doubt that change has happened.
 This statistic may be compared to a standard CUSUM statistic (left panel of \cref{fig:introfigMonitoring}) used in the literature, which does not encode this progressive deviation of data from initial states.  Additionally, martingale properties of the monitoring statistic allow us to locate changes which may be attributable to fatigue  with statistical guarantees. The easy interpretability of the monitoring statistic allows the practitioner (or trainer in our case), to make real-time decisions on the changing nature of the incoming data with the highest form of warning of abnormal movements being the upcrossing of the statistic over the bounds. 
 
 \vspace{0.5cm} 
 \textbf{Our Contributions:}  In this respect, we provide easy to implement statistical methodology for the analysis of a stream of data,  based on martingale theory which is (i) particularly well suited to analyze the biomechanical data at hand, (ii) distribution-free and simple to implement and adjust to different data types/ channels and  combinations thereof, (iii) easily comprehensible, (iv) applicable in online monitoring contexts with the added advantage of being (v) applicable not just to curve data but also on specific features of any dataset. We remark that the precision of the change point while advantageous is not of supreme importance but that points (ii)-(v) are the focus of this study. Additionally advantageous is the inherent robustness regarding the data distribution and comprehensive capturing of a generic form of change (not limited to just changes in mean and variance for example). By means of this approach, we focus on  questions such as, (a) selection of interesting feature for fatigue detection and (b)  aggregation or pooling of measurements and features from multiple sources to improve detection. Finally, our methods have the potential to build personalised \textit{risk profiles} for individual runners summarising the effect of fatigue on the runner.

\vspace{0.5cm}
\textbf{Dataset:} The dataset analysed in this study is quite extensive with devices used in data collection being smartwatches tracking the heart rate, IMUs (inertial measurement units) attached to the body of the athletes and multiple cameras and marker systems, tracking the movement of the runners, force plates under the treadmill (more details in \cref{application_section}). Each data set contains a tremendous amount of information, and we focus here on measurements that are easy to understand for a general audience. More precisely, we are mostly interested in  the analysis of data of the lower extremity joint angles, which can be obtained from the measurements via body-worn sensors (IMUs). These measurements show the progression of the angles over time, creating repetitive patterns from stride to stride (see Figure \ref{fig:sketch} for an example on patterns and a sketch of the angles  measured).  
\\
\par

    This paper is structured as follows. We first introduce the basic model assumptions (Section \ref{Sec:modelling}), the mathematical preliminaries and theoretical results (both \cref{math_methods}), where all technical proofs are deferred to the Appendix . The main theorems forming the basis of our methods are in \cref{thm:BoundLarget} and \cref{lemma:gamma1}, written more concisely in   \cref{alg:LIL} and \cref{alg:Hybrid} applied in the analysis of the sports data presented herein.   Further,  the methods presented in \cref{math_methods} require certain parameters as input, which are chosen based on simulations for different case scenarios. These simulations are conducted in \cref{simulation_section} and provide us with the required parametric inputs for computations in the application of our methods to the biomechanical sports data. \cref{application_section} is devoted to our data analysis. There, we present all relevant details and our main findings and conclusions, with complementary results presented in the Appendix.

\section{Basic model assumptions}\label{Sec:modelling}

In this paper, we assume that independent  data,  $X_i\in\mathcal{X},  i = 1, 2 \ldots$ are obtained in a stream, where the endpoint ($i = N$), i.e., the sample size $N$ is unknown in advance and $\mathcal{X}$ is some (abstract) sampling space. A common approach to establish differences between data sets is the application of statistical hypothesis tests such as the $t$-test. If many such tests are performed over time in order to monitor a data stream for the on-set of a certain change to occur, false positives will inevitably occur in long streams. The proportion of false positives is (on average) given by the level of the tests employed. In order to fairly assess sequentially obtained test results we need to take these occurrences of false positives into account.
To this end, we will set up a sequential hypothesis testing procedure,  applicable to data point $X_i$ (via test result $\Phi_i(X_i)$) as soon as it becomes available (i.e., without waiting for $X_{i+1}$). In this respect, our test procedure may be closed-ended (when all data points uptil $i=N$ are available) or open-ended (full sample not available) and can be seen as a form of meta analysis of a large number of test results.  To specify a mathematical framework, let us assume the following \textit{local test problems} at time points $i$, with null hypothesis ($H_i$) and  alternative $(K_i)$:
 \begin{align}
 H_i : X_i \quad &\text{has  property } \, \mathcal{P} \quad\quad\textbf{vs.}\quad\quad\quad K_i : X_i&\text{does not have property}\,\,  \mathcal{P}. \label{local_null}
 \end{align}
 In our analysis of the movement of running athletes, it is observed that the movement patterns remain stable up to a certain  time-point $T_0$ . More precisely, from time-point $t=1$ to time-point $t= T_0$, collected movement patterns  maintain property $\mathcal{P}$. After this point in time,  the movement patterns change and  deviate progressively from the initial patterns, which corresponds to the data no longer having property $\mathcal{P}$. Following this example, we will make the following model assumption
 \begin{align}
 \label{eq:model}
 X_1,\ldots,X_{T_{0}} \text{ have property } \mathcal{P} \qquad X_i,\,  i>T_0 \quad\text{do not have property }\mathcal{P}, 
 \end{align}
 where  we say the loss of property $\mathcal{P}$ corresponds to the runner showing signs of fatigue. We further assume that a statistical \textit{local level-$\alpha$ test} $\test_i$ is available for each local hypothesis $H_i$. The tests $\test_i$ do not need to be identical. In particular, the tests may look at varying features of the data, which, in our application, may be local peak and trough values of biomechanical knee angles during the course of the run.\\

\indent
Our approach is as follows. For each measurement $X_i$, which in our case could be a single cycle from the  lower extremity joint angles collected from the knee, hip and ankle; we perform the hypothesis test $\Phi_i$, for test problem \eqref{local_null}. Instead of keeping track of all the data, we will only keep track of the number of rejections and store data close to the current point in time (buffering). The number of rejections is monitored over time  and the global null hypotheses at time points $t$, $\mathcal{H}_t:=\bigcap_{i=1}^t H_i$, are accepted until the total number of rejections exceeds a certain time-dependent bound.  
 We will provide two different kinds of bounds that can be used for monitoring the number of rejections in our sequential test problem. Both bounds will be based on  martingale theory (\cite{balsubramani2014sharp,balasubramani2016,howard2020time}) discussed further in Section \ref{Sec:Theory}.

\section{Mathematical Preliminaries }
\par
\label{math_methods}
\begin{definition} (Local test function $\test_i$)\label{def:local}
	For the local (i.e., at fixed time points $i$) null hypotheses defined in \cref{local_null}, we define the local statistical hypothesis tests $\test_i$  by
	\begin{align*}
	\test_i :\begin{cases} \mathcal{X} \rightarrow \{0,1\}, \quad \quad \text{where}\quad  i = 1,2, \dots\\
	X_i\mapsto \test_i(X_i)
	\end{cases}
	\end{align*}
	such that it holds that,  
	\begin{align}
	\PP_{H_i} \big(\test_i (X_i) = 1\big)= \alpha. \label{local_level_alpha}
	\end{align}
\end{definition}
\par

\begin{definition} (Sequential setup)
   We define the \textit{sequential hypotheses}
\begin{align}
\mathcal{H}_t = \bigcap_{i=1}^t H_i, \quad \text{and} \quad \mathcal{K}_t = \bigcup_{i=1}^t K_i,  \label{cum_Hypothesis}
\end{align}
and we denote by $\mathcal{H}_{\infty}$ the \textit{overall null hypothesis}
\begin{align}
\mathcal{H}_{\infty}=\bigcap_{j=1}^{\infty}H_{i}. \label{global_null_inf}
\end{align}
	\textbf{Test statistic: }For given sequential and independent data points $X_i\in\mathcal{X},  i = 1, 2, \dots;$ and the test problem with null hypothesis \eqref{cum_Hypothesis} and local level-$\alpha$ as in in Definition \ref{def:local}, we define the sequential test statistic $M_t$ as
	\begin{equation}
	M_t =M_t(\alpha)= \sum_{i=1}^t \indifunc{\test_i=1} - t\alpha, \quad \forall t\in \N, 
	\label{martingale_stat}
	\end{equation}
	where each $\test_i$ is any appropriately defined local level-$\alpha$ test as in \cref{local_level_alpha} for the local test problem in \cref{local_null}. 
\end{definition}

\par
\begin{remark}
	The sequential test statistic \cref{martingale_stat} has the following properties:
	\begin{itemize}
		\item[(a)] The random variables $\indifunc{\test_i=1}, i=1,2,\ldots$ are independent and identically distributed, even though the local tests applied throughout may differ. 
  
		\item[(b)] Because of (a), for any fixed $\alpha\in(0,1),$ $M_t (\alpha)$ is a centered martingale and $M_t + t \alpha \sim \text{Bin }(t, \alpha)$ under the sequential null hypothesis,  $\mathcal{H}_t$ in \cref{cum_Hypothesis}.
	\end{itemize}  
\end{remark}
In Section \ref{Sec:Theory}, we provide two time-uniform critical thresholds $\Gamma_t$ for $M_t$ in \cref{martingale_stat} such that, 
\begin{align}\label{eq:gamma}
\PP_{\mathcal{H}_\infty} \bigg(\exists t\geq s_{0}:\, \, M_t> \Gamma_t \bigg)\leq \delta,
\end{align}
for some initial time $s_0\geq1$, and a global  level-$\delta$ for sequentially monitoring our data. The different time-uniform thresholds are applicable uniquely or in combination for different time regimes as well as specific applications, see \cref{alg:LILSpec} - \cref{alg:Hybrid} below.\\
\par

Notice that our procedure requires the choice of two significance levels $\alpha, \delta\in(0,1)$. The choice of $\delta$ determines the overall significance level of the sequential procedure under $\mathcal{H}_{\infty}$ and controls the occurrence of false positives, i.e., of early or false detections and is therefore  naturally chosen to be small.  The local level-$\alpha$, on the other hand,  can be seen as a parameter that can, in principle, be set to an arbitrary value in $(0,1)$.  Surprisingly perhaps, choosing $\alpha\approx0.2$ will result in favourable properties, whereas small values of $\alpha$, such as the standard choice of $\alpha=0.05$ lead to inferior performance, as will be demonstrated in our simulations  in \cref{simulation_section} and motivated in the Appendix  in \cref{fig:choice of alpha}.

\subsection{Related literature}
Our approach to sequentially monitoring complex data is related to quite a few relevant topics in the statistics literature. While an exhaustive  review is infeasible and outside the scope of this work, the following closely related topics and literature are covered along with similarities and differences to this work. 

\subsubsection{Statistical Process Control}
Clearly, there is a close connection of our initial question to the topic of statistical process control (SPC, See, e.g, \cite{montgomery2020introduction} for a general introduction to the topic or \cite{qiu2017statistical} for a general discussion of SPC in the context of big data analysis). In SPC control charts  are applied to monitor processes, often in the manufacturing industry, for change that may correspond to a production system being out of control. Our martingale statistic  $M_t(\alpha),\,t=1,2,\ldots$ can be considered as a type of control chart, where the control limits are provided by the time-uniform bounds  $\Gamma_t$. However, the performance of a control chart is usually judged in terms of the \textit{run length distribution}  and its \textit{average run length} (ARL). Our viewpoint is slightly different: 
Using time-uniform bounds, we control the overall probability of having a false alarm uniformly in the sense of \eqref{eq:gamma} for the following reason.
One main issue in the monitoring of sports data is the typical run length under control required from a control chart. An hour of running provides data of $10,000$ or more strides, these are numbers for which typical control charts are not designed. Moreover, the precise detection of change location is not the primary focus as, say, a delay of 100 strides  corresponds to less than a minute of running and can still be considered a timely detection, whereas false alarms need to be ruled out.
Finally, many frequently used, traditional control charts require strong parametric assumptions such as a normal distribution and cannot handle a change in the quantities that are monitored \citep{champ1987exact,shewhart1929control,lowry1992multivariate,crowder1989design,crosier1988multivariate}. While  non-parametric methodology based on ranks, order statistics, signs or general quantiles  (see, e.g., \cite{rankseq,qiu2001rank,qiu2003nonparametric,janacek1997control,SignStat,chakraborti2004class}) have been proposed in the literature as well, these approaches are not the right fit for our data analysis, as they were all setup following the usual vantage point assumed in process control.


\subsubsection{Higher Criticism}
In the context of conducting many independent hypothesis tests with the goal of rejecting the joint null hypothesis, various versions of the test statistic, 
\begin{align*}
    \mathrm{HC}_{\alpha,t}=\frac{M_t(\alpha)}{\sqrt{\alpha(1-\alpha)}},
\end{align*}
are often referred to as Higher Criticism statistic in \cite{HC} and have been proposed in the literature from as early as \cite{HCfirst}. The higher criticism statistic is of particular interest in the context of detecting sparse mixtures, ensuring optimality in a certain regime in a variety of different models (see, e.g., \cite{HC,HCArias-Castro,HCCai} and the references therein). In an applied context, referred to as Sequential-Goodness-of-fit, \citep{HCapplied} propose to use the higher criticism statistic to identify the number of hypotheses that can be rejected in a multiple testing procedure providing some sort of relaxed multiplicity control compared to strong control of the FWER and the FDR (neither of which are ensured by this method). While related by using the same kind of statistic, a direct comparison to the methodology proposed in this paper is difficult, as the objectives and models imposed are fundamentally different. It should be noted that the higher criticism statistic is applied to \textit{needle in a haystack} type of problems, i.e., in unordered settings with few but strong signals. While we expect that using the ordered structure in \eqref{eq:model}  can be explicitly incorporated to improve the accuracy of the detection of change, this will certainly happen at the expense of simplicity of the proposed method and is therefore not pursued further.

\subsubsection{Change Point Detection}
\label{CP_literature review}
The literature on change point (CP) detection is vast and a full review is outside the ambit of this work. However, change point detection can be inferred as either detecting abrupt changes in the the data when a certain property changes \citep{kawahara2009change} or an accumulation of change over time due to external situations, example, remote sensing image data \citep{wen2021change}. A branch of CP- detection, known better as CP- estimation,  models known changes in data and interprets the nature and plausibility of the estimates of change. The focus in these is where the existence of change is known and the degree and explanation of change is to be ascertained \citep{hido2008unsupervised}. Within CP detection, a well known sub-field is curated by CUSUM (cumulative sum)-like techniques. CUSUM methods are designed for an offline application and/or are computationally quite intense. For an introduction to the case of  analysing CP retrospectively, an excellent review can be found in \citep{siegmund1995using}. For online change detection in mean, there exist moving window change-detection algorithms,  where the size of change is to be specified or all possible window sizes are considered, see \citep{kirch2018modified} for a survey article, where   various CUSUM-detector statistics are considered, namely Page-CUSUM,  MOSUM and mMOSUM. Alternatively, \cite{otto2022backward}, consider backward-CUSUM test, extendable to online monitoring scenario after $t>T$. In such cases, asymptotics typically lead to a Wiener process due to a functional central limit theorem. Under various regularity and mixing conditions, quantiles may be approximated by appropriate bootstrap strategies.  Further, parameter choices for these methods affect  the kind of changes that these algorithms have the most power to detect. Another substantial shortcoming is the prerequisite of having the knowledge of pre-change and post-change mean (in Page-CUSUM).  These techniques, therefore, have stringent parametric prerequisites and  misspecifications in these models result in unpredictable results and are hard to interpret. A more detailed discussion of these drawbacks is found in \citep{romano2021fast}. \citep{romano2021fast} and \citep{oskiper2002online} also serve as a reference for using a CUSUM-detector for online CP- detection, with the latter proposing to construct CUSUM matrices, however again with the assumptions on density functions of the underlying data. The entensive advances by now in CUSUM type detectors may translate to precise CP- estimates, but may not be so relevant in applications which do not demand the most precise estimates of change location but rather require controlling early detection and achieving robustness towards model misspecification while also being computationally less intense. Further, methods herein only require change analysis over any chosen property of the dataset and is applicable to any choice of feature. A practitioner has full freedom in feature selection, relevant for our work in human movement data and is not restricted to traditional monitoring statistics like mean and variance.

\subsubsection{Martingale bounds and the law of the iterated logarithm (LIL)}
Concentration bounds for martingales that are uniform over finite times were derived in \cite{balsubramani2014sharp}. 
The bounds presented therein take the form of the LIL upper bound for very large time regimes $t$ and provide CLT-type bound below the LIL-rate for small enough $t$ and thus provide the theoretical basis for our work. The results have been applied in the context of sequential testing in \cite{balasubramani2016} for two sample mean testing. As we are only interested in monitoring one particular, very simple martingale, we adjust and extend the proofs in \citep{balsubramani2014sharp} to formulate a bound tailored more precisely towards $M_t(\alpha)$. Since the LIL-based bounds obtained in this manner are not valid for  small values of $t$, we use results of \citep{howard2020time} to derive time-uniform piece-wise linear bounds which can be applied right from the beginning of a monitoring procedure. The bounds can be hybridized to provide sharp bounds over all time regimes.

\section{Theoretical results}\label{Sec:Theory}
In this section, we will state two theoretical results which will be the basis for different sequential procedures proposed in this work. In Theorem \ref{thm:BoundLargetSpec}, we provide a time-uniform bound $\Gamma_t^{\text{LIL}}$ based on the law of the iterated logarithm. This is a special case of Theorem \ref{thm:BoundLarget} in the Appendix, which is a refined version of a bound derived in \cite{balsubramani2014sharp}. This theorem is applicable only after a stopping time is reached, hereafter denoted by $s_{0,\text{LIL}}$ and is  applicable in a (large-) time regime i.e. for $t \geq s_{0,\text{LIL}}$. 

\begin{theorem}\label{thm:BoundLargetSpec}
	Let the random variables $\mathbbm{1}\{\test_i=1\} \stackrel{\text{i.i.d}}{\sim}$Bin$(1,0.2)$ and  $M_t=M_t(0.2)$ be as defined in \cref{martingale_stat}.  For any significance level $\delta\in(0,1/2]$, 
it holds with probability at least $1-\delta$, for all $t\geq s_{0,\text{LIL}}:=780\log(1/\delta)$ simultaneously, that
	\begin{align*}
	M_t\leq \Gamma_t^{LIL}:=\sqrt{0.7t \bigg( \log\log(0.2t) + \frac{1}{2}\log \bigg(\frac{10}{\delta}\bigg)\bigg)}.
	\end{align*}
	
\end{theorem}
The general version of the above theorem involves several parameters and its
proof is quite technical. Therefore, both are deferred to Section \ref{proof_LIL} in the Appendix for a better readability of the main part of this document.
\begin{remark}
 The initial time $s_{0,\text{LIL}}$ in the general \cref{thm:BoundLarget} can easily reach values of $10,000$ and more, when (un-)desirable values of the parameters are chosen in the general Theorem \ref{thm:BoundLargetSpec}.  
\end{remark}
We propose to use an additional bound before $s_{0,\text{LIL}}$. \cref{lemma:gamma1Spec} establishes a piece-wise linear function that can be used in this context. For the sake of a clearer display, we formulate here a version that holds uniformly in time but is only tight until a fixed time point, say $s_{0,\text{LIL}}$. The technical general version that also holds for an infinite time horizon together with its proof are presented in the Appendix. 
 
\begin{lemma}
\label{lemma:gamma1Spec}
	Fix the number $p$ of time points and a significance level $\delta\in(0,1)$.
  Define further a finite sequence of time points $t_1,\ldots,t_p$ such that $t_1\leq t_2\leq\ldots\leq t_p$. Set
	\begin{align*}
	  \tau_0:=0.4\log(p/\delta),\quad
	  \tau_j:=\sqrt{t_jt_{j+1}}\quad j=1,\ldots,p-1,\, \tau_p=\infty,
	\end{align*}
and define the piece-wise linear function $\Gamma^{\text{Linear}}_t$  by
	\begin{align*}
	  \Gamma_t^{\text{Linear}}= \sum_{j=0}^{p-1}\sqrt{\frac{1}{8}\log(p/\delta)}\left(\frac{1}{\sqrt{t_j}}t+\sqrt{t_j}\right)\indifunc{ t\in[\tau_j,\tau_{j+1})},
	\end{align*}
	 the following holds
	\begin{align*}
	  \PP_{\mathcal{H}_{\infty}}\left(\forall t\in\N\; M_t\leq \Gamma^{\text{Linear}}_t \right)\geq 1-\delta.
	\end{align*}

\end{lemma}	
\subsection{Properties of the bounds}\label{Sec:Comparison}
We will now discuss the properties of the two bounds and provide a comparison. Clearly, the piece-wise linear bounds are sharpest for $\alpha=\frac{1}{2}$, %
whereas the LIL-bound scales with $\alpha$. While $\alpha=\frac{1}{2}$ may not be the most obvious choice for the parameter $\alpha$, the procedures remain valid with (global-) level of the sequential procedures given by $\delta$ (and not $\alpha$). 
 Therefore, in this section, we consider the case of  $\alpha=\frac{1}{2}$. 
  To obtain a better understanding of the methods presented here, we will compute each bound at time-points $t_j=j\cdot K$. Then, the bounds are rewritten in the following form: 
\begin{align*}
    \phi^{-1}\left(1-\gamma_j\right)\sqrt{\alpha(1-\alpha)t_j}
\end{align*}
where $\phi$ denotes the cdf of the standard normal distribution.
This allows for a direct comparison to the asymptotic pointwise bound at time point $t_i$ at level $\delta$ obtained via the CLT, that is, 
\begin{align*}
    \phi^{-1}\left(1-\delta\right)\sqrt{\alpha(1-\alpha)t_j}.
\end{align*}
 At the points $t_j$, the linear bound is given by
\begin{align*}
  \Gamma^{\text{Linear}}_{t_i}=\sqrt{\frac{1}{2}t_i\log \bigg(\frac{1}{\Delta_i}\bigg)}.
\end{align*}
Using Mill's ratio for the normal distribution, we approximate
\begin{align*}
  \sqrt{2\log(1/\Delta_i)}\approx\phi^{-1}\left(1-\frac{\Delta_i}{\sqrt{2\pi}\sqrt{2\log(\frac{1}{\Delta_i})}}\right)
\end{align*}
and find
\begin{align*}
  \Gamma_{t_j}^{\text{Linear}}\approx\phi^{-1}\left(1-\frac{\Delta_j}{\sqrt{2\pi}\sqrt{2\log(\frac{1}{\Delta_j})}}\right)\sqrt{\frac{1}{4}t_j},
\end{align*}
i.e., a moderate adjustment of the level for  reasonably chosen $\Delta_j$.
Setting $\alpha=\frac{1}{2}$, $k=0.1$ and $\kappa=1-k$ yields a relatively moderate starting point of $s_{0,\text{LIL}}=615$ and a bound
\begin{align*}
\Gamma_t^{\text{LIL}}\leq \sqrt{t\Big(2\log(\log(0.66t))+\log\Big(\frac{3.06}{\delta}\Big)\Big)}.
\end{align*} 
Using $t_j=j\cdot K$ we find that the LIL bound is of the following form
\begin{align*}
\Gamma_{t_j}^{\text{LIL}}\leq\phi^{-1}\left(1-\frac{\delta}{C\log(j)^4\sqrt{\log(\log(j))}}\right)\sqrt{\frac{1}{4}t_j},
\end{align*}
for a positive constant $C>0$. 
Due to the requirement that $\sum_{j=1}^{\infty}\Delta_j\leq\delta$, it is clear that for sufficiently large values of $t$ and hence $j$, we have
\begin{align*}
\frac{\Delta_j}{\sqrt{2\pi}\sqrt{2\log(\frac{1}{\Delta_j})}}
\gg
\frac{\delta}{C\log(j)^4\sqrt{\log(\log(j))}}
\end{align*}
Therefore, asymptotically, the bound $\Gamma_{t_j}^{\text{LIL}}$ is superior to the linear bound. However as discussed, this only concerns rather large values of $t$.

\subsection{Monitoring procedures}\label{sec:proceduresSpec}
The bounds presented in \cref{Sec:Theory} provide  multiple possibilities of defining thresholds for monitoring procedures for the martingale process. We provide two general approaches which can be readily used or adjusted according to information on the problem at hand in the Appendix.
Here, we will put the specific bounds from Theorem \ref{thm:BoundLargetSpec} and Lemma \ref{lemma:gamma1Spec} to use and provide two specific algorithms for sequential analysis in our setting. More general versions are provided in the appendix.

\begin{algorithm}{}
	\begin{algorithmic}[1]
		\State \fix $\delta \in (0,1/2)$ 
		\State $s_{0,\text{LIL}}\leftarrow\left\lceil780\log\left(\frac{1}{\delta}\right)\right\rceil.$
		\State $\Gamma_t^{\text{LIL}}\leftarrow \sqrt{0.7t \bigg( \log\log(0.2t) + \frac{1}{2}\log \bigg(\frac{10}{\delta}\bigg)\bigg)}$
		\For{$t=s_{0,\text{LIL}},s_{0,\text{LIL}}+1,\ldots$ }
		\If{ $M_{t}>\Gamma_{t}^{\text{LIL}}$} \State \textbf{return} $\widehat T_{0,\text{LIL}}=t$ \Else \textbf{ set} $t=t+1$ \EndIf
		\EndFor
	\end{algorithmic}
 \caption{Sequential monitoring via LIL bounds; }
 \label{alg:LILSpec}
\end{algorithm}
\noindent 
\newline
\cref{alg:LILSpec} provides a monitoring method for large time points $s_{0,\text{LIL}}\geq\left\lceil780\log\left(\frac{1}{\delta}\right)\right\rceil$ and returns the point of up-crossing if it exists. However, in cases when monitoring the martingale already before point $s_{0,\text{LIL}}$ is desired,  a hybrid approach is proposed in \cref{alg:HybridSpec}, which combines the piece-wise linear bound from Lemma \ref{lemma:gamma1Spec} and the LIL-bound from Theorem \ref{thm:BoundLargetSpec}. In this approach, the algorithm splits the overall level $\delta$ of the sequential procedure equally between early and later times. Before time point $s_{0,\text{LIL}}$, $p$ linear bounds are used whose construction is based on equidistantly spaced time points $\tau_0=t_1\leq t_2\leq\ldots\leq t_p=s_{0,\text{LIL}}.$

\begin{algorithm}
	\begin{algorithmic}[1]
		\State \fix $\delta\in(0,1/2) $, and $p\in \N$ 
		\State $\tau_0=t_1 \leftarrow0.4\log(p/\delta)$
		\State $s_{0,\text{LIL}}\leftarrow\left\lceil780\log(1/\delta)\right\rceil.$
		\State $t_j\leftarrow t_1+(j-1)/(p-1)(s_{0,\text{LIL}}-t_1)$
		\State $\tau_j\leftarrow\sqrt{t_jt_{j+1}},\quad j=1,\ldots,p-1.$

		\State $\Gamma_t^{\text{Linear}}\leftarrow \sqrt{\frac{1}{8}\log(\frac{p}{\delta})\left(\frac{1}{\sqrt{t_j}}t+\sqrt{t_j}\right)}, \quad t\in[\tau_j,\tau_{j+1}),$
        \State  $\widehat T_{0,\text{hybrid}}\leftarrow 0$
		\For{$t=t_0,\ldots,s_{0,\text{LIL}}$ }
		\If{ $M_{t}>\Gamma_{t}^{\text{Linear}}$} \State \textbf{return} $\widehat T_{0,\text{hybrid}}=t$ \Else \textbf{ set} $t=t+1$ \EndIf
		\EndFor
        \If{$T_{0,\text{hybrid}}=0$}
		\State $\Gamma_t^{\text{LIL}}\leftarrow \sqrt{0.7t \bigg( \log\log(0.2t) + \frac{1}{2}\log \bigg(\frac{10}{\delta}\bigg)\bigg)}$
		\For{$t=s_{0,\text{LIL}},s_{0,\text{LIL}}+1,\ldots$ }
		\If{ $M_{t}>\Gamma_{t}^{LIL}$} \State \textbf{return} $\widehat T_{0,\text{hybrid}}=t$ \Else \textbf{ set} $t=t+1$ \EndIf
		\EndFor
        \EndIf
	\end{algorithmic}
 \caption{Sequential testing (Hybrid Algorithm)}
\label{alg:HybridSpec}
\end{algorithm}

\section{Case study: Performance analysis of runners} 
\label{application_section}
  As  mentioned, the focus of our 
  analysis is the performance of athletes during training, in particular in running, which has gained considerable attention in the biomechanical literature in recent years (see, e.g., \cite{Saarakkala} for a review on the use of sensor technology in this context). Our analysis provides a step towards the  ultimate goal of giving feedback for training intervention in real time to avoid problematic movements, which may cause injuries in the long-term. The aim is therefore to detect change in the movement patterns due to fatigue  before the runner notices any tiredness.   \\
  \indent
As is obvious, running entails higher risk of injuries in the lower extremity joints i.e., the hip, knee and ankle joints.  Further, it is known that long-distance and long-duration runs are to be associated with injuries due to the fatigue of the runner \citep{fatigue_factor} and specifically in the knees (see, \citep{van2007incidence}, \citep{Nielsen}) and tracking the onset of changes which may lead to such injuries via  the joint angles can help to stop or alter training appropriately. It is therefore, of particular interest to study  how running kinematics change due to running-induced fatigue (see  \cite{apte2021biomechanical} and \cite{ZANDBERGEN202360} as well as references therein).  This is done by the study of biomechanical data obtained by using inertial measurement units (IMUs), activity trackers like FitBits as well as recorded video data with and without optical motion capture system.\\ \indent
For the application of the methods presented here, we  focus on the data from running, obtained using a fatigue protocol. In other words, the data collection is designed in such a manner that it is ensured that the athlete taking part in the study will surely but steadily get tired during the course of the run. As discussed before, the fatigue protocol for data collection may be designed in different ways, see \citep{apte2021biomechanical}. In our case, the runners ware asked for their average speed in an 8 km run and then the speed of the treadmill would be adjusted to $103\%$ of this speed.\\
\indent
Fatigue  has different effects on  the knee biomechanics of the runner (\cite{harato2021fatigue}).  This is because all lower extremity joints, namely, the hip, knee and ankle joints have to somewhat \textit{adjust} to endure the movement even in a fatigued state and therefore are the joints  undergoing maximum change during the course of the run. The movement of these joints are recorded via the  joint angles, which  are repeated patterns of the form shown in Figure \ref{fig:sketch}.  In sections \ref{joint_ang_single}, 
and 
\ref{pooled_data}, 
 we study these  joint angles in detail and apply different versions of the sequential testing methods given in \cref{alg:LIL} and \cref{alg:Hybrid} for monitoring for change. To complement this study, we also analyze data from an additional source in \cref{app: missing data},  where we look at the ground reaction forces exerted by the foot,  obtained from the force plates embedded in the treadmill. Finally, in  \cref{contact_time}, we look at the contact time of the left and right foot with the ground during the course of the run. 

\subsection{Source devices of data}
As mentioned before, we obtain this dataset by conducting running trials with fatigue protocol. Devices used in data collection are (i) inertial sensors produced by Xsens (Xsens MVN link sensors, sampling at 240 Hz, see \cite{schepers2018xsens}) attached to the body of the athletes,  and (ii) Force plate data obtained when athletes ran on one belt of a dual-belt treadmill with an integrated three-dimensional (3D) force plate (custom Y-mill, Culemborg, The Netherlands).
From the IMUs, one can obtain angles of the hips, knees and ankles (lower extremity joint angles) in 3- dimensions. From a biomechanical point of view, it makes sense to look at the lower extremity joint angles in the sagittal plane (the plane dividing the body into left and right halves). From the force plates, the ground reaction forces that are associated with the strike and take-off of the foot are obtained.
Note further, that for our data collection experiments, 
we had 6 healthy male and female runners, referred to in the following as $R_{0},\ldots,R_{5}$, between the ages of $22- 39$, with no reported prior injuries to the knee joint. Further, the subjects under study in this data analysis are known to have between $1- 11$ years of experience in running. All subjects also reported to be regularly running at least a couple of times  between $10- 60$ km per week as their current training patterns. Runner $R_0$ was the first subject in a trial run of our study. The protocol and lab conditions for this runner slightly differ from the protocol and lab conditions for the other runners, resulting in a slightly less noisy data set that we use to compare the analyses of the other runners to.

\subsection{Modelling the data}
In this section, we model the data on the example of  the progression of the joint angles obtained from the runners. Figure \ref{fig:sketch} shows a few exemplary cycles of such data for each of the three joints (hip, knee and ankle) under consideration. For each run, several thousand of such stride curves, varying around the respective mean curves, are available for each joint. In such biomechanical data, one stride is basically one cycle, or one functional observation, from a functional data set. Therefore, in the following, we implicitly impose a simple functional signal plus noise model to describe the joint angle data mathematically per stride.
Let $Y_i^k (t)$ denote the i-th stride of runner $k$, $i=i,\ldots, N_k$, $k=0,\ldots,5,$ where
\begin{align}\label{eq:functdata}
Y_i^k (t)&= \mu^k (t) + \epsilon_i^k(t) \quad \quad \text{for all} \; t\in[0,1]\; i = 1, \dots, N_k,\; k\in\{R_0,\ldots,R_5\}, 
\end{align}
where $\mu^k (t)$ is the mean function (stride) of the  runner $k$ and $\epsilon_i^k(t)$ are zero mean stochastic process independent of the \textit{true} signal $\mu^k (t)$.  Such problems are considered as a pre-processing step, (see \cite{hormann2022consistently}). Post-processing, there is a tremendous amount of appealing methods from functional data analysis to apply to such data.  Excellent surveys of functional data analysis (FDA) methods are found in \citep{wang2016functional}, \citep{hsing2015theoretical} and \citep{horvath2012inference}, while a recent advanced work also incorporates the relevant size of the change in the banach space of continuous functions (see \cite{dette2020functional}).  While in these references appealing functional data analysis methods exist to deal with such functional models, to avoid technicalities and to demonstrate the appealing simplicity of our approach, even for such complex data, we will refrain from pursuing this direction in this example and introduce a simple pipeline for data analysis in the following section, which also works for other data models than  \eqref{eq:functdata}.

\subsection{Typical procedure for the data analysis}\label{sec:procedure}
To put our monitoring approach to use, we perform the following steps:
\begin{tcolorbox}
\it\small
\begin{enumerate}
    \item Use the first $n_1$ data points (or prior data/information) to estimate a reference stride profile for each of the joints in a rested state of the runner. More precisely, we estimate the reference mean 
    \begin{align}
        \hat \mu_{n_1} (t) = \frac{1}{n_1} \sum_{i = 1}^{n_1} Y_i (t).
    \end{align}
    \item Use the subsequent $n_2$ data points (strides) to compute the distances $D_1,\ldots,D_{n_2}$ of each stride to the reference pattern. We use the $L^2$-distance as follows, 
    \begin{align}
        D_j = \|Y_j (t)- \hat{\mu}_{m_1} (t)\|_2^2, \quad \quad j= 1, \dots, n_2.
    \end{align}
    \item  From the $n_2$ computed differences $D_1,\ldots,D_{n_2}$ , estimate the $(1-\alpha)$-quantile $\hat q_{1-\alpha}$, e.g., the median for a choice of $\alpha=0.5$. 
    \item Start monitoring afterwards: Set $S_0:=0$ and for stride $i$ after the initial $n_0 \coloneqq n_1+n_2$ strides:
\vspace{-0.2cm}
 \begin{itemize}
	\item Add 0 if  $D_{n_2+i}<$ $\hat q_{1-\alpha}$: $S_i=S_{i-1}$
 \vspace{-0.1cm}
	\item Add 1 else: $S_i=S_{i-1}+1$
 \end{itemize}
\end{enumerate}
\end{tcolorbox}
\begin{remark}
    In step 2 above, it is possible to instead work with, 
    \begin{align}
        D_j = \|Y_j (t) - \hat \mu_{n_1} (t)\|_\infty \quad \text{or} \quad  D_j = \|Y_j (t) - \hat \mu_{n_1} (t)\|_1, 
    \end{align}
    as other projections of the behaviour of the curve data into point data. Indeed, in some small simulations this has shown to work well. However, we do not pursue this further in this work and note it as a possible choice for the practitioner based on the problem of interest. 
\end{remark}
We look for changes in the data using the martingales $M_i(\alpha)=S_i-\alpha i,$  using the sequential monitoring scheme provided in \cref{alg:Hybrid}. Note that if not mentioned otherwise, we set local level $\alpha = 0.22$, the global level to $\delta = 0.1$ and $p=10$ (number of partitions for linear bounds, see \cref{lemma:gamma1}) and compute the martingale by taking the $(1- \alpha)$- quantile of the initial 200 points of the dataset $S_i$.  Further, since the lengths of the datasets are not too large, we mostly make use of the linear bounds as a threshold for the monitoring scheme. 
 Further, note that in all of the following plots, the $x$-axis always corresponds to the sequence of strides during the course of the run, unless otherwise stated. 

\subsection{Results}
In this section we present the main results of our data analysis. For this, we focus on specific examples and present additional results and material in the Appendix. As discussed above, our data set consists of the data of the six runners $R_0,\ldots,R_5$. Therefore, due to this limited sample size,  the conclusions drawn here are seeking confirmation in a follow-up study but provide nonetheless several interesting and meaningful pointers and directions for future study in the biomechanics of running.

\subsubsection{Joint angle movement patterns indicate significant changes at an early stage of a fatiguing run}
\label{joint_ang_single}

\begin{figure}[h]
    \centering
    
        \includegraphics[width=4cm, height =7.9cm]{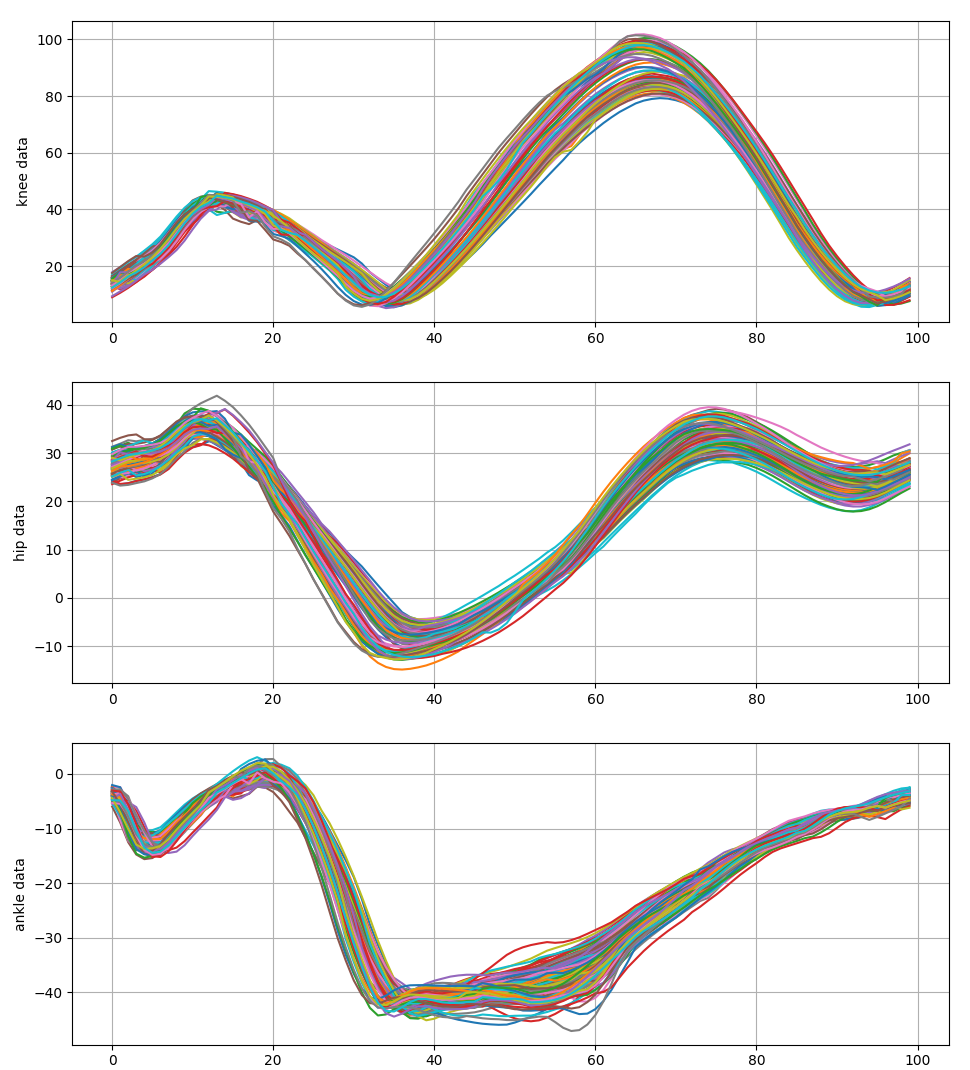}
       \includegraphics[width=9cm, height = 8cm]{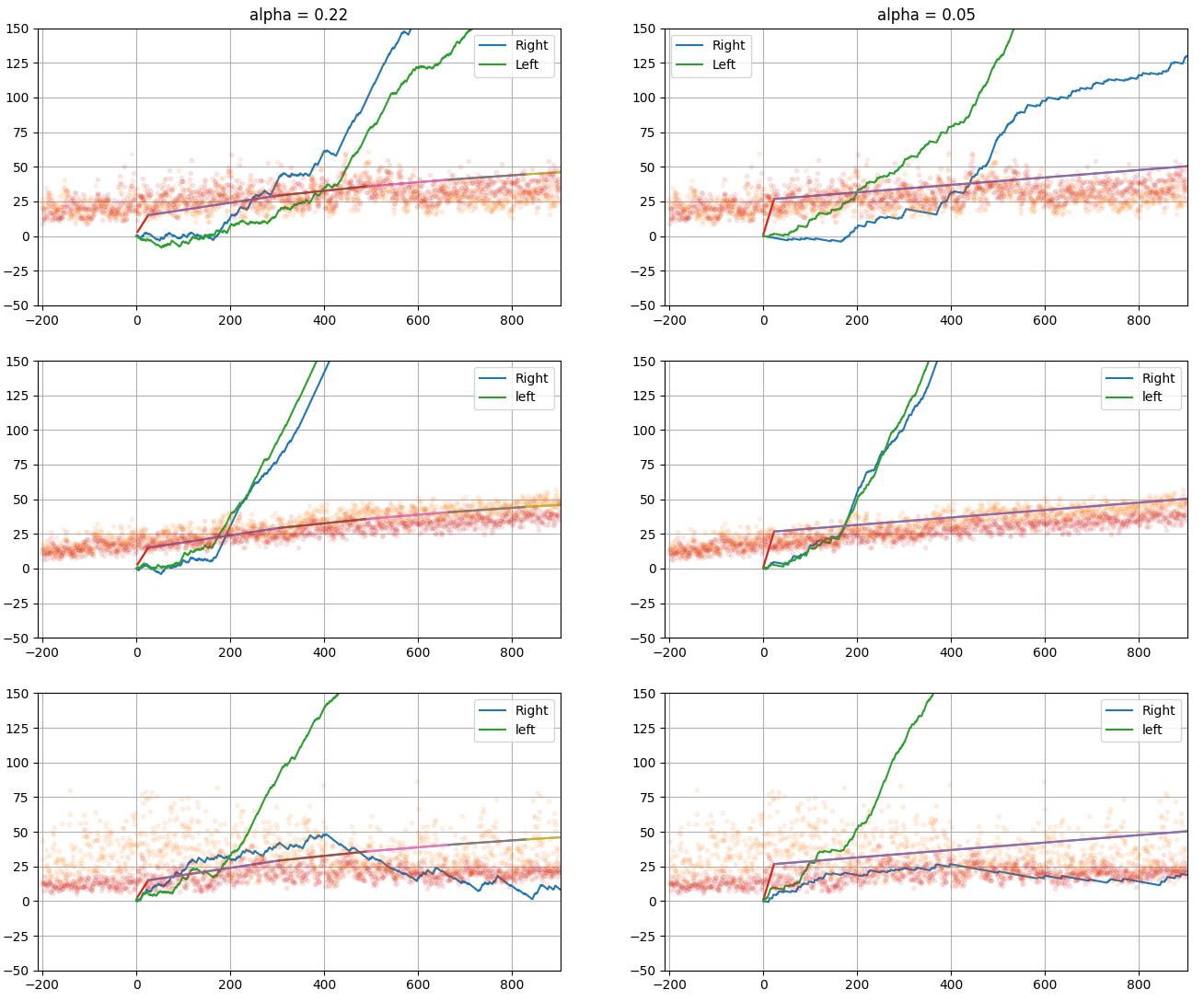}
    
      \caption{ Single runner $R_0$ profile for lower joint angles. Stacked curves from each stride (left col.) and the corresponding martingale statistic for $\alpha = 0.22$ (middle col.) and for $\alpha = 0.05$ (right col.). Total sample size $N_{R_0}= 1103$ and length of training data $n_0 = 200$. }
    \label{fig:single_runner}
\end{figure}

As an initial proof of concept, in this section we first use the hip, knee and ankle joint angles from runner $R_0$,
\begin{align}\label{eq:pilot}
    Y_{i, \text{hip}}^{R_0} (t),  Y_{i, \text{knee}}^{R_0} (t),  Y_{i, \text{ankle}}^{R_0} (t), \quad i = 1, \dots, N,
\end{align}
where $N_{R_0} = 1103$, the full sample size known in advance and recorded over time with IMUs during the course of the fatiguing protocol. 
\vspace{0.05cm}
\cref{fig:single_runner} shows 12 martingale trajectories (left and right knee, hip, ankle for $\alpha\in\{0.05,0.22\}$) and linear bounds that have been obtained  up to step 14 via \cref{alg:HybridSpec}. \\


A couple of first interesting results that we find are: \\

\begin{figure}[h]
    \centering
    \includegraphics[width=0.8\textwidth]{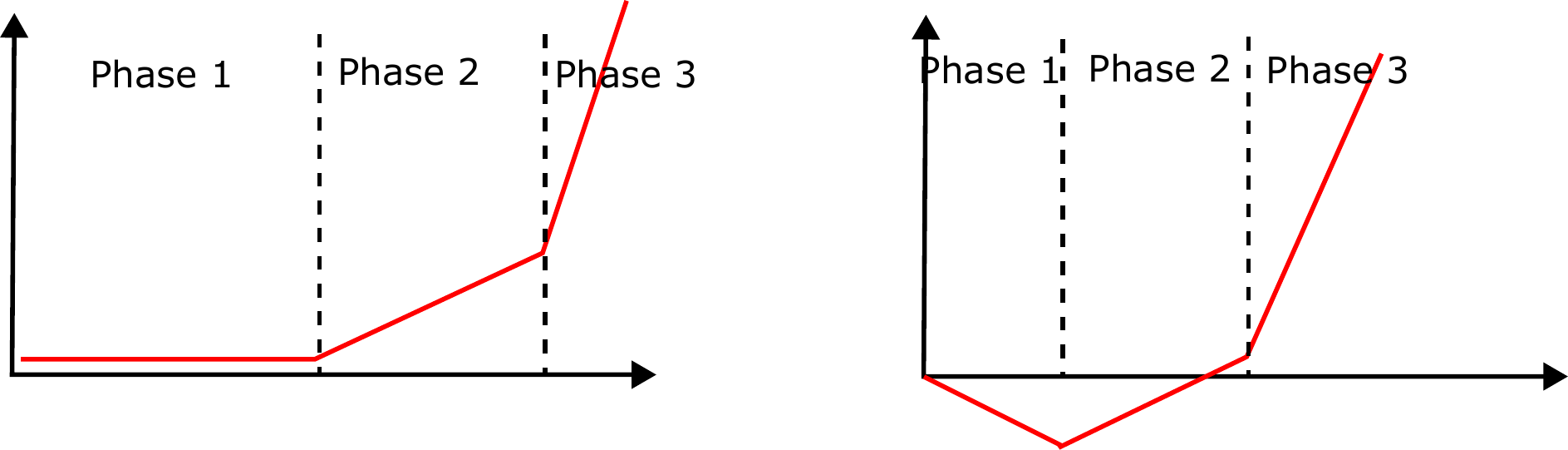}
    \caption{Typical schematic shapes of observed martingale trajectories in a fatiguing run indicating three phases of a run. Left: stable phase, mild increase, steep increase. Right: decrease, mild increase, steep increase (\textquotedblleft {\it check mark}\textquotedblright).  }
    \label{fig:martingale_phases}
\end{figure}

\begin{enumerate}[(a)]
\item Significant change, i.e., upcrossing of the bound by the trajectory,  is established for 11 out of 12 examples. In all cases, the point in time is clearly before fatigue is experienced and reported by the runner, indicating that a fatiguing run typically comprises more than the two phases of pre- and post fatigue. In fact, most shapes of observed martingale trajectories in our study can be reasonably segmented into three phases as indicated in Figure \ref{fig:martingale_phases}.\\

    \item The martingale statistic of the right ankle angle is significantly different from the rest of the runner profile, as it does not clearly show a steady deviation from the initial reference movement pattern. This may be connected to the observation that fatigue  impacts the lower joint angles such that increased asymmetry in the load of hip, knee and ankle joints may be seen (see \cite{gao2022effects}).
\end{enumerate}

\vspace{0.5cm}
\noindent
Qualitatively, the progression of the martingale trajectories is such that before the onset of a steep increase, indicating a drastic change in the movement pattern of the runner, there is a state of stability, indicating a  phase preceding the onset of fatigue, which is consistent with biomechanical modelling of typical phases during a fatiguing run.  The steep increase on the other hand suggests that fatigue does in fact manifest in altered movement patterns and therefore larger deviations from the initial reference pattern (property $\mathcal P$ in \cref{eq:model}). This pattern of the martingale trajectory is sketched in Figure \ref{fig:martingale_phases} and is also seen and discussed in the following analyses of other runners as well as measurements other than joint angles.\\



The data from runner $R_0$ went through some significant post-processing, leading to nicely smooth functional curves from each stride. For a study involving several runners, extensive post processing for each runner accounting for individualistic characteristics is infeasible. In the following, we therefore study a noisier version of the joint angle data recorded for runners $R_1,\ldots,R_5$. In general we see that the methodology of this paper adapts well even to noisier datasets.\\

\begin{figure}[h]
\hspace{-0.05\textwidth}
 \includegraphics[scale = 0.24]{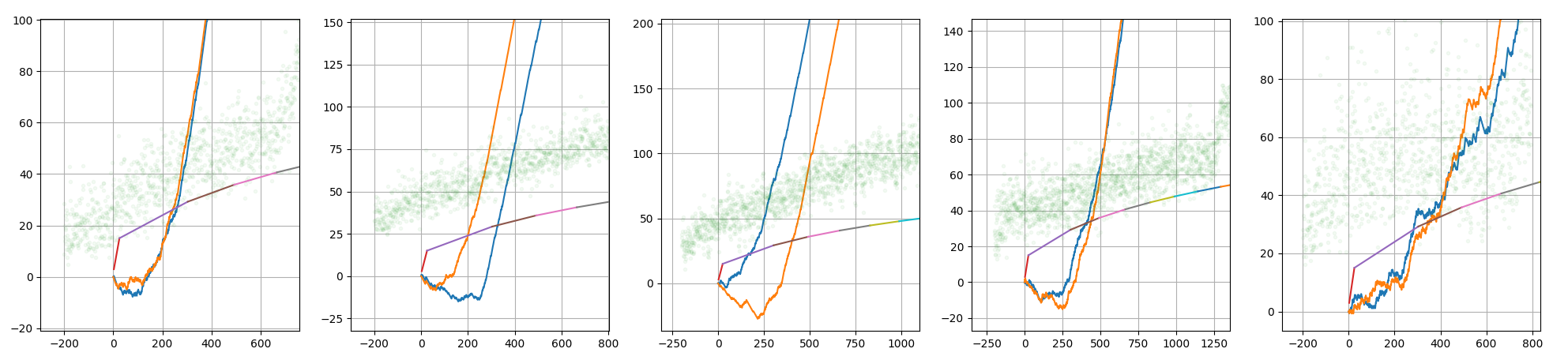}
    \caption{Martingale trajectories and distance data points for the runners $R_1,\ldots,R_5$ (from left to right) based on IMU measurements of left hip (blue) and right hip (orange). } 
    \label{fig:kneeAll}
\end{figure}

\vspace{0.25cm}
Figure \ref{fig:kneeAll} shows martingale trajectories for the runners $R_1,\ldots,R_5$  based on IMU measurements of left  and right hip. For all runners we observe a steady moving away from the reference showing in an increase of the points and the martingale trajectories.  Changes in movement patterns  early on (long before fatigue is perceived by the runners) are confirmed by up-crossings of the bounds by the martingale trajectories. We further observe characteristic check mark patterns as sketched in the right panel  of Figure \ref{fig:martingale_phases}. The points added to the plots show the progression of the computed distances from the reference. There, we can also see clear change at the very end of the run when fatigue kicks in.

\begin{figure}[h]
 \centering
    \includegraphics[width=1\textwidth]{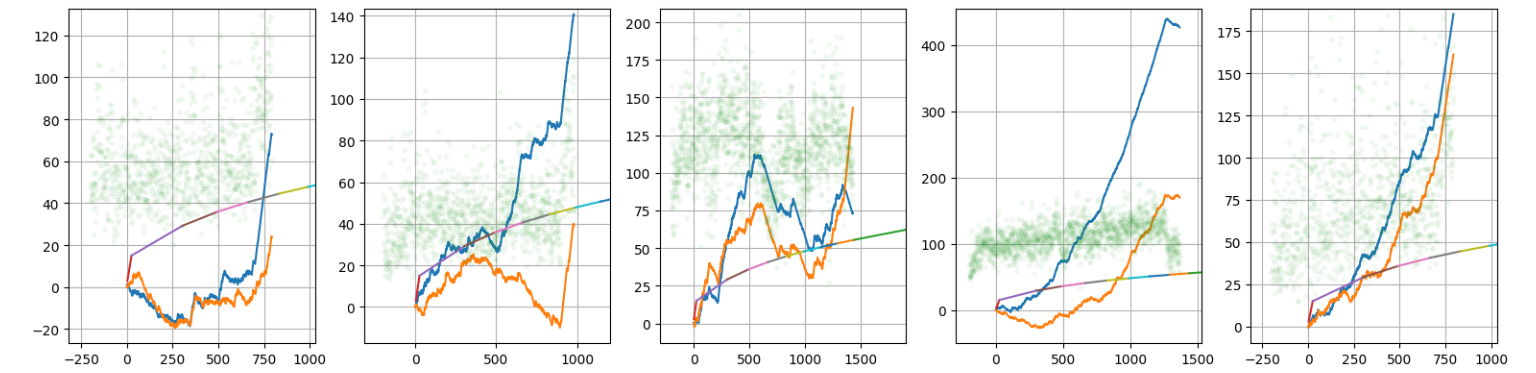}
    \caption{Martingale trajectories and distance data points for the runners $R_1,\ldots,R_5$ (from left to right) based on IMU measurements of left knee (blue) and right knee (orange).}
    \label{fig:IMUknee}
\end{figure}

\vspace{0.5cm}
While the martingale trajectories of runner $R_0$ in Figure \ref{fig:single_runner} based on IMU knee angle data are relatively smooth and all share common features, using noisier versions of the knee angle data of the runners $R_1,\ldots,R_5$ results in less clear results (see Figure \ref{fig:IMUknee}), where we see that for runners $R_2$ and $R_3$ the trajectory shows some deviation from the characteristic \textit{check-mark pattern} that we mostly see in such data. 
We discuss simple ways of pooling data from different channels (e.g., from left and right joint angles but also from different measuring devices) in Sections \ref{pooled_data} and \ref{sec:pooled_different_sources}, where we show that pooling data  typically yields more stable results. Overall,  the observation of an early significant change can be made throughout runners, data sources and parameter settings, albeit with obvious individualistic differences. We believe that the martingale trajectories have the potential to reveal individualistic profiles of how runners compensate fatigue by adjusting their movement. This path will be explored in Section \ref{sec:individual} below.\\

To conclude this section and to substantiate our claim in the title of this section, we now have a look at the perceived level of fatigue of the runners $R_1,\ldots,R_5$.\\
\begin{table}[h!]
\small
    \centering
\begin{tabular}{|c|c|c|c|c|c|c|}
\hline
\% of run  & $R_1$ & $R_2$ & $R_3$ & $R_4$ &$R_5$
\\
\hline
0.0 &   1  &   2 &  1 &  1 & 2     
\\
\hline
25.0 &  4 &  5   &  3 &  6 & 7    
\\
\hline
50.0 &  6 &  6    &  6 &  7 & 8    
\\
\hline
75.0 &  7 &  8    &  7 & 8 & 8   
\\
\hline
100.0 & 8 &  9   &  9 &  9 & 9  
\\
\hline
\end{tabular}
\caption{ Reported Borg scales of all subjects,  corresponding to percent of run. }
\label{tab:reportedBorgs}
\end{table}

\begin{figure}
    \centering
    \includegraphics[height= 9cm, width = 12cm]{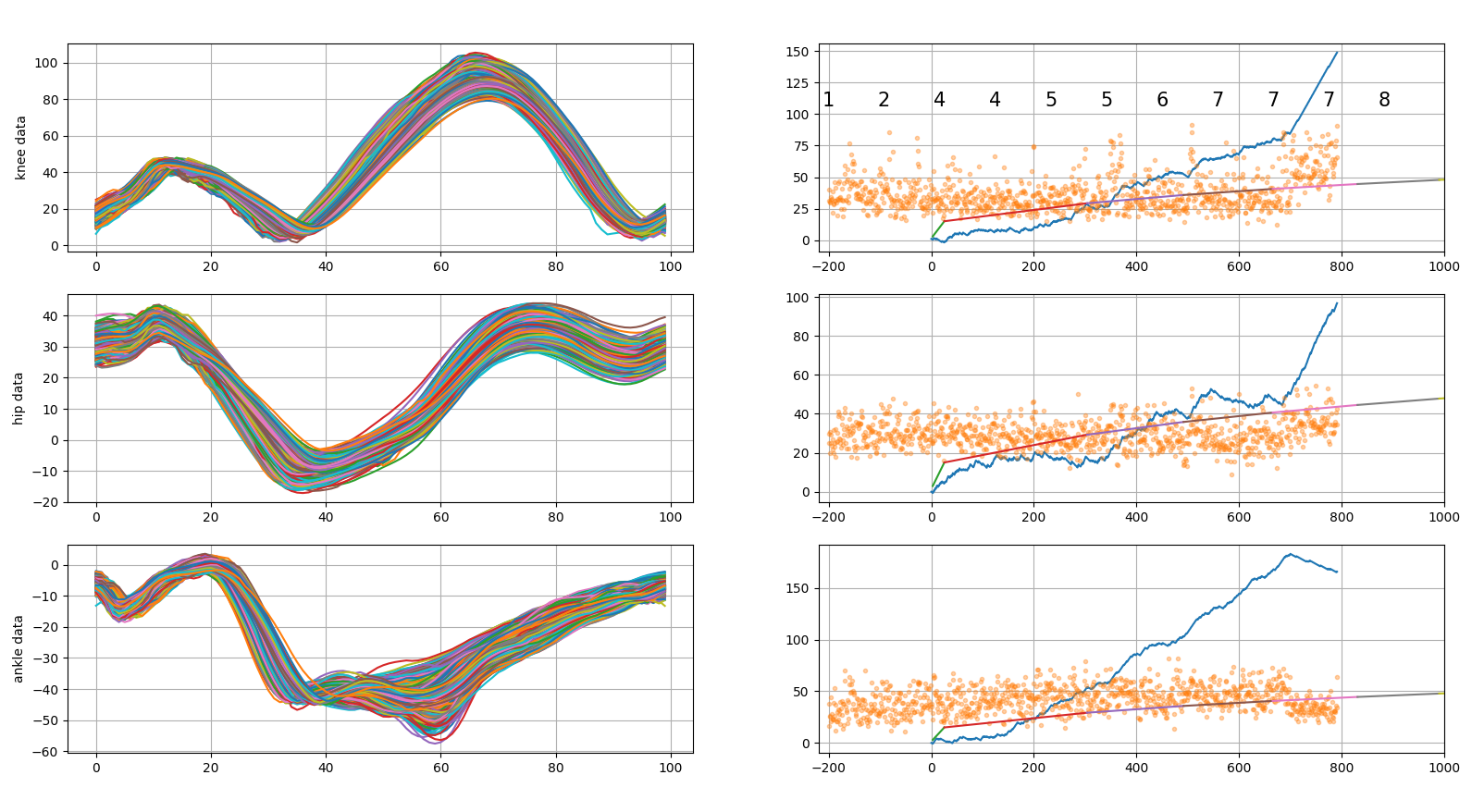}
    \caption{Borg analysis of  runner $R_1$ with sample size being $N = 992$ and length of the reference sample $n_0 = 200.$ }
    \label{fig:borg_single_runner}
\end{figure}

\begin{table}[h!]
    \centering
    \begin{tabular}{|c|c|c|c|c|c|c|}
    \hline
          $R_0$ & $R_1$ & $R_2$ & $R_3$ & $R_4$ & $R_5$ \\
           \hline
          $1103$ & 992 & 1081& 1631 & 1569 & 995\\
          \hline
    \end{tabular}
    \caption{Sample sizes for the IMU data of all the runners used in the dataset.}
    \label{tab:my_label}
\end{table}
For this, we use the response of the runners during the course of the run which was taken in the Borg Scale \citep{borg1982psychophysical}, a scale from 1 to 10, from relaxed to total exhaustion.   The recorded Borg scales of the runners from our dataset are given in \cref{tab:reportedBorgs}. For illustration   we showcase our methodology of fatigue detection with the corresponding perceived level of fatigue of runner $R_1$ in \cref{fig:borg_single_runner}.  It can be seen that up to at least a reported/\textit{perceived} level of fatigue of 5 (the transition from light activity to moderate activity) a stable phase can be consistently observed for all the lower extremity joints. Furthermore, change is typically detected while moving from moderate to vigorous activity (the latter starting at 7). Therefore, we conclude that the martingale trajectory captures the progression of the perceived level of fatigue consistently.

\subsubsection{The martingale trajectories give rise to individual characteristic movement and fatigue profiles}\label{sec:individual}
A comparison of martingale trajectories of different runners of the same type of measurements shows that the process of fatiguing
over the course of a run seems to manifests in a different way for every runner. \\

\begin{figure}[h]
    \centering
    \includegraphics[width=0.95\linewidth]{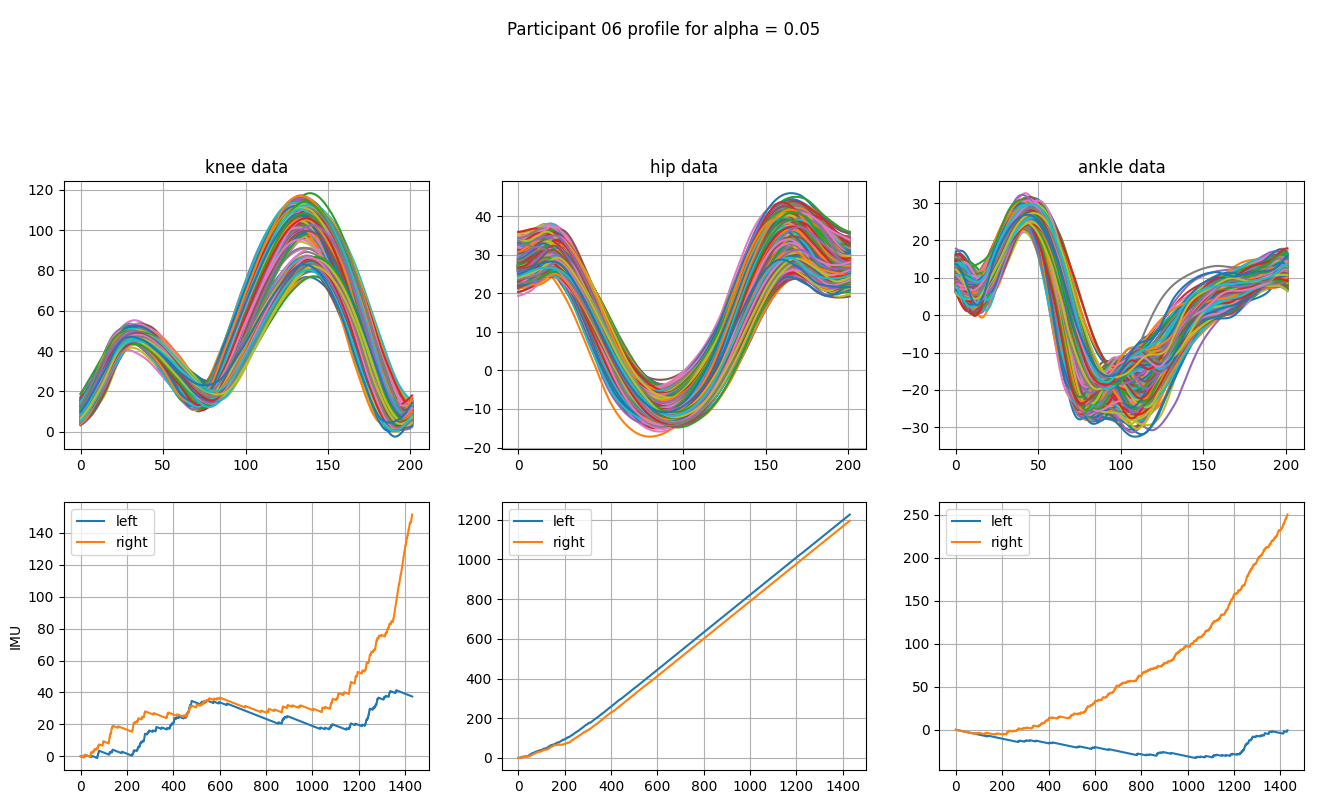}
    \caption{Upper panel: Stacked curves over the course of the run. Left panel: Martingale trajectories for runner $R_3$ for IMU knee, hip and ankle angles (from left to right)  for $\alpha = 0.05$.  }
    \label{fig:R06alljoints}
\end{figure}

\begin{figure}[h]
    \centering
    \includegraphics[width=0.95\linewidth, height = 5cm]{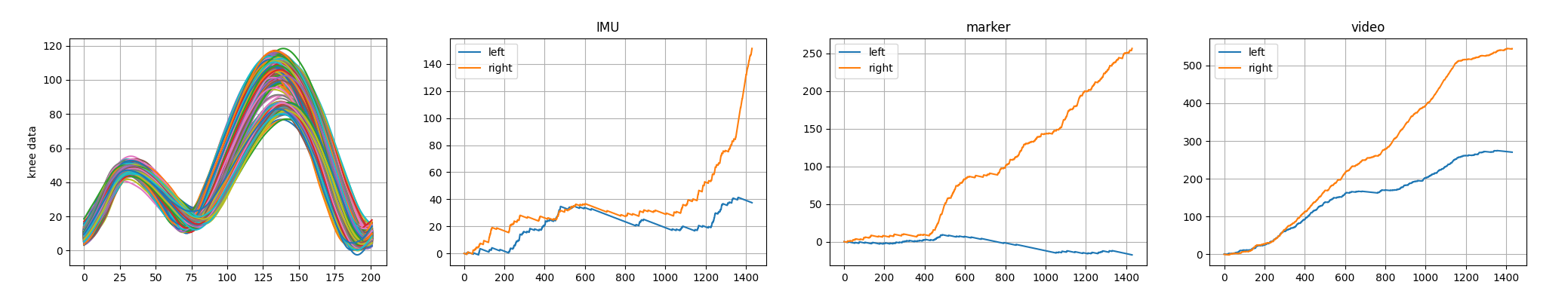}
    \caption{Martingale trajectories of knee angle data of runner $R_3$ measured with IMUs, optical markers and video (from left to right). 
 }
    \label{fig:R06allsources}
\end{figure}

For example, Figure \ref{fig:R06alljoints} shows that runner $R_3$ seems to have multiple phases within the run which are showing in all joints measured with the same measuring system (IMU) simultaneously. The same observation is confirmed when looking at, e.g., the knee angle data across measurement systems as done in Figure \ref{fig:R06allsources}. For this particular runner none of the joint angle sources provide a martingale of the typical forms sketched in Figure \ref{fig:martingale_phases}. Further, a certain asymmetry between left and right joints is observed throughout. It is worthwhile to mention that the results for this runner $R_3$ are consistent also with respect to different  parameter values chosen. 

\begin{figure}[h]
    \centering
    \includegraphics[width=0.95\linewidth, height= 9cm]{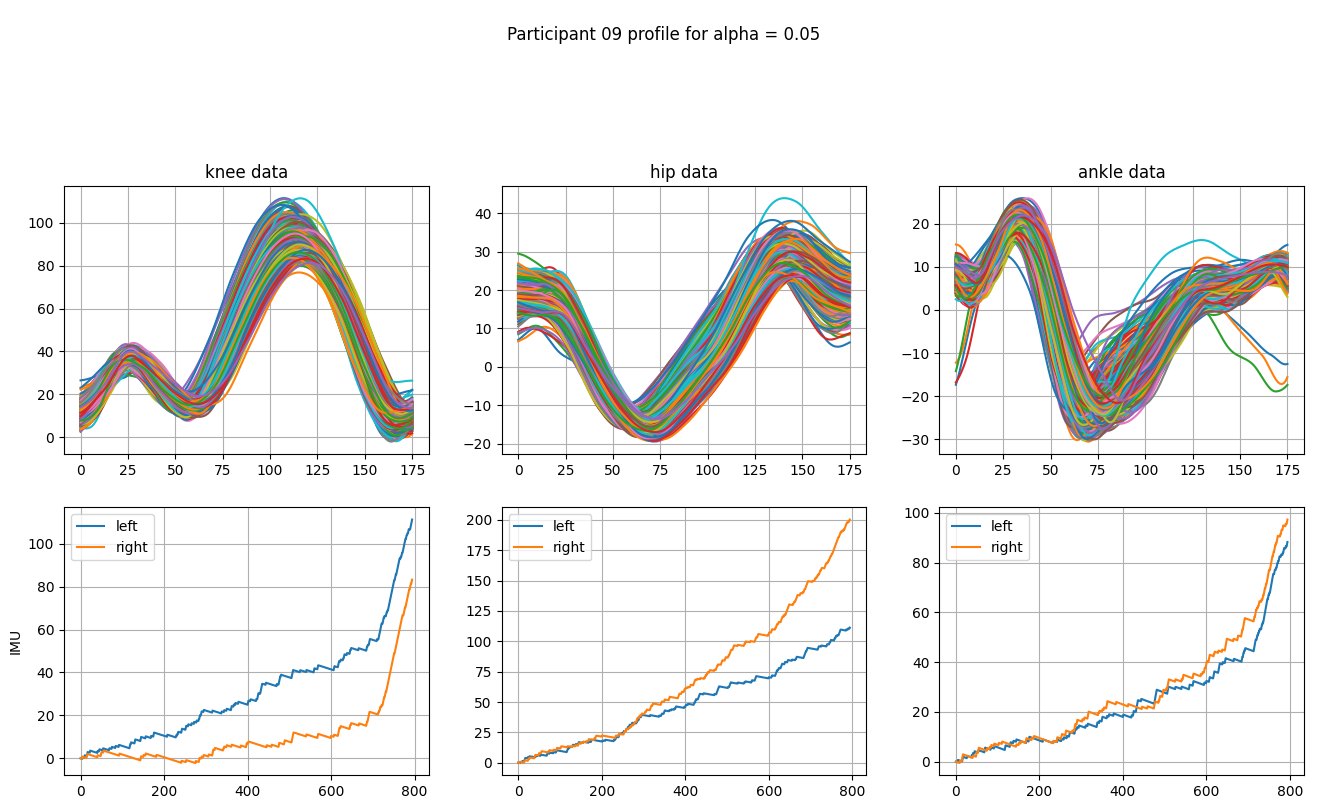}
    \caption{Upper panel: Stacked curves over the course of the run. Left panel: Martingale trajectories for runner $R_5$ for IMU knee, hip and ankle angles (from left to right). }
    \label{fig:R09alljoints}
\end{figure}

\vspace{0.25cm}
For runner $R_5$ on the other hand, we obtain trajectories matching exactly what is sketched in Figure \ref{fig:martingale_phases} for all joints (Figure \ref{fig:R09alljoints}) and, most prominently for the knee, for all measurement systems. Especially the onset of fatigue is nicely pronounced by the onset of a steep increase shortly after time point 700.  Again, the results for runner $R_5$  are stable also with respect to different  parameter values chosen.\\

\begin{figure}[h]
    \centering
    \includegraphics[width=0.95\linewidth, height = 5cm]{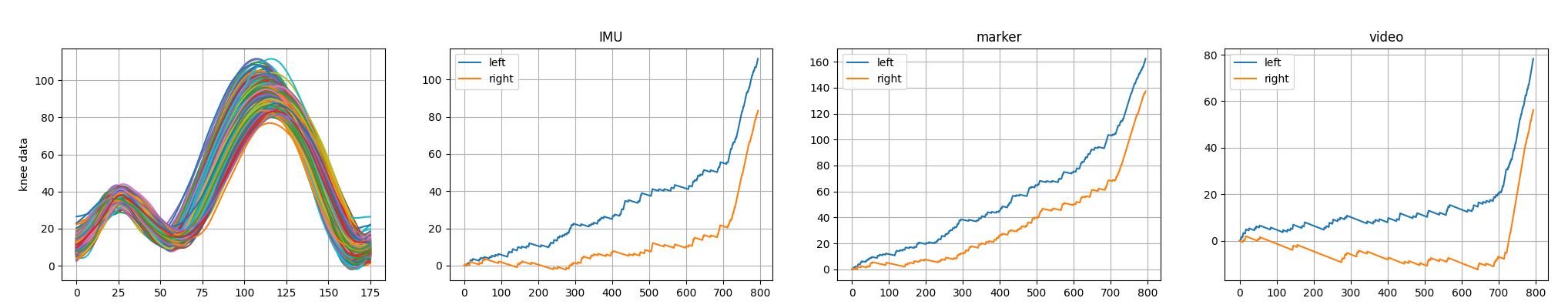}
    \caption{Martingale trajectories of knee angle data of runner $R_5$ measured with IMUs, optical markers and video (from left to right).}
    \label{fig:R09allsources}
\end{figure}

The information on movement characteristics of different individuals  gathered by this kind of analysis can be used to create a profile of a runner. This profile could indicate which joints are in general more indicative of change, whether change shows asymmetrically between left and right, or if fatigue increases or decreases variability of the distances to reference movement patterns. More results and discussions are provided in Section \ref{sec:different_systems} in the appendix.

\subsubsection{Pooling features from left and right joint angles yields more reliable results}
\label{pooled_data}

\begin{figure}
    \centering
    \includegraphics[height= 8.7cm, width = 12cm]{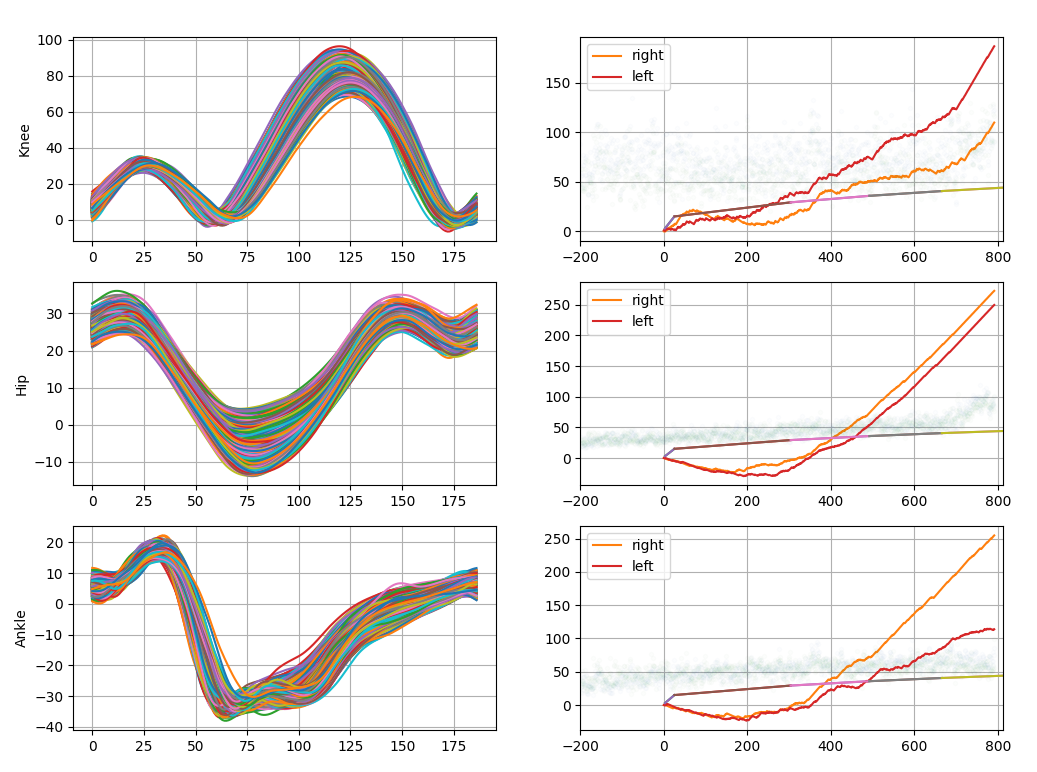}
    \caption{Martingales from the maximum of point data across systems for the right and left (right panel) joint angle data from indoor run for $R_1$. Plots for other runners may be found in \cref{fig:PooledPart04}, \cref{fig:PooledPart06}, \cref{fig:PooledPart07}, \cref{fig:PooledPart09}. }
    \label{fig:PooledSysI}
\end{figure}

\begin{figure}[h]
\includegraphics[width=\textwidth]{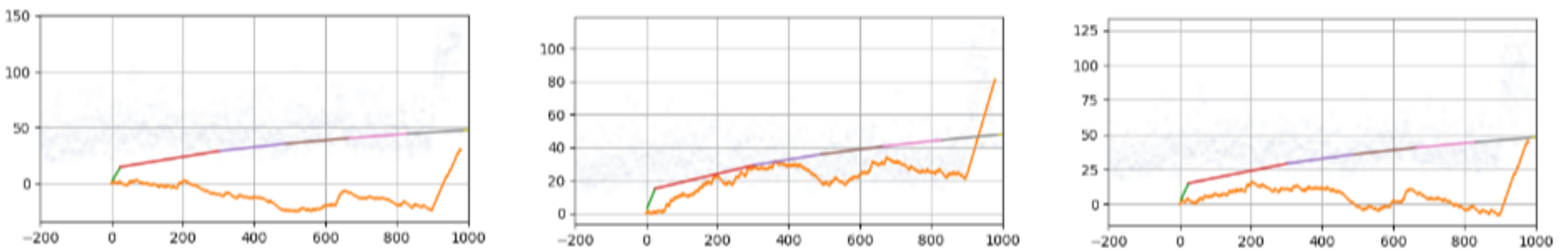}
\caption{Linear bounds and martingale trajectories for knee angle IMU data of runner $R_2$ based on ${D}_{i, \text{joint}}^{\text{max}},{D}_{i, \text{joint}}^{\text{min}},$ and ${D}_{i, \text{joint}}^{\text{ave}}$ (from left and right).}
\label{fig:R2leftrightknee}
\end{figure}

If recorded data are noisier than under ideal lab conditions, the martingale trajectories based on single features might be less stable as seen in Section \ref{application_section}.
From a practical point of view, it is therefore interesting to look at data integrated, e.g., across left and right joints. Recall that we have the following multivariate functional IMU data for samples $i = 1, \dots, N_k$ given by, 
\begin{align}
    Y_{i, \text{joint}} (t) = (Y_{i, \text{Left joint}} (t), Y_{i , \text{Right joint}}),  
\end{align}
where joint$\in\{\text{knee, hip,ankle}\}$.
Using the notation introduced in the box in Section \ref{sec:procedure}, we take the distances for these bivariate functional data as
\begin{align}
    D_{i, \text{joint}} = (D_{i, \text{Left joint}}, D_{i, \text{Right Joint}})
\end{align}
Simple ways of combining left and right distances are given by using maximum, minimum, or mean: 
\begin{align}
{D}_{i, \text{joint}}^{\max} = \max \{ D_{i, \text{Left joint}}, D_{i, \text{Right joint}} \},
\end{align}

\begin{align}
{D}_{i, \text{joint}}^{\min} = \min \{ D_{i, \text{Left joint}}, D_{i, \text{Right joint}} \},
\end{align}
or
\begin{align}
{D}_{i, \text{joint}}^{\text{ave}} = \frac{D_{i, \text{Left joint}}+D_{i, \text{Right joint}}}{2}.
\end{align}
For example, Figure \ref{fig:R2leftrightknee} shows martingale trajectories for knee angle IMU data of runner $R_2$ based on ${D}_{i, \text{joint}}^{\text{max}},{D}_{i, \text{joint}}^{\text{min}},$ and ${D}_{i, \text{joint}}^{\text{ave}}$. All trajectories are smoother and clearly agree that around time point $T=1100$, a drastic change occurs as is visible by the kink and detected with a delay for the minimum-aggregated data. This is also consistent with the reported perceived level of fatigue of this runner, which progresses from 7 (vigorous activity) to 9 (very hard activity close to total exhaustion).

\subsection{Conclusion }\label{results_data}
In this section, we summarize the main findings of our case study and provide ideas for future research. It is observed throughout our case study that a (local-)level of $\alpha = 0.22$ produces good results simultaneously for all data, especially joint angle data but also  force plate measurements, even in non-standard situations like  missing data, as studied in \cref{app: missing data} in the Appendix. \\
A first conclusion to be reached from our analysis of the pilot runner $R_0$ (see Figure \ref{fig:single_runner}) is that for well-processed data measured in a controlled (indoor) environment, results are extremely stable and change points are detected consistently at the same point throughout features (see Figure \ref{fig:single_runner}), which indicate an early change at the beginning of the fatiguing run. This may even be the case if much noisier data is used for the analysis as, e.g., the IMU hip data (Figure \ref{fig:IMUknee}). In order to increase stability, available data can be pooled in simple ways (see Section \ref{pooled_data}). Further, when monitoring joint angles, our analysis showed a strong variability in symmetry within individuals (Section \ref{sec:individual}), which can be reduced by a combination of different features in cases where the main focus is on fatigue detection. For simplicity and proof of concept, in this study we  considered aggregation of left and right joints (Section \ref{pooled_data}) and pooling joint ange data measured by different systems (Section \ref{sec:pooled_different_sources}), as these are automatically on the same scale. However, aggregation of a larger number of features promises even better results, but needs to be carefully designed and studied, potentially in future research. \\
\indent
It is to be also noted that even though upcrossing of the martingale may not always take place (see \cref{fig:icto_all}), the martingale process contains information in its trajectory, which can be used to analyze a run retrospectively and to extract individual movement characteristics (see Section \ref{sec:individual}). 
 In this context, it is worth mentioning that most trajectories created here suggest a certain stability and then show a steady increase (with CP at around 40\% of the run) from which an inference can be made that most local tests at these (later) time-points were rejected. We remark that qualitative information can be deciphered from the use of the sequential martingale statistic, yielding the positions of increase, where it follows that incoming data are progressively deviating from the reference. \\

    Our data analysis shows that the use of the martingale statistic is well suited for the analysis of sensor and related data from running. Future research and extension towards a quantitative analysis of the whole martingale trajectory along-with the development of an adjusted procedure incorporating a lower bound of the martingale (follows due to symmetry), is foreseen to be advantageous in multiple change detection. In our example, this would be particularly beneficial for the definition of different phases in a run with respect to levels of fatigue. From an application point of view, such a refined approach would allow a practitioner to track subtle changes, where even slight departures from reference movements during the course of the activity are monitored. This is expected to be particularly helpful in human movement analysis not just in fatigue detection and injury prevention but also in rehabilitation.\\

\noindent
\textbf{Ethics statement: } The local ethics committee (Ethical Committee EEMCS (EC-CIS), University of Twente, ref.: RP 2022-20) approved the experimental protocol of this study. The code used in this paper is available on git at \href{https://github.com/rupsabasu2020/sequence_test_marti}{rupsabasu2020/sequence\_test\_marti}. Data may be made available on request.\\

\noindent
\textbf{Acknowledgements:} 
The authors thank all members of the \textit{Sports, Data, and Interaction project}, in particular Dennis Reidsma and Jasper Reenalda, for helpful discussions and valuable feedback and  Robbert van Middelaar and Aswin Balasubramaniam for meticulously collecting, pre-processing and providing the data used in this work.  RB gratefully acknowledges the support by the Deutsche Forschungsgemeinschaft
(DFG) with project number 511905296 titled: \textit{Modeling functional time series with dynamic factor structures and points of impact. }Further, the authors report there are no competing interests to declare.

\bibliographystyle{apalike}

\begin{thebibliography}{}

\bibitem[Amin et~al., 1995]{SignStat}
Amin, R.~W., Jr., M. R.~R., and Saad, B. (1995).
\newblock Nonparametric quality control charts based on the sign statistic.
\newblock {\em Communications in Statistics-Theory and Methods},
  24(6):1597--1623.

\bibitem[Apte et~al., 2021]{apte2021biomechanical}
Apte, S., Prigent, G., St{\"o}ggl, T., Mart{\'\i}nez, A., Snyder, C.,
  Gremeaux-Bader, V., and Aminian, K. (2021).
\newblock Biomechanical response of the lower extremity to running-induced
  acute fatigue: a systematic review.
\newblock {\em Frontiers in physiology}, 12:646042.

\bibitem[Arias-Castro and Ying, 2019]{HCArias-Castro}
Arias-Castro, E. and Ying, A. (2019).
\newblock {Detection of sparse mixtures: higher criticism and scan statistic}.
\newblock {\em Electronic Journal of Statistics}, 13(1):208 -- 230.

\bibitem[Balsubramani, 2014]{balsubramani2014sharp}
Balsubramani, A. (2014).
\newblock Sharp finite-time iterated-logarithm martingale concentration.
\newblock {\em arXiv preprint arXiv:1405.2639}.

\bibitem[Balsubramani and Ramdas, 2015]{balasubramani2016}
Balsubramani, A. and Ramdas, A. (2015).
\newblock Sequential nonparametric testing with the law of the iterated
  logarithm.

\bibitem[Bhattacharya and Jr., 1981]{rankseq}
Bhattacharya, P.~K. and Jr., D.~F. (1981).
\newblock {A Nonparametric Control Chart for Detecting Small Disorders}.
\newblock {\em The Annals of Statistics}, 9(3):544 -- 554.

\bibitem[Borg, 1982]{borg1982psychophysical}
Borg, G.~A. (1982).
\newblock Psychophysical bases of perceived exertion.
\newblock {\em Medicine \& science in sports \& exercise}.

\bibitem[Brozek and Tiede, 1952]{HCfirst}
Brozek, J. and Tiede, K. (1952).
\newblock Reliable and questionable significance in a series of statistical
  tests.
\newblock {\em Psychological Bulletin}, 49:339–341.

\bibitem[Buist et~al., 2010]{runner_injury}
Buist, I., Bredeweg, S.~W., Lemmink, K. A. P.~M., van Mechelen, W., and
  Diercks, R.~L. (2010).
\newblock Predictors of running-related injuries in novice runners enrolled in
  a systematic training program: A prospective cohort study.
\newblock {\em The American Journal of Sports Medicine}, 38(2):273--280.

\bibitem[Carvajal-Rodriguez et~al., 2009]{HCapplied}
Carvajal-Rodriguez, A., de~Una-Alvarez, J., and Rolan-Alvarez, E. (2009).
\newblock A new multitest correction (sgof) that increases its statistical
  power when increasing the number of tests.
\newblock {\em BMC Bioinformatics}, 10:209.

\bibitem[Chakraborti et~al., 2004]{chakraborti2004class}
Chakraborti, S., Van~der Laan, P., and Van~de Wiel, M. (2004).
\newblock A class of distribution-free control charts.
\newblock {\em Journal of the Royal Statistical Society: Series C (Applied
  Statistics)}, 53(3):443--462.

\bibitem[Champ and Woodall, 1987]{champ1987exact}
Champ, C.~W. and Woodall, W.~H. (1987).
\newblock Exact results for shewhart control charts with supplementary runs
  rules.
\newblock {\em Technometrics}, 29(4):393--399.

\bibitem[Crosier, 1988]{crosier1988multivariate}
Crosier, R.~B. (1988).
\newblock Multivariate generalizations of cumulative sum quality-control
  schemes.
\newblock {\em Technometrics}, 30(3):291--303.

\bibitem[Crowder, 1989]{crowder1989design}
Crowder, S.~V. (1989).
\newblock Design of exponentially weighted moving average schemes.
\newblock {\em Journal of Quality technology}, 21(3):155--162.

\bibitem[Dette et~al., 2020]{dette2020functional}
Dette, H., Kokot, K., and Aue, A. (2020).
\newblock Functional data analysis in the banach space of continuous functions.

\bibitem[Donoho and Jin, 2004]{HC}
Donoho, D. and Jin, J. (2004).
\newblock {Higher criticism for detecting sparse heterogeneous mixtures}.
\newblock {\em The Annals of Statistics}, 32(3):962 -- 994.

\bibitem[Gao et~al., 2022]{gao2022effects}
Gao, Z., Fekete, G., Baker, J.~S., Liang, M., Xuan, R., and Gu, Y. (2022).
\newblock Effects of running fatigue on lower extremity symmetry among amateur
  runners: From a biomechanical perspective.
\newblock {\em Frontiers in Physiology}, page 1792.

\bibitem[Harato et~al., 2021]{harato2021fatigue}
Harato, K., Morishige, Y., Niki, Y., Kobayashi, S., and Nagura, T. (2021).
\newblock Fatigue and recovery have different effects on knee biomechanics of
  drop vertical jump between female collegiate and recreational athletes.
\newblock {\em Journal of Orthopaedic Surgery and Research}, 16:1--7.

\bibitem[Hido et~al., 2008]{hido2008unsupervised}
Hido, S., Id{\'e}, T., Kashima, H., Kubo, H., and Matsuzawa, H. (2008).
\newblock Unsupervised change analysis using supervised learning.
\newblock In {\em Advances in Knowledge Discovery and Data Mining: 12th
  Pacific-Asia Conference, PAKDD 2008 Osaka, Japan, May 20-23, 2008 Proceedings
  12}, pages 148--159. Springer.

\bibitem[H{\"o}rmann and Jammoul, 2022]{hormann2022consistently}
H{\"o}rmann, S. and Jammoul, F. (2022).
\newblock Consistently recovering the signal from noisy functional data.
\newblock {\em Journal of Multivariate Analysis}, 189:104886.

\bibitem[Horv{\'a}th and Kokoszka, 2012]{horvath2012inference}
Horv{\'a}th, L. and Kokoszka, P. (2012).
\newblock {\em Inference for functional data with applications}, volume 200.
\newblock Springer Science \& Business Media.

\bibitem[Howard et~al., 2020]{howard2020time}
Howard, S.~R., Ramdas, A., McAuliffe, J., and Sekhon, J. (2020).
\newblock Time-uniform chernoff bounds via nonnegative supermartingales.
\newblock {\em Probability Surveys}, 17:257--317.

\bibitem[Hsing and Eubank, 2015]{hsing2015theoretical}
Hsing, T. and Eubank, R. (2015).
\newblock {\em Theoretical foundations of functional data analysis, with an
  introduction to linear operators}, volume 997.
\newblock John Wiley \& Sons.

\bibitem[Janacek and Meikle, 1997]{janacek1997control}
Janacek, G. and Meikle, S. (1997).
\newblock Control charts based on medians.
\newblock {\em Journal of the Royal Statistical Society: Series D (The
  Statistician)}, 46(1):19--31.

\bibitem[Kawahara and Sugiyama, 2009]{kawahara2009change}
Kawahara, Y. and Sugiyama, M. (2009).
\newblock Change-point detection in time-series data by direct density-ratio
  estimation.
\newblock In {\em Proceedings of the 2009 SIAM international conference on data
  mining}, pages 389--400. SIAM.

\bibitem[Kirch and Weber, 2018]{kirch2018modified}
Kirch, C. and Weber, S. (2018).
\newblock Modified sequential change point procedures based on estimating
  functions.
\newblock {\em Electronic Journal of Statistics}, 12:1579--1613.

\bibitem[Lowry et~al., 1992]{lowry1992multivariate}
Lowry, C.~A., Woodall, W.~H., Champ, C.~W., and Rigdon, S.~E. (1992).
\newblock A multivariate exponentially weighted moving average control chart.
\newblock {\em Technometrics}, 34(1):46--53.

\bibitem[Lu and Chang, 2012]{lu2012biomechanics}
Lu, T.-W. and Chang, C.-F. (2012).
\newblock Biomechanics of human movement and its clinical applications.
\newblock {\em The Kaohsiung journal of medical sciences}, 28:S13--S25.

\bibitem[Montgomery, 2020]{montgomery2020introduction}
Montgomery, D.~C. (2020).
\newblock {\em Introduction to statistical quality control}.
\newblock John Wiley \& Sons.

\bibitem[Morin et~al., 2011]{morin2011changes}
Morin, J.-B., Samozino, P., and Millet, G.~Y. (2011).
\newblock Changes in running kinematics, kinetics, and spring-mass behavior
  over a 24-h run.
\newblock {\em Medicine \& Science in Sports \& Exercise}, 43(5):829--836.

\bibitem[Nielsen et~al., 2012]{Nielsen}
Nielsen, R., Buist, I., Sørensen, H., Lind, M., and Rasmussen, S. (2012).
\newblock Training errors and running related injuries: a systematic review.
\newblock {\em Int J Sports Phys Ther.}, 7(9):58--75.

\bibitem[Oskiper and Poor, 2002]{oskiper2002online}
Oskiper, T. and Poor, H.~V. (2002).
\newblock Online activity detection in a multiuser environment using the matrix
  cusum algorithm.
\newblock {\em IEEE Transactions on Information Theory}, 48(2):477--493.

\bibitem[Otto and Breitung, 2022]{otto2022backward}
Otto, S. and Breitung, J. (2022).
\newblock Backward cusum for testing and monitoring structural change with an
  application to covid-19 pandemic data.
\newblock {\em Econometric Theory}, pages 1--34.

\bibitem[Postema et~al., 1997]{postema1997energy}
Postema, K., Hermens, H.~J., De~Vries, J., Koopman, H.~F., and Eisma, W.
  (1997).
\newblock Energy storage and release of prosthetic feet part 1: Biomechanical
  analysis related to user benefits.
\newblock {\em Prosthetics and Orthotics International}, 21(1):17--27.

\bibitem[Qiu, 2017]{qiu2017statistical}
Qiu, P. (2017).
\newblock Statistical process control charts as a tool for analyzing big data.
\newblock {\em Big and Complex Data Analysis: Methodologies and Applications},
  pages 123--138.

\bibitem[Qiu and Hawkins, 2001]{qiu2001rank}
Qiu, P. and Hawkins, D. (2001).
\newblock A rank-based multivariate cusum procedure.
\newblock {\em Technometrics}, 43(2):120--132.

\bibitem[Qiu and Hawkins, 2003]{qiu2003nonparametric}
Qiu, P. and Hawkins, D. (2003).
\newblock A nonparametric multivariate cumulative sum procedure for detecting
  shifts in all directions.
\newblock {\em Journal of the Royal Statistical Society: Series D (The
  Statistician)}, 52(2):151--164.

\bibitem[Robbins and Siegmund, 1970]{robbins1970boundary}
Robbins, H. and Siegmund, D. (1970).
\newblock Boundary crossing probabilities for the wiener process and sample
  sums.
\newblock {\em The Annals of Mathematical Statistics}, pages 1410--1429.

\bibitem[Romano et~al., 2021]{romano2021fast}
Romano, G., Eckley, I., Fearnhead, P., and Rigaill, G. (2021).
\newblock Fast online changepoint detection via functional pruning cusum
  statistics.
\newblock {\em arXiv preprint arXiv:2110.08205}.

\bibitem[Saarakkala et~al., 2020]{Saarakkala}
Saarakkala, S., Taborri, J., Keogh, J., Kos, A., Santuz, A., Santuz, A., Umek,
  A., Urbanczyk, C., van~der Kruk, E., and Rossi, S. (2020).
\newblock Sport biomechanics applications using inertial, force, and emg
  sensors: A literature overview.
\newblock {\em Applied Bionics and Biomechanics}, 2020:18 pages.
\newblock Article ID 2041549.

\bibitem[Scheerder et~al., 2015]{running_popularity}
Scheerder, J., Breedveld, K., and Borgers, J. (2015).
\newblock Who {Is} {Doing} a {Run} with the {Running} {Boom}?
\newblock In {\em Running across {Europe}: {The} {Rise} and {Size} of {One} of
  the {Largest} {Sport} {Markets}}, pages 1--27. Palgrave Macmillan UK, London.

\bibitem[Schepers et~al., 2018]{schepers2018xsens}
Schepers, M., Giuberti, M., Bellusci, G., et~al. (2018).
\newblock Xsens mvn: Consistent tracking of human motion using inertial
  sensing.
\newblock {\em Xsens Technol}, 1(8).

\bibitem[Shafer and Vovk, 2005]{shafer2005probability}
Shafer, G. and Vovk, V. (2005).
\newblock {\em Probability and finance: it's only a game!}, volume 491.
\newblock John Wiley \& Sons.

\bibitem[Shewart, 1931]{shewhart1929control}
Shewart, W. (1931).
\newblock Economic control of quality of manufactured product.
\newblock {\em Bull. Amer. Soc. Qual. Control}.

\bibitem[Siegmund and Venkatraman, 1995]{siegmund1995using}
Siegmund, D. and Venkatraman, E. (1995).
\newblock Using the generalized likelihood ratio statistic for sequential
  detection of a change-point.
\newblock {\em The Annals of Statistics}, pages 255--271.

\bibitem[Tam et~al., 2017]{fatigue_factor}
Tam, N., Coetzee, D.~R., Ahmed, S., Lamberts, R.~P., Albertus-Kajee, Y., and
  Tucker, R. (2017).
\newblock Acute fatigue negatively affects risk factors for injury in trained
  but not well-trained habitually shod runners when running barefoot.
\newblock {\em European Journal of Sport Science}, 17(9):1220--1229.

\bibitem[Tony~Cai et~al., 2011]{HCCai}
Tony~Cai, T., Jessie~Jeng, X., and Jin, J. (2011).
\newblock {Optimal Detection of Heterogeneous and Heteroscedastic Mixtures}.
\newblock {\em Journal of the Royal Statistical Society Series B: Statistical
  Methodology}, 73(5):629--662.

\bibitem[Van~Gent et~al., 2007]{van2007incidence}
Van~Gent, R., Siem, D., van Middelkoop, M., Van~Os, A., Bierma-Zeinstra, S.,
  and Koes, B. (2007).
\newblock Incidence and determinants of lower extremity running injuries in
  long distance runners: a systematic review.
\newblock {\em British journal of sports medicine}, 41(8):469--480.

\bibitem[Wang et~al., 2016]{wang2016functional}
Wang, J.-L., Chiou, J.-M., and M{\"u}ller, H.-G. (2016).
\newblock Functional data analysis.
\newblock {\em Annual Review of Statistics and its application}, 3:257--295.

\bibitem[Wen et~al., 2021]{wen2021change}
Wen, D., Huang, X., Bovolo, F., Li, J., Ke, X., Zhang, A., and Benediktsson,
  J.~A. (2021).
\newblock Change detection from very-high-spatial-resolution optical remote
  sensing images: Methods, applications, and future directions.
\newblock {\em IEEE Geoscience and Remote Sensing Magazine}, 9(4):68--101.

\bibitem[Yoshioka et~al., 2009]{yoshioka2009biomechanical}
Yoshioka, S., Nagano, A., Hay, D.~C., and Fukashiro, S. (2009).
\newblock Biomechanical analysis of the relation between movement time and
  joint moment development during a sit-to-stand task.
\newblock {\em Biomedical engineering online}, 8(1):1--9.

\bibitem[Zandbergen et~al., 2023]{ZANDBERGEN202360}
Zandbergen, M.~A., Marotta, L., Bulthuis, R., Buurke, J.~H., Veltink, P.~H.,
  and Reenalda, J. (2023).
\newblock Effects of level running-induced fatigue on running kinematics: A
  systematic review and meta-analysis.
\newblock {\em Gait \& Posture}, 99:60--75.

\bibitem[Zandbergen et~al., 2022]{zandbergen2022drift}
Zandbergen, M.~A., Reenalda, J., van Middelaar, R.~P., Ferla, R.~I., Buurke,
  J.~H., and Veltink, P.~H. (2022).
\newblock Drift-free 3d orientation and displacement estimation for
  quasi-cyclical movements using one inertial measurement unit: Application to
  running.
\newblock {\em Sensors}, 22(3):956.

\end{thebibliography}

\newpage
\normalsize
\bigskip
\begin{center}
{\large\bf Appendix }
\end{center}
\vspace{0.5cm}
This section of the manuscript consists of all the simulations and the results therein in \cref{simulation_section}. This includes the change-point detection in various scenarios and the comparison of the methods presented in this manuscript with a standard CUSUM- detector statistic. This is followed by the \cref{Fig_and_tab_section} which consists of all the figures and tables from the main part of the manuscript, comprising of the results of the data analysis. Finally the proofs of \cref{thm:BoundLarget} and \cref{lemma:gamma1} are provided in 
\cref{sec:proofs}.

\section{Simulations}\label{simulation_section}

In this section, we evaluate our methods on simulated data. This is done in two settings to   understand (1) the qualitative estimation of the change point with respect to the true change point in various combinations of distributions and (2)   the impact of the local-level $\alpha$ and global-level $\delta$ in the sequential testing scenario and finally (3) a comparison in change point detection with respect to a standard CUSUM-like detector  statistic. The mathematical model for the simulated data is as follows. We choose a sequence of points $X_1, X_2, \dots, X_n\in \R$ such that an underlying model ensures exactly one  \textit{true} change point.  A martingale is then constructed such that, 
\begin{align}\label{eq:martingale_supplement}
    M_t = \sum_{i=1}^t \mathbbm{1}\{ X_i > \gamma_{1-\alpha}\}- t\alpha,
\end{align}
where $\alpha$ and $\gamma_{1-\alpha}$ are such that
$P(X_i>\gamma_{1-\alpha})=\alpha$ before the change point and $P(X_i>\gamma_{1-\alpha})>\alpha$  after the change point.
 In what follows, $\gamma_{1-\alpha}$ is the estimated $(1-\alpha)\times100\%$-quantile from the first $10\%$ of data points in the sequence $X_i$, as described in the box in Section \ref{sec:procedure}. Naturally, this is only possible in a retrospective setting, at least until the required initial fraction of the data is available. In a sequential setting, an appropriate choice of $\gamma_{\alpha}$ 
 can be done on prior data or based on a pre-determined number of initial data points.
 Further, the martingale bounds are then constructed with this choice of $\alpha$ along with the other essential parameters.

\subsection{Change point detection in different scenarios }
\begin{figure}[h]
    \centering
    \includegraphics[width=0.9\textwidth]{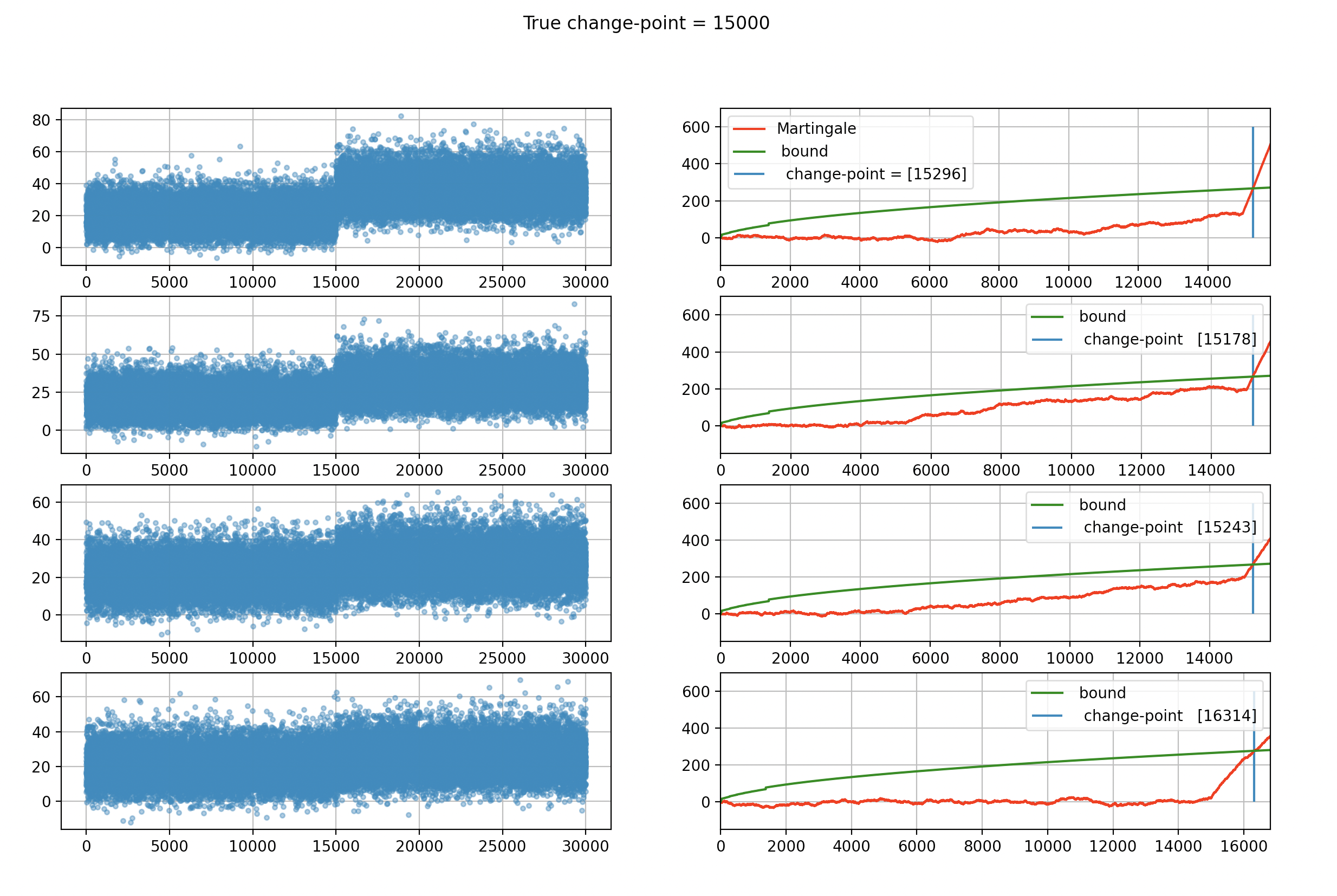}
    \caption{Left: Exemplary simulated data following model \eqref{model_sims} with increasing level of difficulty from top to bottom (for the exact degrees of freedom of the $\chi^2$-distributions see \cref{tab:setting_sims}). Left: Corresponding martingale and bounds  and exemplary martingale trajectories of the martingales \eqref{eq:martingale_supplement}.}
    \label{fig:Simulations1}
\end{figure}

In this section, we explore the abilities of our martingale statistic as a change point detector in dependence on the parameter $\alpha$. To this end, we choose simple models following distributions where it gets increasingly harder to detect the change point i.e., 
\begin{align}
    Y_{ij} = \chi_{kj}^2, \quad \quad i = 1, \dots, N,  \quad j=1,\ldots,4,\label{model_sims}
\end{align}
where the index $i$ denotes the samples drawn from a $\chi^2$-distribution with $k_j$-degrees of freedom in different scenarios numbered by the index $j$.  A visual representation of the simulated data is provided in  \cref{fig:Simulations1}, where $j$ varies along the left panels from top to bottom with increasing level of difficulty 
and $i$ corresponds to the $x-$axis of each panel i.e., the stream of data under study. Note that the fatigue protocol is designed to induce fatigue rather quickly and thus generates relatively short data streams. Here, we simulated longer data streams as compared to our data sets to evaluate the feasibility of the procedure in more realistic training settings. We work with the situation where the  true change point lies at the location $i = \frac{n}{2}$ and $j=1,\dots, 4$. When this is not the case, it is explicitly mentioned. The degrees of freedom for $\chi_{kj}^2$ for each $j$ are given in \cref{tab:setting_sims}. 
\subsection{Choice of $\alpha$}\label{sec:choice of alpha}
\begin{figure}[h]
\centering
\includegraphics[width=0.9\textwidth]{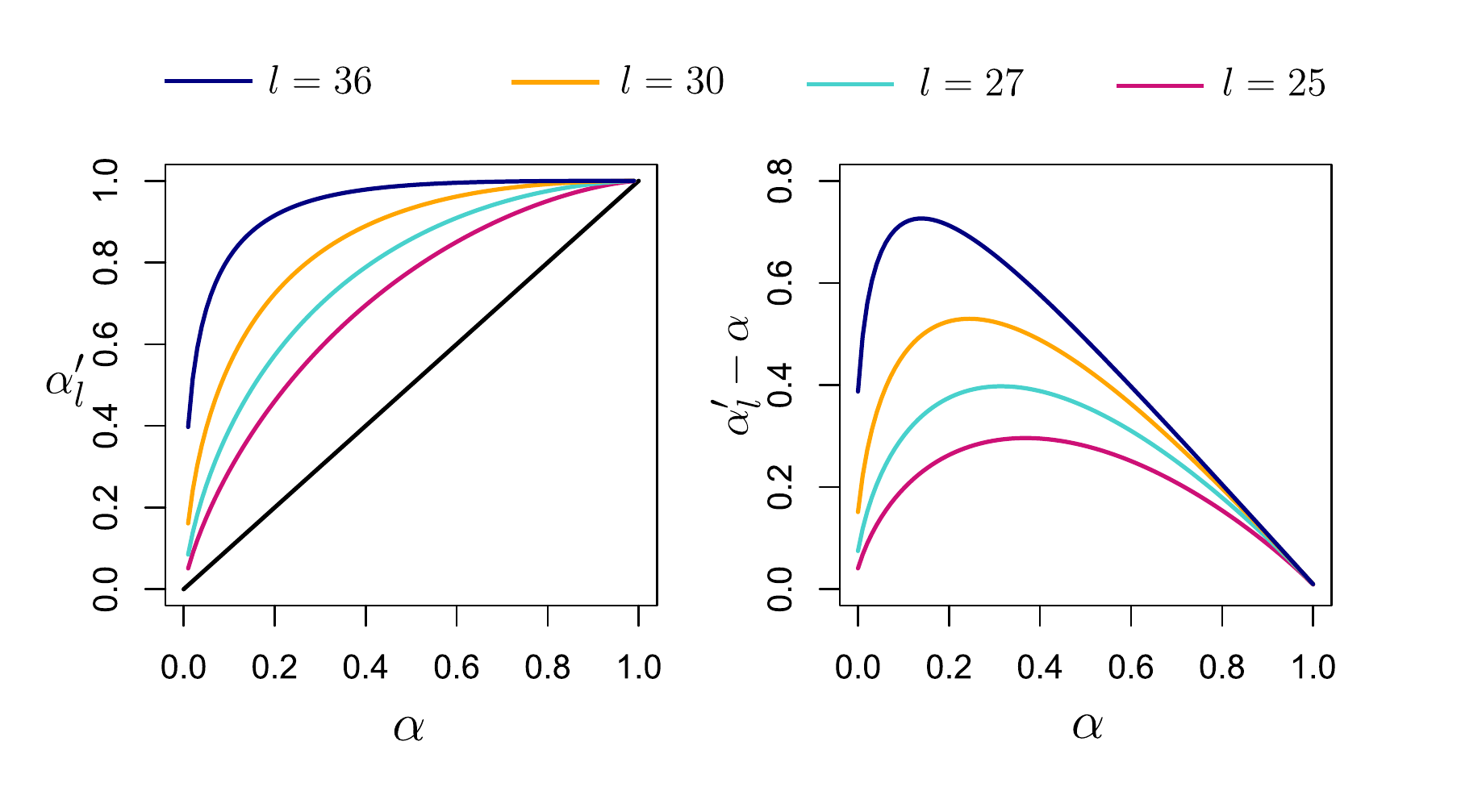}
\caption{Left: $\alpha$ such that $q_{\chi^2_{20},1-\alpha}$ is $(1-\alpha)$- quantile of the $\xi^2_{20}$ distribution against $\alpha_l'$ with  $q_{\chi^2_{20},1-\alpha}$ being the $(1-\alpha_l')$-quantile of the $\chi^2_{l}$-distribution for $l\in\{25,27,30,36\}$. Right: $\alpha$ versus $\alpha_l'-\alpha$, where $\alpha$ and $\alpha_l'$ are defined as on the left. }
\label{fig:choice of alpha}
\end{figure}

The parameter $\alpha$, which is the level in the local tests is merely a parameter in our monitoring context as the level of the overall procedure, i.e., the global martingale test, is given by $\delta$. Therefore, the value of $\alpha$ can be chosen such that the power of the procedure is maximized without changing the level. Recall that throughout our simulations $\Phi_i=1$ if $X_i>q_{\chi^2_{20},1-\alpha}$, where $q_{\chi^2_{20},1-\alpha}$ denotes the $1-\alpha$ quantile of the $\chi^2_{20}$-distribution. As soon as the distribution of the data changes towards larger values of degrees of freedom, $q_{\chi^2_{20},1-\alpha}$ is no longer the $1-\alpha$ quantile but the $1-\alpha_l'$ quantile of the $\chi^2_{l}$-distribution for $l\in\{25,27,30,36\}$. Notice that  $\alpha'_l>\alpha$ means that more data will fall above this quantity on average after the change. The left plot of Figure \ref{fig:choice of alpha} shows the values of $\alpha'_l$ against the values of $\alpha$. It shows that the largest net yield (in terms of expected additional rejections) can be obtained around $\alpha=0.5$. However, the number of additional expected points crossing the threshold is given by $(\alpha'-\alpha)\cdot t$, which reaches its maximum for smaller values of $\alpha$ and suggests a choice of $\alpha\approx0.2$ (see right panel of Figure \ref{fig:choice of alpha}).
Therefore, we now compute martingale bounds for $\alpha \in \{0.05, 0.25, 0.5 \}$ and discuss the change point detection results in Tables \ref{tab:sims_alpha_05}-\ref{tab:sims_alpha_005}, for the different local-levels $\alpha$ respectively. \\
We chose a simple change point model to provide a general proof of concept. The $\chi^2$-distribution was chosen as, in the case of the analysis of the joint angle data, we expect the the observed $L^2$-distances to follow $\chi^2$-type distributions. However, in contrast to model \eqref{model_sims} the real data typically show a more complicated behaviour of gradual and multiple changes and the presence of outliers, which is easily handled by our martingale approach, but more challenging for standard change point detection strategies.

\begin{table}[h]
    \centering
    \begin{tabular}{|c|c|c|c|}
    \hline
         & panel number & degrees of freedom (before cp) & degrees of freedom (after cp)  \\
         \hline
         & $j = 1$ & 20 & 36\\
         & $j = 2$ & 20 & 30\\
         & $j = 3$ & 20 & 27 \\
         & $j = 4$ & 20 & 25 \\
         \hline
    \end{tabular}
    \caption{Setting for running simulations, making change point detection increasingly harder by bringing degrees of freedom of the $\chi^2$ distributions closer to each other. (See \cref{fig:Simulations1} for four exemplary data sets and corresponding martingale trajectories).}
    \label{tab:setting_sims}
\end{table}

\begin{table}[h]
    \centering
    \begin{tabular}{|c|c|c|c|c|c|c|c|}
    \hline
          & $\alpha = 0.5, \delta = 0.1$  &  True CP & Est. CP & std dev. & std dev$^*$ &Number false positive \\
         \hline
         & $j = 1$ & 15000&  15556.598 & 800.347&  230.683 & 149\\
         \hline
         & $j= 2$ & 15000 & 15632.6317& 812.636 & 261.8841 & 150\\
         \hline
         &$j = 3$ & 15000 &  15769.035 & 869.8& 327.269 & 142\\
         \hline
         &$j=4$  & 15000 & 15999.196 & 991.52 &  428.065 & 169 \\
         \hline
    \end{tabular}
    \caption{Estimated average change points, standard deviations and number of false positives for $\alpha=0.5$ and $\delta=0.1$ based on 10000 simulation runs for $j=1$ (easiest) and $j =4$ (hardest) (see \cref{tab:setting_sims} for the exact parameter settings).}
    \label{tab:sims_alpha_05}
\end{table}
\begin{table}[h]
    \centering
    \begin{tabular}{|c|c|c|c|c|c|c|c|}
    \hline
          & $\alpha = 0.25, \delta = 0.1$  &  True CP & Est. CP & std dev. & std dev$^*$ & Number false positive \\
         \hline
         & $j=1$ & 15000&  15332.092 &   528.614 & 139.63 & 185\\
         \hline
         & $j=2$  & 15000 &  15439.279 & 566.91 & 187.589  & 185 \\
         \hline
         & $j=3$& 15000 &  15595.733 & 566.919 & 263.598  & 198\\
         \hline
         &$j=4$ & 15000 & 15850.357 &   738.53 & 381.25 & 178 \\
         \hline
    \end{tabular}
    \caption{Estimated average change points, standard deviations and number of false positives for $\alpha=0.25$ and $\delta=0.1$ based on 10000 simulation runs for $j=1$ (easiest) and $j =4$ (hardest) (see \cref{tab:setting_sims} for the exact parameter settings).}
    \label{tab:sims_alpha_025}
\end{table}

\begin{table}[h]
    \centering
    \begin{tabular}{|c|c|c|c|c|c|c|c|}
    \hline
    & $\alpha = 0.05, \delta = 0.1$ &  True CP & Est. CP & std. dev. & std. dev$^*$ & Number false positive \\
    \hline
    & $j=1$ & 15000&  15175.351 &  650.412 &  77.481 & 249 \\
    \hline
    & $j=2$ & 15000 & 15327.588 &  646.26 & 154.4 & 244 \\
    \hline
    & $j=3$ & 15000 &  15565.188 & 708.175 &  269.389 & 268 \\
    \hline
    & $j=4$  & 15000 &  15930.514 & 849.81 & 462.32& 246\\
    \hline
    \end{tabular}
    \caption{Estimated average change points, standard deviations and number of false positives for $\alpha=0.05$ and $\delta=0.1$ based on 10000 simulation runs for $j=1$ (easiest) and $j =4$ (hardest) (see \cref{tab:setting_sims} for the exact parameter settings).\cref{tab:setting_sims}.}
    \label{tab:sims_alpha_005}
\end{table}

\begin{table}[]
    \centering
    \begin{tabular}{|c|c|c|c|c|c|c|c|}
    \hline
          & local-, global-, level &  True CP & Est. CP & std. dev & std. dev$^*$ & Number false positive \\
          \hline
         & $\alpha$ = 0.2 & 15000 & 15845.278 & 920 &  384.462 & 317\\
         \hline
         & $\alpha$ = 0.22 & 15000 &  15850.392 & 467.875& 358.826 & 163\\
         \hline
         & $\alpha = 0.25, \delta = 0.1$& 15000 & 15832 & 466.671 & 372.87&  66\\
         \hline
         &$\alpha = 0.3, \delta = 0.1$ & 15000 & 15850.993 & 554.145 & 363.214 &   91\\
         \hline
         &$\alpha = 0.36, \delta = 0.1$  & 15000 &  15895 & 1027.735 & 382.822 & 159\\
         \hline
         &$\alpha = 0.45, \delta = 0.1$  & 15000 & 15916 & 811.16 & 416.933 & 134\\
         \hline
         &$\alpha = 0.5, \delta = 0.1$  & 15000 & 16004 & 808.657&  443.6289 & 69\\
         \hline
    \end{tabular}
    \caption{Narrowing down the choice of $\alpha$: Estimated average change points, standard deviations and number of false positives for $\alpha$ between $0.2$ and $0.5$ and $\delta=0.1$ based on 10000 simulation runs for the scenario given in $j= 4$ i.e.,  situations where the change point is hardest to detect.}
    \label{tab:diff_levels}
\end{table}

\subsubsection{Discussion : Result analysis} 
As mentioned before  change point detection is harder as we go from panel $j=1$ to $j=4$. Therefore it also makes sense that the (mean-) change point detection (along-with observed standard deviations,) makes better estimates for panel $j=1$ as compared to $j = 4$. Further, observing the results for the (local-) level $\alpha$,  it is seen that a small-local level $\alpha = 0.05$ produces somewhat acceptable estimates for change points only in the most obvious case of $j = 1$, albeit with alarge number of false positives (see \cref{tab:sims_alpha_005}) compared to larger values of $\alpha$. change point estimates for $j=4$, the case where change point is harder to detect and the situation we are most interested in; are best for $\alpha= 0.25, 0.5$. In the latter case, the observed standard deviation in the estimated change point location is quite high (possibly due to some very early false detections) while the overall number of false positives is relatively lower. When compared to the case of $\alpha = 0.25 $ a slightly better performance in terms of the standard deviations can be observed. The choice of the local level is therefore narrowed down to the range $\alpha \in (0.2, 0.5)$. In order to find an optimal value for change point detection, especially in the case of subtle change as in $j= 4$, we pursue simulations with focus on a grid of local levels $\alpha \in [0.2, 0.5]$ and present the results in \cref{tab:diff_levels}. We notice that in terms of having the least number of false positives, the values $\alpha = 0.25, 0.3, 0.5$ work the best. When considered together with the quality of the change point estimation, it appears only the first two are best performing for the (most interesting-) case $j= 4$.

\subsection{Comparison to CUSUM detector}
In this section, the objective is to compare change point detection via our methods as compared to a standard CUSUM statistic. We assume again that $Y_{ij}, \, \, i = 1, \dots, n,$ $j=1,\ldots,4,$ are independent samples with the underlying model as in \cref{model_sims} for $j=4$. A typical CUSUM statistic is given by, 
\begin{align}
    \widetilde{Y}_t = \sqrt{\frac{(s_2 -t)(t- s_1)}{s_2- s_1}} \big( \hat{\mu}_1 - \hat{\mu}_2\big); \quad \quad s_1 < s_2 -1 \quad \text{and}  \quad t= s_1 +1, \dots, s_2 -1,  \label{cusum_stat}
\end{align}
where $\hat{\mu}_1 = \frac{1}{t-s_1} \sum_{i= s_1 +1}^t X_i$ and $\hat{\mu}_2 = \frac{1}{s_2 -t} \sum_{i = t+1}^{s_2} X_i$. More specifically, the statistic $\widetilde{Y}_t$ scans through time  $t \in (s_1, s_2)$ and checks for the maximum deviation in means before and after a (scanning-) time point $t$. Further, the estimated change point is given by, 
\begin{align}
    t_{cp} = \argmax_t |\widetilde{Y}_t|. \label{cusum_cp}
\end{align}
In this way, the time point showing maximum deviation to the mean before and after it is considered the time point of change. Results of the analysis are presented in Tables \ref{tab:cusum_init}-\ref{tab:cusum_end}, where  the underlying model has a change point at an initial part, at the mid-point and towards the end respectively. 
As reflected in the precise estimation results 
of the  CUSUM detector, our model provides a very simple task for such classical approaches. In more realistic situations where change sets in gradually, multiple change points as well as outliers may be present, such methods are certainly more deeply affected than our martingale approach. Since this becomes evident in the comparison of the performances of both approaches in our data analysis, we refrain from  investigating this issue further in our simulations.\\

 Despite that, it is important to note the following, (a) Scan-statistic: this means computations are repeated over the scanning index $t$. This results in the CUSUM being more memory intensive as already discussed in \cref{Sec:Comparison}. (b) While online variations of the CUSUM-detector exist, it is important to note that our method only makes minimal assumptions on the underlying model and distribution of the data which is a clear advantage over CUSUM (see again, \cref{Sec:Comparison}).  (c) Finally, we note that our data analysis does not focus on recovering precise change points but rather on robust, widely applicable sequential analysis of CP-detection while controlling an overall error of false alarms, i.e.\ early detection.  In order to provide a more complete picture, Tables \ref{tab:cusum_init}- \ref{tab:cusum_end}, in addition to  the sample standard deviation (std dev) of the estimated change point location, we also list the standard deviation after removing the false positives (std dev$^*$). 
\begin{table}[]
    \centering
    \resizebox{\textwidth}{!}{
    \begin{tabular}{c|c|c|c|c|c|c|c|}
    \hline
         & (local-) (global-) level &True CP  & CUSUM  & Martingale CP (mean) & overall std & std$^*$ & false positives  \\
         \hline & $\alpha = 0.25, \delta = 0.1$ & 5000 & 5000.1484 &  5121.9596 & 25.07376    2538 &- & 0\\
         \hline
         & $\alpha = 0.1, \delta = 0.1$ & 5000 & 5000.1498 & 5121.1066 &   22.97& - & 0 \\
         \hline
         & $\alpha = 0.5, \delta = 0.1$ & 5000 & 5000.14 &   5127.3645  & 38.65057360   & - & 0 \\
         \hline
    \end{tabular}}
    \caption{Comparison to standard CUSUM statistic based on 10000 simulations of scenario $j= 4$, where the true CP lies at an initial part of a sequence of in total 30000 data points for different values of $\alpha$ and $\delta=0.1$.}
    \label{tab:cusum_init}
\end{table}

\begin{table}[]
    \centering
    \resizebox{\textwidth}{!}{
    \begin{tabular}{c|c|c|c|c|c|c|c|}
    \hline
         & (local-) (global-) level &True CP  & CUSUM & Martingale CP (mean) & overall std & std$^*$ & false positives  \\
         \hline & $\alpha = 0.25, \delta = 0.1$ & 15000 & 15000.1626&  15467.4207 &   954.0    1&  225.934&148\\
         \hline
         & $\alpha = 0.1, \delta = 0.1$ & 15000 &  15000.1432 &  15459.1764 & 864.26 & 236.69 &  148\\
         \hline
        & $\alpha = 0.5, \delta = 0.1$ & 15000 &  15000.1432& 15477.0819 & 876.81 & 228.02 & 163\\
         \hline
    \end{tabular}
  }
\caption{Comparison to standard CUSUM statistic based on 10000 simulations of scenario $j= 4$, where the true CP lies in the middle of a sequence of in total 30000 data points for different values of $\alpha$ and $\delta=0.1$.}
\label{tab:cusum_mid}
\end{table}

\begin{table}[]
    \centering
     \resizebox{\textwidth}{!}{
    \begin{tabular}{c|c|c|c|c|c|c|c|}
    \hline
         & (local-) (global-) level &True CP  & CUSUM & Martingale CP (mean) & overall std & std dev$^*$ & false positives  \\
         \hline & $\alpha = 0.25, \delta = 0.1$ & 25000 & 25000.1659 & 24781.1254 &   2279.    1626 & 157.053089 &  678\\
         \hline
         & $\alpha = 0.1, \delta = 0.1$ & 25000 & 25000.1454 & 24745.155 & 2393.28& 153.81 & 691\\
         \hline
         & $\alpha = 0.5, \delta = 0.1$ & 25000 &  25000.136 & 24783.6333 & 2265.55 & 169.579 & 687\\
         \hline
    \end{tabular}
    }
    \caption{Comparison to standard CUSUM statistic based on 10000 simulations of scenario $j= 4$, where the true CP lies at 5/6 of a sequence of in total 30000 data points for different values of $\alpha$ and $\delta=0.1$.}
    \label{tab:cusum_end}
\end{table}

\newpage

\section{Additional results from our case study in \cref{application_section}}
\label{Fig_and_tab_section}

In this section we provide additional  results from our case study in \cref{application_section} that are not contained in the main part of this paper.
We illustrate in Section \ref{app: missing data} that missing data are handled well by our method using force plate data of runner $R_0$. In Section \ref{contact_time} we demonstrate that non-standard features such as the contact time of the foot contain valuable information and produce nice check-mark type martingale trajectories.

\subsection{ Missing data are handled well by our method}
\label{app: missing data}
\begin{figure}[]
    \centering
    \includegraphics[scale = 0.23]{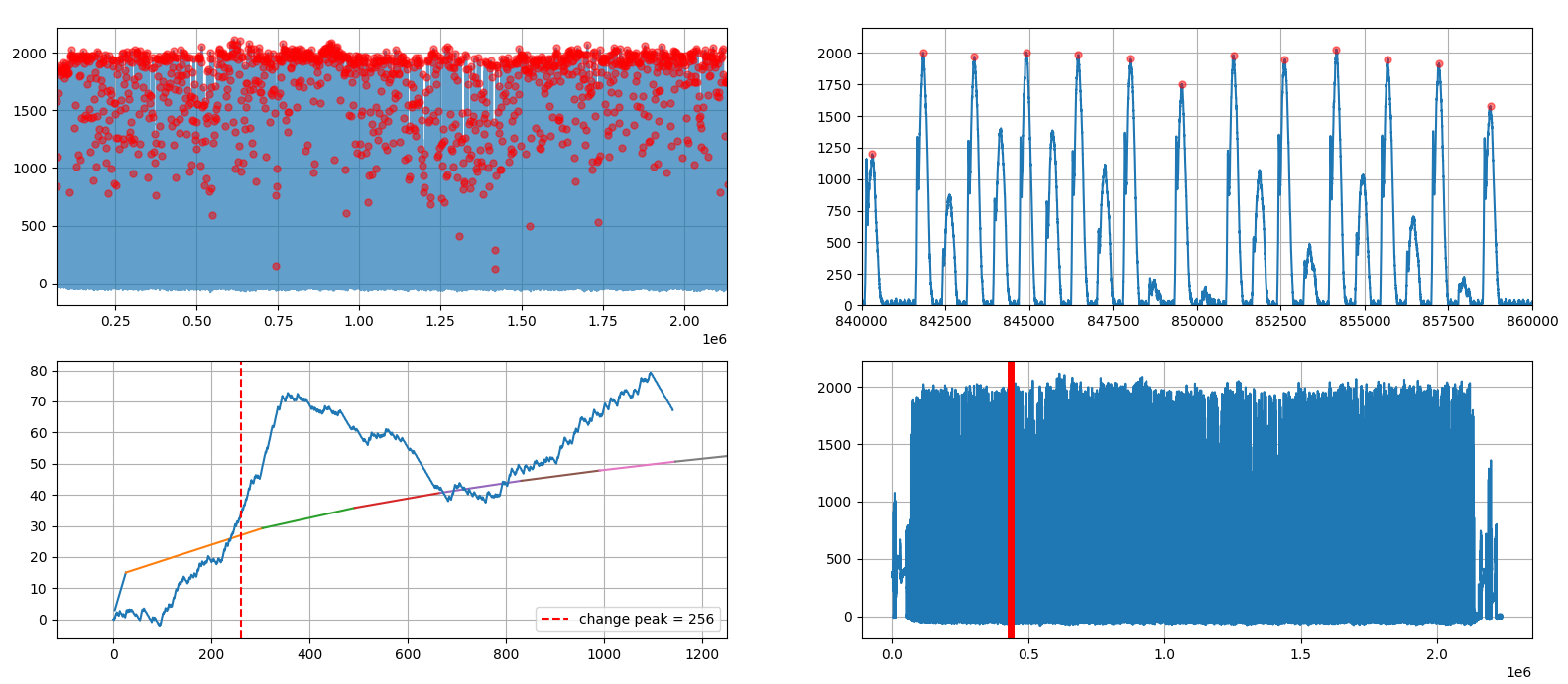}
    \caption{Force plate data from single runner $R_0$. (top left) Data from full run with peaks marked by red dots, (top right) zoomed in plot, (bottom left) martingale and corresponding bounds, (bottom right) change detected by dashed red line with respect to data from full run.}
    \label{fig:forceplate_init}
\end{figure}

In this application, we consider the data collected via force plates (FP) denoted by $ Y_{i,\text{FP}}^{R_0} (t), i = 1, \dots, N_{R_0, FP}$,  where the force plates are embedded in the treadmill. In particular, in this case, our point data is, 
\begin{align}
   \| Y_{i,\text{FP}}^{R_0} (t)\|_{\infty} = \sup_{t\in[0,1]} |Y_{i,\text{FP}}^{R_0} (t)|, \quad \quad  i = 1, \dots, N_{R_0, FP},  
\end{align}
i.e., the peak forces exerted by the foot during the times of contacts with the band of the treadmill and hence captured by the integrated plates underneath. In practise, these are the values corresponding to the take-off of the foot during the run. Further we mimic the possibility of missing values in the data. In practise, this would help in case a measurement system does not detect a few signals due to some external influence. In \cref{fig:forceplate_init}, we use the force plate data from the single runner $R_0$ from the pilot study, with the corresponding biomechanical joint angle data analysis shown in \cref{fig:single_runner}. As can be seen, the force plate data  detects changes at around 32\% of the run, which is slightly before than in the case of joint angles. This makes sense as sustained changes in forces applied by the foot would impact changes to biomechanical joint angle data in the case of activities like running. This propels the use of our methodologies as a potential application in the case of missing data. 

\subsection{Non-standard features may contain valuable information}
\label{contact_time}
\begin{figure}[h!]
\centering
\includegraphics[scale= 0.63]{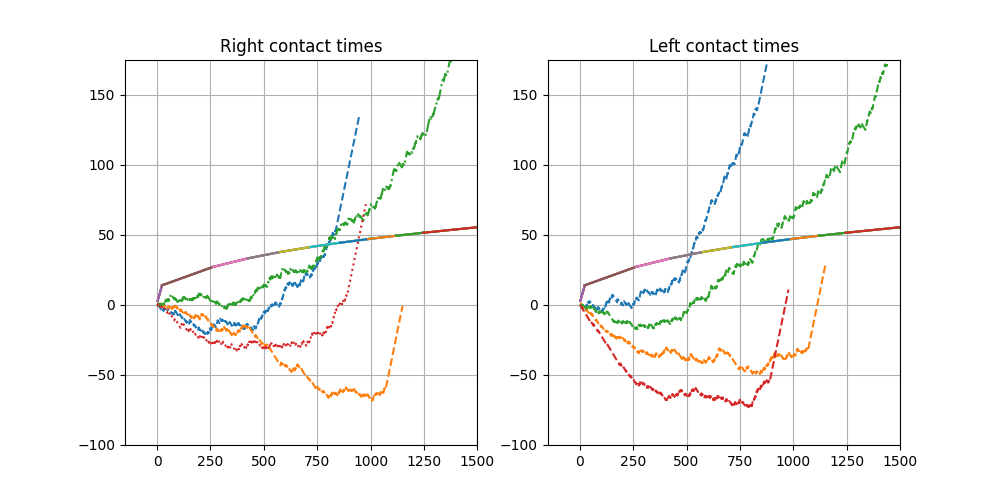}
\caption{Martingales for contact times of the foot with ground for select group of runners $R_1, R_2, R_4, R_5$. Data from $R_3$ was not available. }
\label{fig:icto_all}
\end{figure}

In this section, we analyse the time duration of contact of the left and right foot with the ground during the course of the indoor run, i.e., with the runner on the treadmill. 
The data stream for the contact time of the foot may be denoted as, \begin{align}
    Y_{i, \text{CT}}^{k}, \quad \quad i = 1, \dots, N_k, 
\end{align}
for the corresponding stride $i$ of the joint angle data. In \cref{fig:icto_all}, the analysis for 4 runners is shown. Quantitative analysis for multiple runners is hard due to the fact that data is only available in a laboratory setting for a few runners. Evidently, martingale upcrossing over the bounds is rare or much later in the dataset. However, as before, qualitative analysis shows the sudden onset of a  steep increase in all trajectories, reiterating that information on change is contained in the recorded contact times. More clearly than in all the other analyses, we see that initially, an adverse effect plays a role: there seems to be a more distinct stable phase before the onset of fatigue, resulting again in \textit{check-mark-shaped} martingale trajectory. In order to capture such a stable phase as well, a segmentation using first a lower martingale bound followed by the given upper bound could be applied resulting in more than one CP, also discussed  in \cref{results_data}. Finally, as an added remark, we note that current literature records conflicting results in terms of contact times (a) decreasing during fatigue \citep{morin2011changes} while  (b) another study by \citep{morin2011changes} records no changes with (c) \citep{apte2021biomechanical} recording a significant increase. Due to these difficulties, we propose the possibility of this application for our methodology without further discussion. 

\subsection{Additional results comparing different measurement systems}
As introduced in Section \ref{application_section}, our data set consists of data from a variety of sources, including inertial measurement units (IMUs) as well as recorded video data with and without optical motion capture system. Implementing such extensive measurement processes is of course not feasible in training practice. Therefore,  the question of comparability of measurements of the same quantity by different sources arises. In particular because  optical marker data is considered the golden standard in the biomechanics.
In this section we will compare and pool analyses based on measurements from the three  different systems. While the comparison does not provide a clear answer, we will show again in Section \ref{sec:pooled_different_sources} that pooling data by the approaches provided in Section \ref{pooled_data} significantly stabilizes results.

\subsubsection{Comparison of joint angle analysis based on different measurement systems}\label{sec:different_systems}
\begin{figure}[h]
    \centering
    \includegraphics[width=0.95\textwidth, height = 5cm]{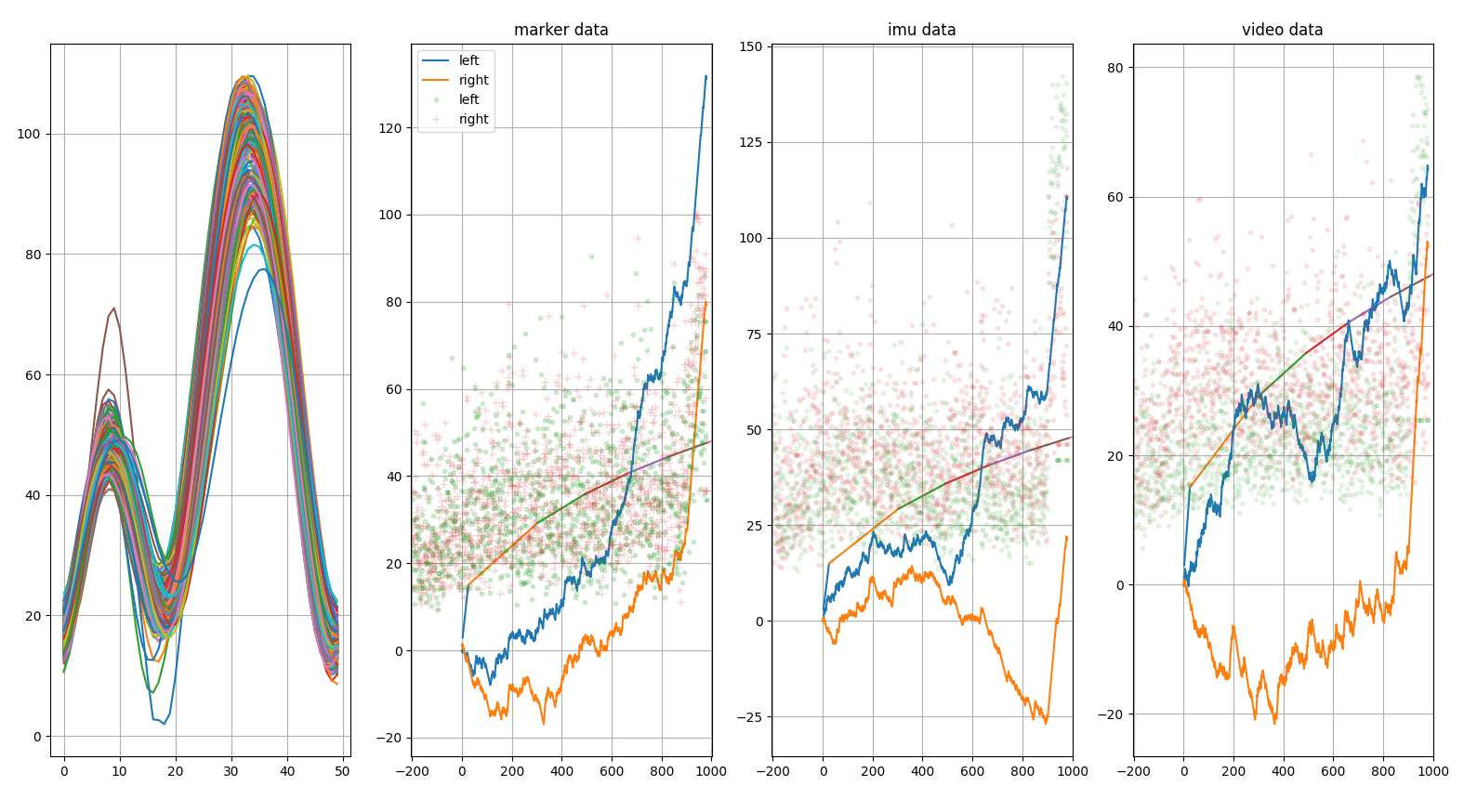}
    \caption{Knee angle data from $R_2$ different systems $s \in \{$marker, IMU, video$\}$. Potential benchmarking  of newer sources of data. Sample size of data is $ 1081$, length of training data is $m_0 = 200$.} 
    \label{fig:kneeBenchmark}
\end{figure}

\begin{figure}
    \centering
    \includegraphics[width=0.95\textwidth, height = 5cm]{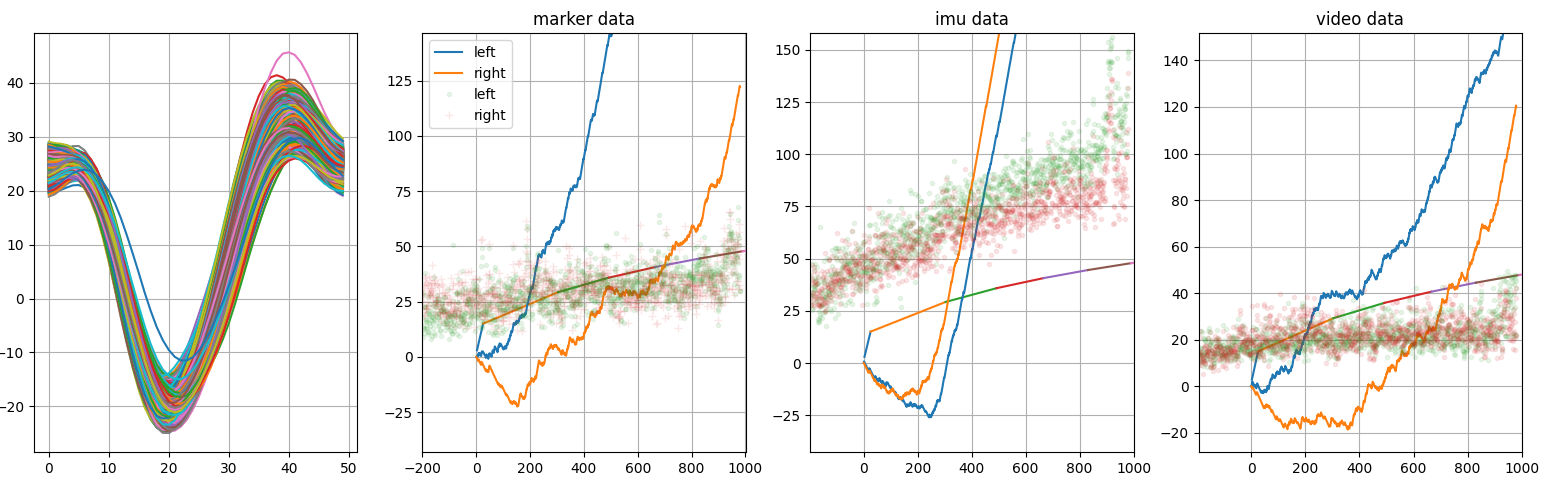}
    \caption{Hip angle data from runner $R_2$ different systems  $s \in \{$marker, IMU, video$\}$. Potential benchmarking  of newer sources of data. Sample size of data $N = 1081$, length of training data is $m_0 = 200$. Some plots for other runners are provided in \cref{fig:hipPart01}, \cref{fig:hipPart06}, \cref{fig:hipPart07}, \cref{fig:hipPart09}. }
    \label{fig:ehipBenchmark}
\end{figure}

   As a first example we look at the knee angle data of runner $R_2$ visualised in \cref{fig:kneeBenchmark} and given by, 
\begin{align}
    Y_{i, \text{knee}}^{R_2, \text{Sys}} (t), \quad \quad i = 1, \dots, N_{R_2}, 
\end{align}
where $N_{R_2} = 1081$
obtained simultaneously from three different sources $\text{Sys} \in \{$marker, IMU, video$\}$.   An interesting aspect, parallel to the pilot runner $R_0$ in \cref{joint_ang_single} is the \textit{check-mark} like pattern, where one can see an initial stable phase of the martingale with deviations from the reference data (with property $\mathcal P $) being observed more in later period of the run.  A runner specific- asymmetry is seen between the left and right knee angle data with the left knee angle martingales showing fatigue detection at about 62\% of the run from the marker and IMU systems.  Alternatively, the same runner shows slight different characteristics in the hip angle data $Y_{i, \text{hip}}^{R_2, \text{Sys}} (t), \, i = 1, \dots, N_{R_2}$  in  \cref{fig:ehipBenchmark}. 
While the \textit{check-mark} like pattern of the martingale is to be seen, a stark difference is how the martingale shows a steeper increase throughout which brings to attention that for this specific runner the adjustment of the hip during the course of the run is significantly higher, corroborating our findings from Section \ref{sec:individual} in the main part of the manuscript. \\


\begin{figure}
   \centering
    \includegraphics[width=0.95\textwidth, height = 5cm]{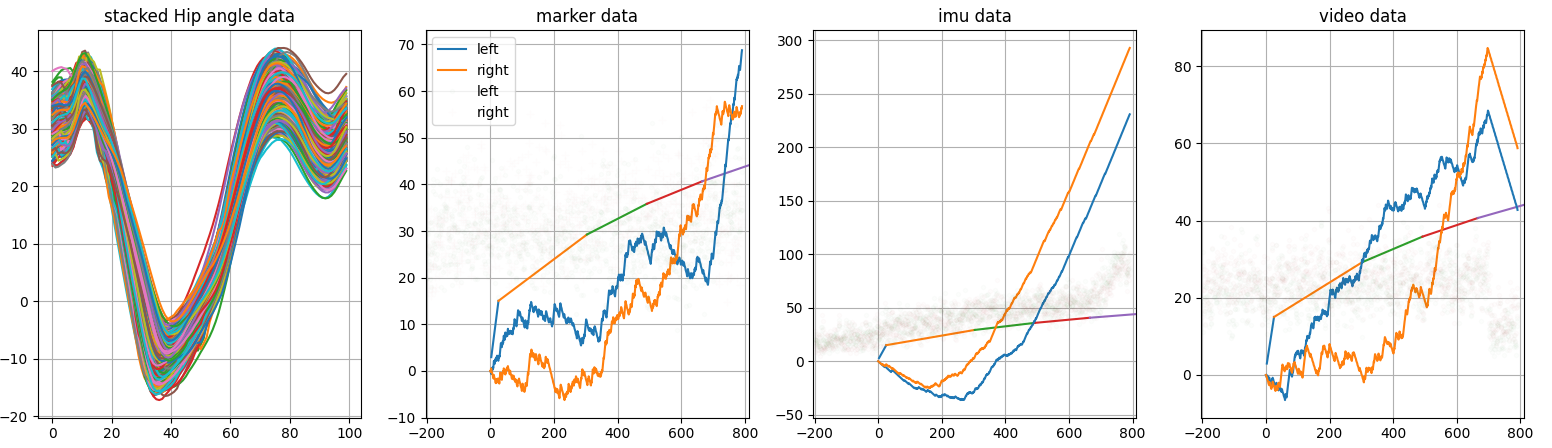}
    \caption{Hip angle data from $R_1$ different systems  $s \in \{$marker, IMU, video$\}$. Potential benchmarking  of newer sources of data. Sample size of data $N = 992$, length of training data is $n_0 = 200$.}
    \label{fig:hipPart01}
\end{figure}

\begin{figure}
\centering
    \includegraphics[width=0.95\textwidth, height = 5cm]{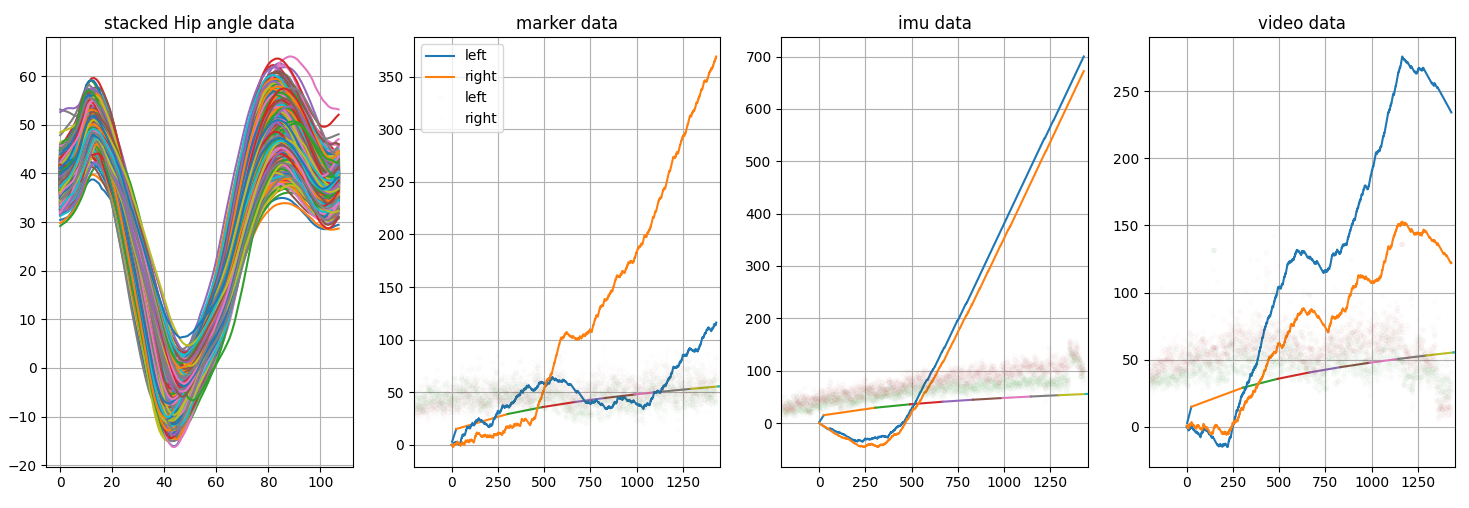}
    \caption{Hip angle data from $R_3$ different systems  $s \in \{$marker, IMU, video$\}$. Potential benchmarking  of newer sources of data. Sample size of data $N = 1631$, length of training data is $n_0 = 200$.}
    \label{fig:hipPart06}
\end{figure}

\begin{figure}
   \centering
    \includegraphics[width=0.95\textwidth, height = 5cm]{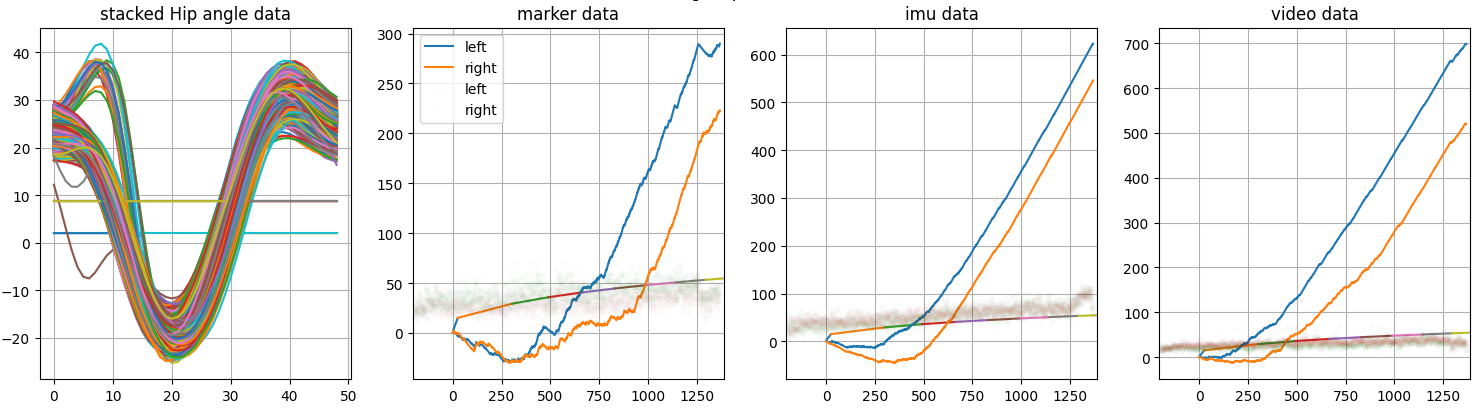}
    \caption{Hip angle data from $R_4$ different systems  $s \in \{$marker, IMU, video$\}$. Potential benchmarking  of newer sources of data. Sample size of data $N = 1569$, length of training data is $n_0 = 200$.}
    \label{fig:hipPart07}
\end{figure}

\begin{figure}
\centering
    \includegraphics[width=0.95\textwidth, height = 5cm]{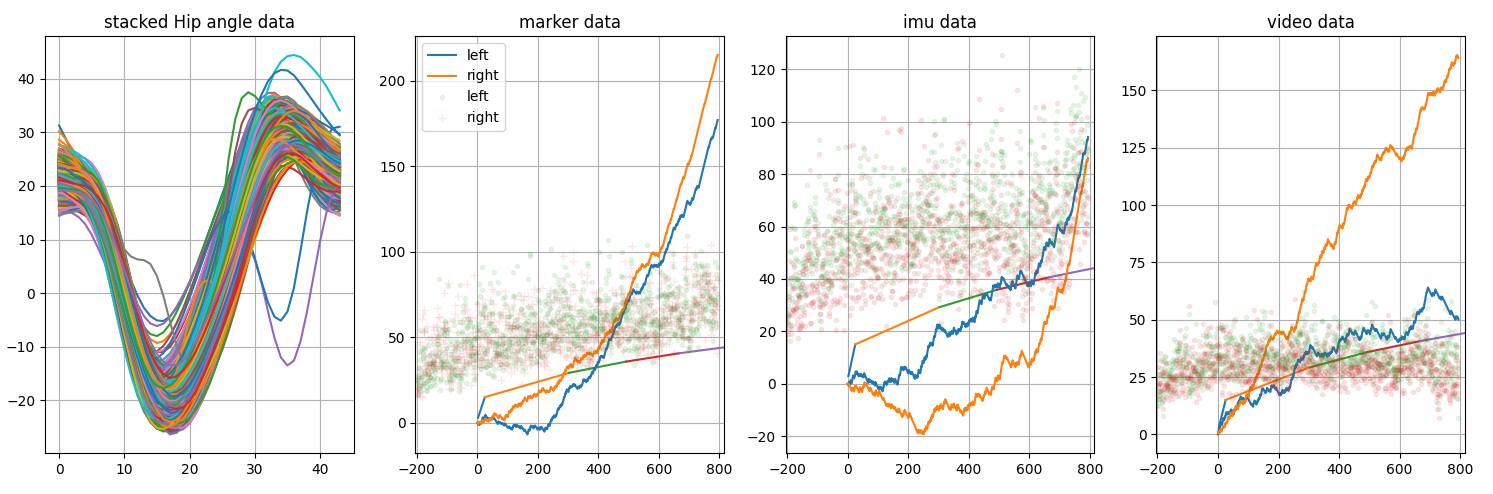}
    \caption{Hip angle data from $R_5$ different systems  $s \in \{$marker, IMU, video$\}$. Potential benchmarking  of newer sources of data. Sample size of data $N = 995$, length of training data is $n_0 = 200$.}
    \label{fig:hipPart09}
\end{figure}



\subsubsection{Pooling data from different measurement systems stabilizes results}\label{sec:pooled_different_sources}

From a practical point of view, it is interesting to look at data integrated across various systems. In our case for the indoor datasets, we have the following multivariate functional data for samples $i = 1, \dots, n$ given by, 
\begin{align}
    Y_{i, \text{RKnee}}^{\text{Sys}} (t) &= (Y_{i, \text{RKnee}}^{\text{IMU}} (t), Y_{i , \text{RKnee}}^{\text{marker}} (t)),  Y_{i , \text{RKnee}}^{\text{video}} (t)), \\
        Y_{i, \text{LKnee}}^{\text{Sys}} (t) &= (Y_{i, \text{LKnee}}^{\text{IMU}} (t), Y_{i , \text{LKnee}}^{\text{marker}} (t)),  Y_{i , \text{LKnee}}^{\text{video}} (t)), 
\end{align}

\begin{figure}
    \centering
    \includegraphics[height= 8.7cm, width = 12cm]{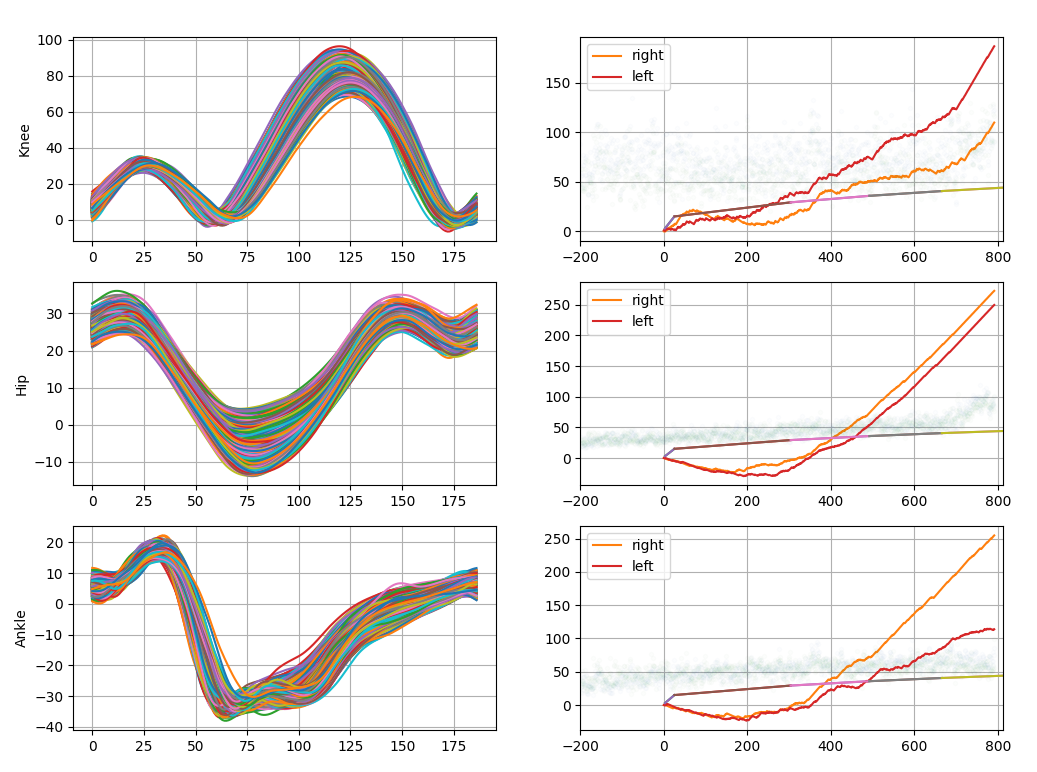}
    \caption{Martingales from the maximum of point data across systems for the right and left (right panel) joint angle data from indoor run for $R_1$. Plots for other runners may be found in \cref{fig:PooledPart04}, \cref{fig:PooledPart06}, \cref{fig:PooledPart07}, \cref{fig:PooledPart09}.}
    \label{fig:PooledSys}
\end{figure}

\vspace{10pt}
where the subscripts ``RKnee'' and ``LKnee'' stand for the right and left knee respectively and IMU, video and marker are the three different systems which measure the corresponding knee angles. Using the notation introduced in the box in Section \ref{sec:procedure}, we take the distances for the 3-variate functional data, $(Y_{i, \text{RKnee}}^{\text{IMU}} (t), Y_{i , \text{RKnee}}^{\text{marker}} (t)),  Y_{i , \text{RKnee}}^{\text{video}} (t)) $ as,
\begin{align}
    D_{i, \text{RKnee}}^{\text{Sys}} = (D_{i, \text{RKnee}}^{\text{IMU}}, D_{i, \text{RKnee}}^{\text{marker}}, D_{i, \text{RKnee}}^{\text{video}}).
\end{align}
The point data in this case for the construction of the martingale at local level $\alpha$ is given by 
\begin{align}
{D}_{i, \text{RKnee}}^{\max} = \max \{ D_{i, \text{RKnee}}^{\text{IMU}}, D_{i, \text{RKnee}}^{\text{marker}}, D_{i, \text{RKnee}}^{\text{video}} \},
\end{align}
which can of course be similarly done for the left knee. The result of this  feature selection across systems is shown in corresponding plot in \cref{fig:PooledSys}. This analysis is done for runner $R_1$ (also seen in \cref{fig:borg_single_runner}).  Comparing the analysis for this single runner in \cref{fig:borg_single_runner} and \cref{fig:PooledSys}, which is respectively for the right joint angle data and the integrated data across systems, we see that the statistic in case of the pooled data has a more clear interpretation of a stable phase in the beginning  and a progressive increase due to fatigue. 
We consider here aggregation using the maximum across systems, as it is known that increased muscle stiffness can cause reduced range of motion of lower extremity joints and hence a pronounced effect on the peak values. 

\begin{figure}
    \centering
    \includegraphics[width=0.7\textwidth]{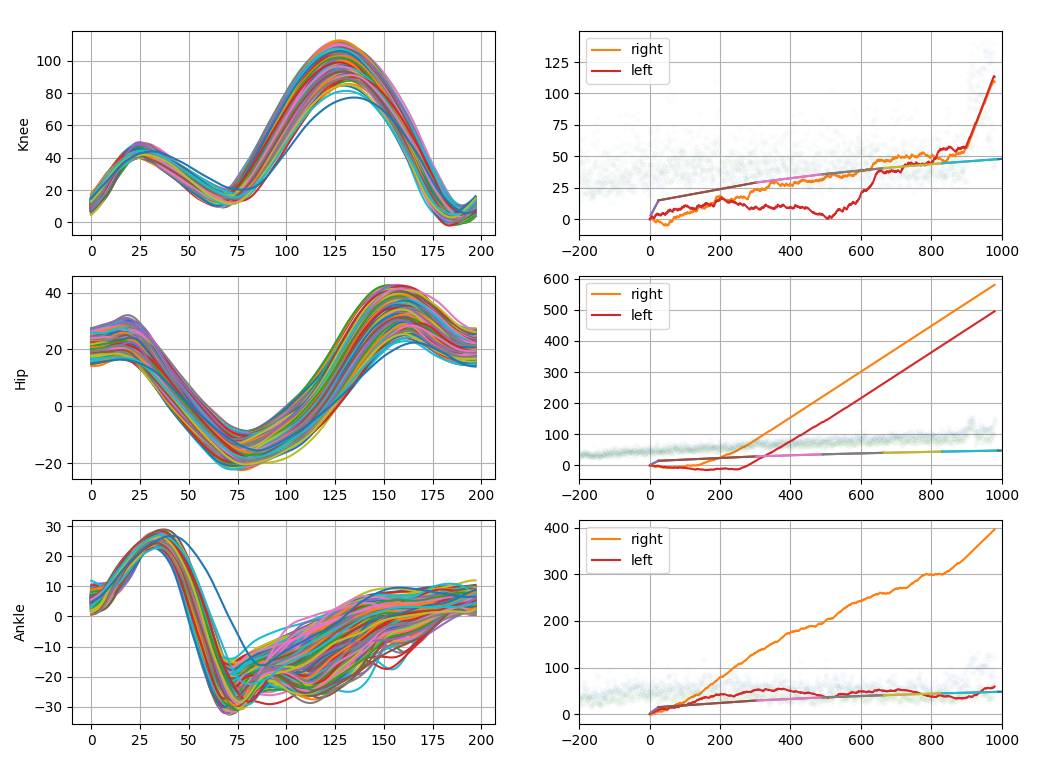}
    \caption{Martingales from the maximum of point data across systems for the right and left (right panel) joint angle data from indoor run for $R_2$.}
    \label{fig:PooledPart04}
\end{figure}

\begin{figure}
    \centering
    \includegraphics[width=0.7\textwidth]{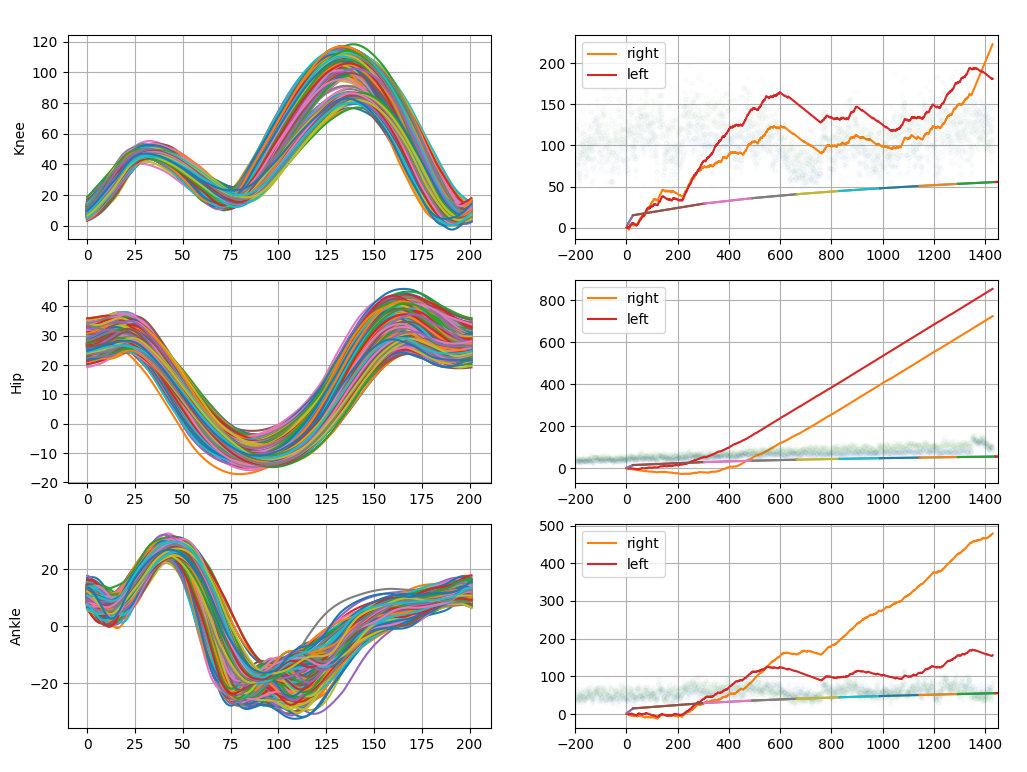}
    \caption{Martingales from the maximum of point data across systems for the right and left (right panel) joint angle data from indoor run for $R_3$.}
    \label{fig:PooledPart06}
\end{figure}

\begin{figure}
    \centering
    \includegraphics[width=0.7\textwidth]{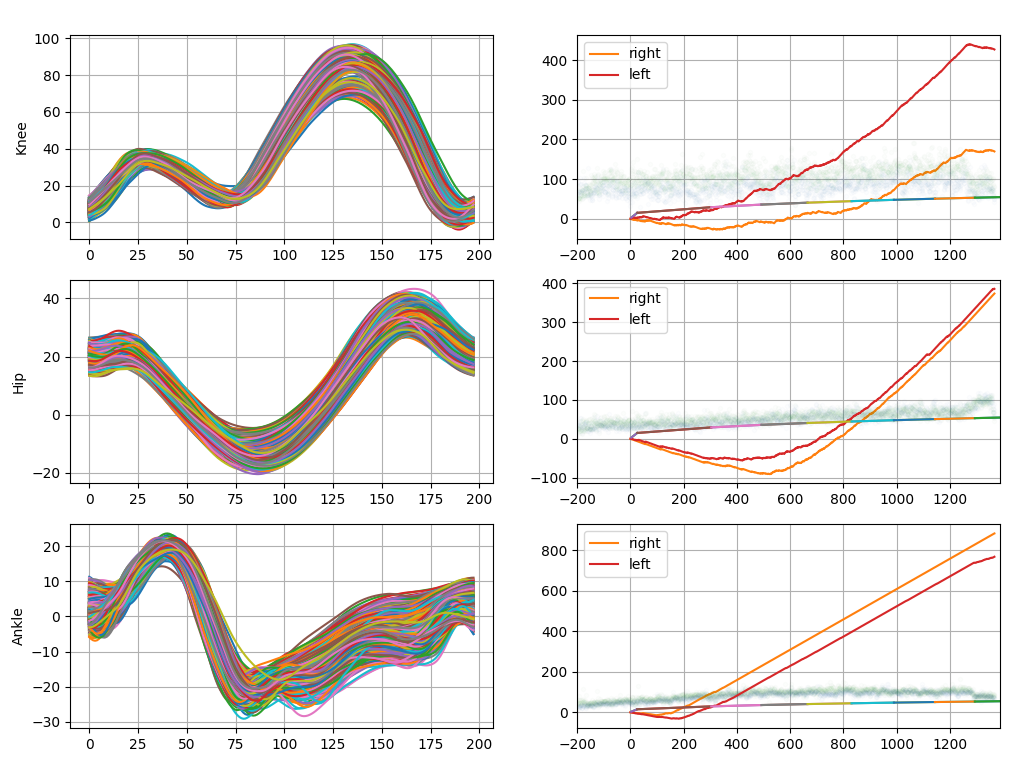}
    \caption{Martingales from the maximum of point data across systems for the right and left (right panel) joint angle data from indoor run for $R_4$.}
    \label{fig:PooledPart07}
\end{figure}

\begin{figure}
    \centering
    \includegraphics[width=0.7\textwidth]{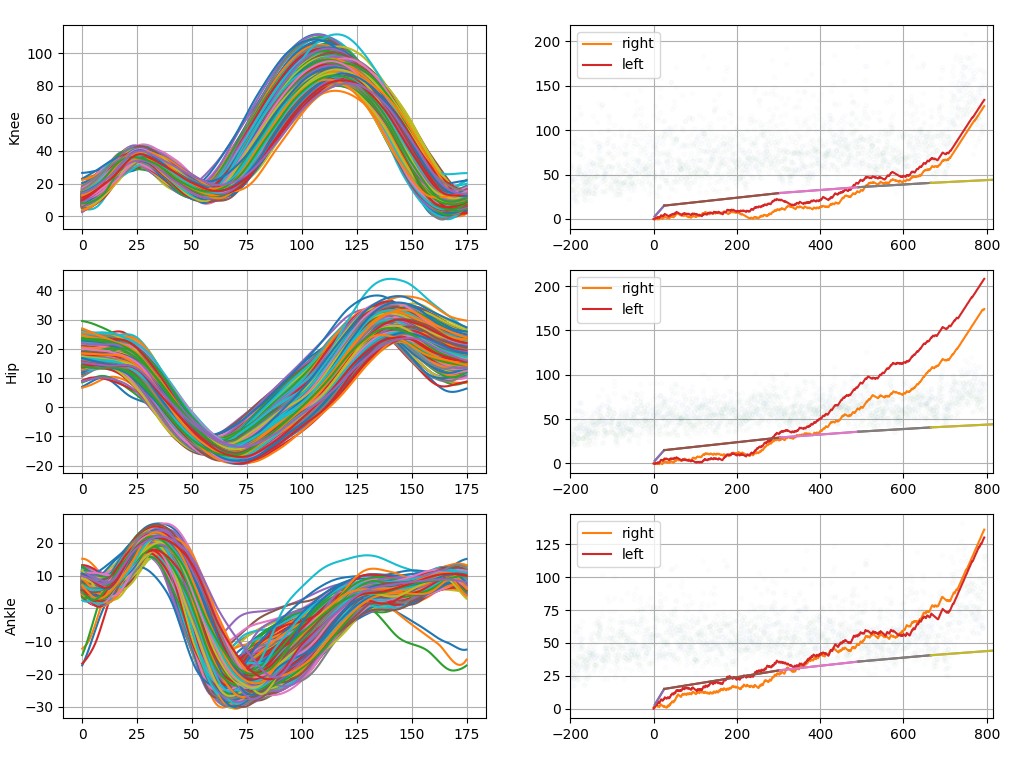}
    \caption{Martingales from the maximum of point data across systems for the right and left (right panel) joint angle data from indoor run for $R_5$.}
    \label{fig:PooledPart09}
\end{figure}

\newpage

\section{Statement and proofs of the general versions of the theoretical results}\label{sec:proofs}

In this section, we will state two theoretical results which will be the basis for different sequential procedures proposed in this work. In Theorem \ref{thm:BoundLarget}, we provide the time-uniform bound $\Gamma_t^{\text{LIL}}$ based on the law of the iterated logarithm, which is a refined version of a bound derived in \cite{balsubramani2014sharp}. This theorem is applicable only after a stopping time is reached, hereafter denoted by $s_{0,\text{LIL}}$ and is  applicable in a (large-) time regime i.e. for $t \geq s_{0,\text{LIL}}$. 

\begin{theorem}\label{thm:BoundLarget}
	Let, for $\alpha\in(0,1/2]$, the random variables $\mathbbm{1}\{\test_i=1\} \stackrel{\text{i.i.d}}{\sim}$Bin$(1,\alpha)$ and  $M_t=M_t(\alpha)$ be as in \cref{martingale_stat}.  For any $\delta\in(0,1/2]$, any $k\in(0,1)$, and $\kappa$ with
	\begin{align*}
	\kappa\geq\frac{\frac{1}{2}+\frac{1}{20e^{8}}-0.4\alpha+\max\{\frac{1}{6e^4}-0.1\alpha,0\}}{1-\alpha}\quad
	\text{and}\quad
	s_{0,\text{LIL}}=\left\lceil\frac{e^{4}(1+\sqrt{k})^2}{\kappa\alpha(1-\alpha)}\log\left(\frac{1}{\delta}\right)\right\rceil,
	\end{align*}
	it holds with probability at least $1-\delta$, for all $t\geq s_0$ simultaneously, that
	\begin{align*}
	M_t\leq \Gamma_t^{LIL},
	\end{align*}
	where
	\begin{align*}
	 \Gamma_t^{LIL}=
	\sqrt{\frac{4}{1-k}\kappa \alpha(1-\alpha)t \bigg(2 \log\log \frac{2\kappa \alpha(1-\alpha)t}{ (1-\sqrt{k})} + \log \frac{2}{\delta\log(\frac{1+\sqrt{k}}{1-\sqrt{k}})}}\bigg)\wedge\frac{2\kappa\alpha(1-\alpha)t}{e^2}\vee1.
	\end{align*}

\end{theorem}
The proof of the above theorem is quite technical and therefore deferred to Section \ref{proof_LIL} in the Appendix .
\begin{remark}
  The initial time $s_{0,\text{LIL}}$ in the general \cref{thm:BoundLarget} can easily reach values of $10,000$ and more, when (un-)desirable values of the parameters such as small values of $\kappa, \alpha$ are chosen.  As, on the other hand, a small value of $\kappa$  decreases the bound $\Gamma^{LIL}$ by a factor of $\sqrt{\kappa}$, whereas too large values render the bound useless in practice. We will, as a compromise, accept a larger starting point $s_{0,\text{LIL}}$ and use  another bound for time points $1\leq t\leq s_{0,\text{LIL}}$ i.e., in time-regimes prior to stopping time $s_{0,\text{LIL}}$.  
\end{remark}
To this end, \cref{lemma:gamma1} establishes a piece-wise linear function as a time-uniform bound with an initial time $s_{0,\text{Linear}}= 1$ and is therefore applicable in small-time regimes. 
\begin{lemma}
\label{lemma:gamma1}
	Fix $p\in\N\cup\{\infty\}$, $\alpha\in(0,1)$, $\delta\in(0,1)$, $\Delta_1,\ldots,\Delta_p\in(0,1)$ such that $\sum_{j=1}^{p}\Delta_i\leq\delta$. Define further a sequence of time points $t_1,\ldots,t_p$ such that $t_1\leq t_2\leq\ldots\leq t_p$. Set
	\begin{align}\label{eq:tauj}
	  \tau_0&:=2\alpha\log(\frac{1}{\Delta_1}),\\
	  \tau_j&:=\sqrt{t_jt_{j+1}}\frac{\sqrt{\log(\frac{1}{\Delta_{j+1}})t_{j+1}}-\sqrt{\log(\frac{1}{\Delta_{j}})t_{j}}}{\sqrt{\log(\frac{1}{\Delta_{j}})t_{j+1}}-\sqrt{\log(\frac{1}{\Delta_{j+1}})t_{j}}},\quad j=1,\ldots,p.
	\end{align}
	If $\Delta_j$ and $t_j$, $j=1,\ldots,p$ are such that $\tau_j, j=1,\ldots,p-1$ satisfies 
\begin{align}\label{eq:tjtauj}
t_1\leq\tau_1\leq t_2\leq\tau_2\leq\ldots\leq\tau_p\leq t_p,
\end{align}
and the piece-wise linear function $\Gamma^{\text{Linear}}_t$ is defined by
	\begin{align}\label{eq:linboundlemma}
	  \Gamma_t^{\text{Linear}}= \sum_{j=0}^{p-1}\sqrt{\frac{1}{8}\log(\frac{1}{\Delta_j})}\left(\frac{1}{\sqrt{t_j}}t+\sqrt{t_j}\right)\indifunc{ t\in[\tau_j,\tau_{j+1})},
	\end{align}
	 the following holds
	\begin{align}\label{eq:linbound}
	  \PP_{\mathcal{H}_{\infty}}\left(\forall t\in\N\; M_t\leq \Gamma^{\text{Linear}}_t \right)\geq 1-\delta.
	\end{align}

\end{lemma}	
\begin{remark}(Properties of \cref{lemma:gamma1})~\\
	\begin{enumerate}
		\item It is clear that using one linear bound in $t$ to monitor the martingale $M_t$ over time provides an undesirable (linear) asymptotic rate in $t$. Nonetheless, locally, at the time points $t_j$, we obtain the following bound
		\begin{align*}
		  \Gamma^L_{t_j}=\sqrt{\frac{t_j}{2}\log\bigg(\frac{1}{\Delta_j}\bigg)}\quad\text{for}\quad j=1,\ldots,p,
		\end{align*} 
		i.e., the correct asymptotic behaviour in $t$, up to a logarithmic factor. Away from the time points $t_j$, for $p<\infty$, $\Delta_j\equiv\frac{\delta}{2p}$ and $t_{j+1}=t_j+\frac{s_{0,\text{Linear}}}{p}$ for some initial value $s_{0,\text{Linear}}>0$, the additional error for $t_j\leq t\leq t_{j+1}$ is bounded from above by the difference
		\begin{align*}
		\Gamma^L_{t_{j+1}}-\Gamma^L_{t_j} \approx \sqrt{\frac{s_{0,\text{Linear}}\log(\frac{2p}{\delta})}{8pj}},
		\end{align*}
		i.e., a moderate deviation, in particular if $p$ and $j$ assume large values. 
		This issue will be discussed in more detail in Section \ref{Sec:Comparison}. 
\vspace{0.5cm}
\item Notice that each $\tau_j$ defined in \eqref{eq:tauj} reduces to $\tau_j=\sqrt{t_jt_{j+1}}$ if $\Delta_j\equiv\frac{\delta}{2p}$ and \eqref{eq:tjtauj} is satisfied automatically.
	\end{enumerate}
\end{remark}

\subsection{Monitoring procedures}\label{sec:procedures}
The bounds presented in \cref{Sec:Theory} provide  multiple possibilities of defining thresholds for monitoring procedures for the martingale process. We provide two general approaches which can be readily used or adjusted according to information on the problem at hand.
To this end, we will put Theorem \ref{thm:BoundLarget} and Lemma \ref{lemma:gamma1} to use and provide two algorithms for sequential analysis in our setting.

\begin{algorithm}{LIL}
	\begin{algorithmic}[1]
		\State \fix $\alpha,\delta \in (0,1)$ and $k\in (0,1)$
		\State $\kappa\leftarrow \frac{\frac{1}{2}+\frac{1}{20e^{8}}-0.4\alpha+\max\{\frac{1}{6e^4}-0.1\alpha,0\}}{1-\alpha}$
		\State $s_{0,\text{LIL}}\leftarrow\left\lceil\frac{e^{4}(1+\sqrt{k})^2}{\kappa\alpha(1-\alpha)}\log\left(\frac{1}{\delta}\right)\right\rceil.$
		\State $\Gamma_t^{\text{LIL}}\leftarrow \sqrt{\frac{4}{1-k}\kappa \alpha(1-\alpha)t \bigg(2 \log\log \frac{2\kappa \alpha(1-\alpha)t}{ (1-\sqrt{k})} + \log \frac{2}{\delta\log(\frac{1+\sqrt{k}}{1-\sqrt{k}})}}\bigg)$
		\For{$t=s_{0,\text{LIL}},s_{0,\text{LIL}}+1,\ldots$ }
		\If{ $M_{t}>\Gamma_{t}^{\text{LIL}}$} \State \textbf{return} $\widehat T_{0,\text{LIL}}=t$ \Else \textbf{ set} $t=t+1$ \EndIf
		\EndFor
	\end{algorithmic}
 \caption{Sequential testing via LIL bounds}
 \label{alg:LIL}
\end{algorithm}
\noindent 
\newline
\cref{alg:LIL} provides a monitoring method for large time points. However, in cases when monitoring the martingale already before point $s_{0,\text{LIL}}$ is desired,  a hybrid approach in \cref{alg:Hybrid} is proposed, which combines the piece-wise linear bound from Lemma \ref{lemma:gamma1} and the LIL-bound from Theorem \ref{thm:BoundLarget}. In this approach, the algorithm splits the overall level $\delta$ of the sequential procedure equally between early and later times. Before time point $s_{0,\text{LIL}}$, $p$ linear bounds are used whose construction is based on equidistantly spaced time points $\tau_0=t_1\leq t_2\leq\ldots\leq t_p=s_{0,\text{LIL}}.$

\begin{algorithm}
	\begin{algorithmic}[1]
		\State \fix $\alpha,\delta \in (0,1)$, $p\in \N$ and $k\in (0,1)$
		\State \textbf{Choose} $\Delta_1,\ldots,\Delta_p\in(0,1)$ such that $\sum_{j=1}^{p}\Delta_i\leq\delta$.
		\State $\tau_0=t_1 \leftarrow 2\alpha\log(\frac{1}{\Delta_1})$
		\State $\kappa\leftarrow \frac{\frac{1}{2}+\frac{1}{20e^{8}}-0.4\alpha+\max\{\frac{1}{6e^4}-0.1\alpha,0\}}{1-\alpha}$
		\State $s_{0,\text{LIL}}\leftarrow\left\lceil\frac{e^{4}(1+\sqrt{k})^2}{\kappa\alpha(1-\alpha)}\log\left(\frac{1}{\delta}\right)\right\rceil.$
		\State $t_j\leftarrow t_1+(j-1)/(p-1)(s_0-t_1)$
		\State $\tau_j\leftarrow\sqrt{t_jt_{j+1}}\frac{\sqrt{\log(\frac{1}{\Delta_{j+1}})t_{j+1}}-\sqrt{\log(\frac{1}{\Delta_{j}})t_{j}}}{\sqrt{\log(\frac{1}{\Delta_{j}})t_{j+1}}-\sqrt{\log(\frac{1}{\Delta_{j+1}})t_{j}}},\quad j=1,\ldots,p-1.$

		\State $\Gamma_t^{\text{Linear}}\leftarrow \sqrt{\frac{1}{8}\log(\frac{1}{\Delta_j})}\left(\frac{1}{\sqrt{t_j}}t+\sqrt{t_j}\right), \quad t\in[\tau_j,\tau_{j+1}),$
        \State  $\widehat T_{0,\text{hybrid}}\leftarrow 0$
		\For{$t=t_0,\ldots,s_{0,\text{LIL}}$ }
		\If{ $M_{t}>\Gamma_{t}^{\text{Linear}}$} \State \textbf{return} $\widehat T_{0,\text{hybrid}}=t$ \Else \textbf{ set} $t=t+1$ \EndIf
		\EndFor
        \If{$T_{0,\text{hybrid}}=0$}
		\State $\Gamma_t^{\text{LIL}}\leftarrow \sqrt{\frac{4}{1-k}\kappa \alpha(1-\alpha)t \bigg(2 \log\log \frac{2\kappa \alpha(1-\alpha)t}{ (1-\sqrt{k})} + \log \frac{2}{\delta\log(\frac{1+\sqrt{k}}{1-\sqrt{k}})}}\bigg)$
		\For{$t=s_{0,\text{LIL}},s_{0,\text{LIL}}+1,\ldots$ }
		\If{ $M_{t}>\Gamma_{t}^{LIL}$} \State \textbf{return} $\widehat T_{0,\text{hybrid}}=t$ \Else \textbf{ set} $t=t+1$ \EndIf
		\EndFor
        \EndIf
	\end{algorithmic}
 \caption{Sequential testing (Hybrid Algorithm)}
\label{alg:Hybrid}
\end{algorithm}
\subsection{Proof of Lemma \ref{lemma:gamma1}}

\begin{proof}[Proof of Lemma \ref{lemma:gamma1}] 
We will start with deriving one linear bound that holds uniformly for all $t\in\N$.
  It follows from  Example 2, equation (2.29) in \cite{howard2020time} 
  \begin{align}\label{concBoundLinear}
\PP_{\mathcal{H}_\infty} \bigg(  \exists t \in \mathbb{N} :  M_t \geq \underbrace{ x + \frac{x}{2 \alpha (1- \alpha)t_0} (t- \alpha (1-\alpha)t_0)}_{\text{Linear bound}} \bigg) \leq \exp \bigg\{ - \frac{2 x^2}{ \alpha (1- \alpha) t_0} \bigg\},
\end{align}
for any $t_0\in\N$ and $x\in(0,(1-\alpha)t_0)$.
In \cref{concBoundLinear}, we focus on the linear bound as in the following:
\begin{align*}
M_t &\geq x + \frac{x}{2 \alpha (1- \alpha)t_0} (t- \alpha (1-\alpha)t_0)\\
& =\frac{x}{2 \alpha(1-\alpha)t_0} t + \frac{x}{2}\\
& = \frac{c}{\alpha (1- \alpha)t_0}t + c,
\end{align*}
when for simplification of notation,  $c = x/2$ denotes the intercept of the line and  the slope is given by $c/ (\alpha (1- \alpha)t_0)$.  Further, notice that it follows as a consequence of  Example 2.29 of    \cite{howard2020time} that $c \in \big(0, \frac{(1-\alpha) t_0}{2}\big)$.  With the appropriate substitutions, \cref{concBoundLinear} becomes,
  \begin{equation}
  \PP_{\mathcal{H}_\infty} \bigg(t \in \N, M_t < \frac{c}{\alpha(1-\alpha)t_0}t + c \bigg) \geq 1-  \exp \bigg\{ - \frac{8c^2}{\alpha (1- \alpha) t_0}\bigg\}. 
  \label{linearBound}
  \end{equation}
 We now set $\Delta=\exp \bigg\{ - \frac{8c^2}{\alpha (1- \alpha) t_0}\bigg\}$ 	and solve for $c$. This gives
 \begin{align*}
   c=\sqrt{\frac{\alpha(1-\alpha)t_0}{8}\log\Big(\frac{1}{\Delta}\Big)},
 \end{align*}
 and therefore
 \begin{align*}
    \PP_{\mathcal{H}_\infty} \bigg(t \in \N, M_t < \sqrt{\tfrac{\log(\frac{1}{\Delta})}{8\alpha(1-\alpha)t_0}}t+ \sqrt{\tfrac{\alpha(1-\alpha)t_0\log(\frac{1}{\Delta})}{8}}\bigg) \geq 1-  \Delta,
 \end{align*}
 whenever 
 \begin{align}\label{eq:condc}
    c=\sqrt{\frac{\alpha(1-\alpha)t_0}{8}\log(\frac{1}{\Delta})}<\frac{1-\alpha}{2}t_0.
 \end{align}
 Condition \eqref{eq:condc} is equivalent to the following condition on the parameter $t_0$
 \begin{align}\label{eq:boundt0}
 t_0>\frac{\alpha}{2(1-\alpha)}\log\bigg(\frac{1}{\Delta}\bigg).
 \end{align}
 The piece-wise linear function is now constructed from the linear bound 
 \begin{align*}
  L_{t_0}(t)=\sqrt{\frac{\log(\frac{1}{\Delta})}{8\alpha(1-\alpha)t_0}}t+ \sqrt{\frac{\alpha(1-\alpha)t_0\log(\frac{1}{\Delta})}{8}},
 \end{align*} 
   where the linear parts are chosen (i.e., the parameter $t_0$ is adjusted) such that $ L_{t_0}$ is minimal at each of the  time points $t=t_1,\ldots,t=t_p$.
 To achieve this optimality property, we now fix $t=t_j$, $j\geq1$ and and minimise $L_{t_0}(t_j)$ with respect to $t_0$. Elementary calculus yields one local minimum at $t_{0,\min,j}:=t_j/(\alpha(1-\alpha))$, where $L_{t_{0,\min,j}}(t_j)=\sqrt{\frac{1}{2}t_j\log(\frac{1}{\Delta_j})}$. We now establish that $t_{0,\min,j}$ is a global minimum as well for each fixed value of $j$. Equation \eqref{eq:boundt0} provides the lower bound on $t_0$ of $\frac{\alpha}{2(1-\alpha)}\log(\frac{1}{\Delta})$. We find
 \begin{align*}
     \lim_{t_0\to\frac{\alpha}{2(1-\alpha)}\log(\frac{1}{\Delta_j})}L_{t_0}(t_j)=\frac{1}{2\alpha}t_j+\frac{\alpha}{4}\log(\frac{1}{\Delta_j}),
 \end{align*}
and
\begin{align*}
    \sqrt{\frac{1}{2}t_j\log\Big(\frac{1}{\Delta_j}\Big)}\leq\frac{1}{2\alpha}t_j+\frac{\alpha}{4}\log\Big(\frac{1}{\Delta_j}\Big)\quad
    \Longleftrightarrow\quad
    0\leq\left(t_j-\frac{\alpha^2}{2}\log\Big(\frac{1}{\Delta_j}\Big)\right)^2.
\end{align*}
In addition
\begin{align*}
     \lim_{t_0\to\infty}L_{t_0}(t_j)=\infty,
 \end{align*}
establishing that the global minima are indeed reached at $t_{0,\min,j}$. Finally, condition \eqref{eq:boundt0} needs to be met for all $t_{0,\min,j}$. This is the case if
\begin{align*}
    t_{0,\min,j}=\frac{t_j}{\alpha(1-\alpha)}>\frac{\alpha}{2(1-\alpha)}\log(\frac{1}{\Delta_j})\quad\Longleftrightarrow\quad t_{0,\min,j}\geq\frac{\alpha^2}{2}\log(\frac{1}{\Delta_j}).
\end{align*}
We now consider two consecutive linear bounds $L_{t_{0,\min,j}}$ and $L_{t_{0,\min,j+1}}$. Both are valid bounds for all $t\in \N$, however, by construction, $L_{t_{0,\min,j}}\leq L_{t_{0,\min,j+1}}$ for all $t\leq\tau_j$, where $\tau_j$ denotes the the time point at which both lines intersect (see Figure \ref{fig:lines}).
\begin{figure}[H]
    \centering
    \includegraphics[width=0.8\textwidth]{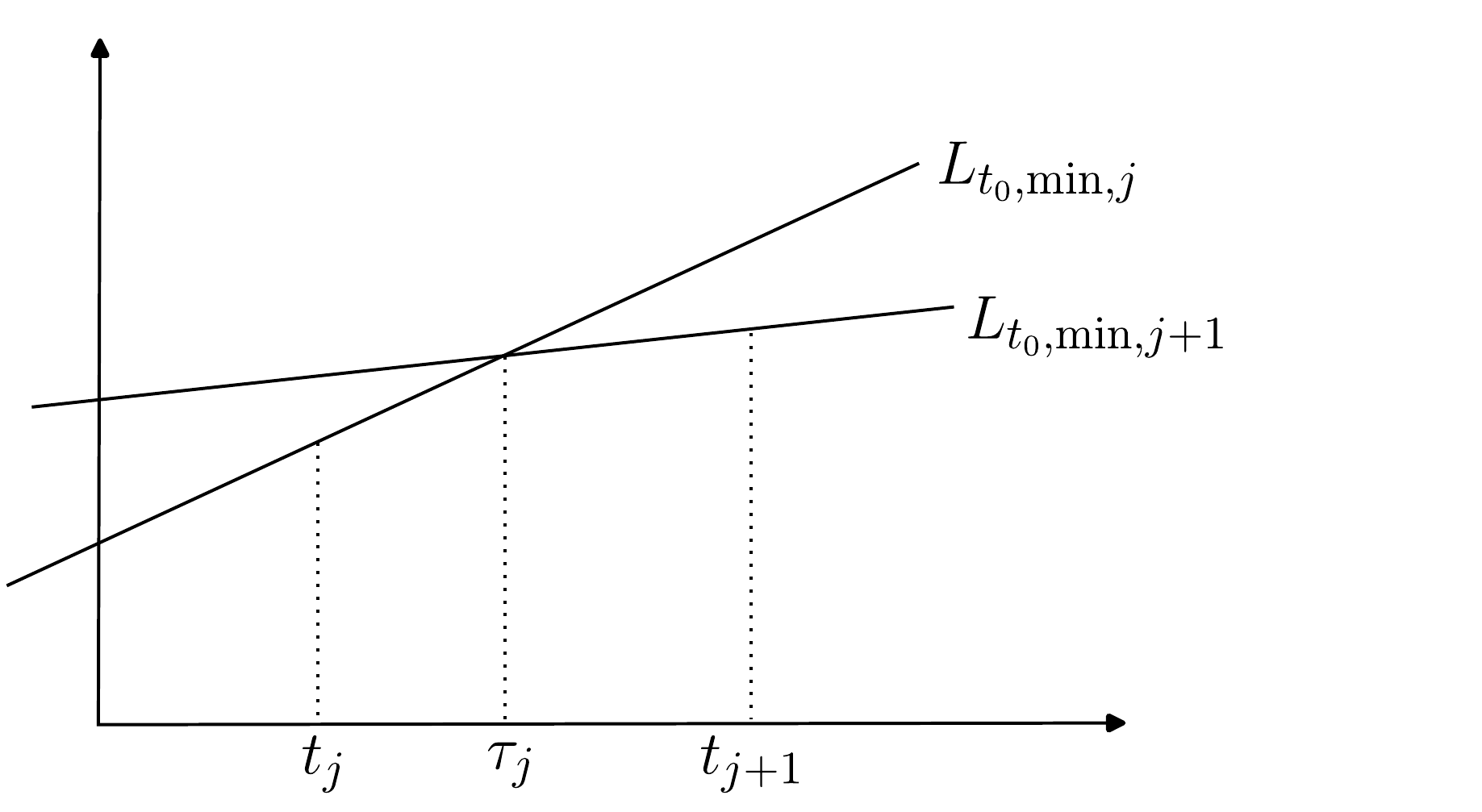}
    \caption{Sketch of the construction of the piece-wise linear bound: Between $\tau_{j-1}$ and $\tau_j$ the line $L_{t_{0,\min,j}}$ is the smallest of the pre-selected ones. Thereby, choosing $t_{0,j}=t_{0,\min,j}$ and $\gamma^{\text{Linear} }$ as in \eqref{eq:linboundlemma}, we obtain the sharpest possible bound by this way of construction. The $\tau_j$'s are obtained by solving the equation $L_{t_{0,\min,j}}(t)=L_{t_{0,\min,j+1}}(t)$ for $t$.}
    \label{fig:lines}
\end{figure}
The piece-wise linear bound is now constructed successively: From the intersection points $\tau_{j-1}$ to $\tau_j$, the smallest linear function in this construction is $L_{t_{0,\min,j}}$, which is therefore used as a bound in this range.\\
 
 We have now established that for each $j$, the following holds
\begin{align*}
  \PP(t\in\N,\quad M_t\leq L_{t_{0,\min,j}}(t))\geq1-\Delta_j,    
\end{align*} 
which is equivalent to
 \begin{align*}
  \PP(t\in\N,\quad M_t>L_{t_{0,\min,j}(t)}\quad\text{for one }j)\leq\sum_{j=1}^p\Delta_j\leq\delta   
\end{align*}
which implies that the piece-wise linear function $t\mapsto\Gamma_t^{\text{Linear}}$ is a time-uniform bound in the sense of \eqref{eq:linbound}.
\begin{remark}
Obviously, the linear bounds can be fine-tuned in a multitude of ways by adjusting the parameters $t_j$ and $\Delta_j$. For instance, splitting $\delta$ into $\Delta_j$'s in such a way that $\Delta_j$ is small for small values of $j$ and large for large values of $j$ makes a later detection easier and vice versa. Moreover, choosing a non-equidistant grid of grid points $t_j$ can place more emphasis/accuracy on any domain of choice. 
\end{remark}

\end{proof}
\subsection{Proof of Theorem \ref{thm:BoundLarget}}
\label{proof_LIL}
Before delving into the lengthy proof of Theorem \ref{thm:BoundLarget}, we provide an outline of the proof strategy in the following steps, also intended to provide a map to the general proof strategy, which is similar to the proof strategy provided for the proof of Theorem 5 in \cite{balsubramani2014sharp}.
\subsubsection{Outline of the proof of Theorem \ref{thm:BoundLarget}}
The proof of Theorem \ref{thm:BoundLarget} is split into five lemmas (Lemma \ref{le:supermartingale} - Lemma \ref{lemma:Bernstein}) as outlined below.\\

\begin{enumerate}
    \item \textbf{Choice of super-martingale (Lemma \ref{le:supermartingale}):} We first choose an appropriate super-martingale $X_t^\lambda$ with  additional parameter $\lambda$. Such a construction was proposed in \citep{shafer2005probability} in Chapter 5 where they take a game-theoretic approach and $M_t$ is the returns of a player. The strategy was to make the players capital unbounded by defying the law of asymptotic logarithm by choosing appropriate values of $\lambda$.\\
    \item \textbf{Intuition on additional parameter $\lambda$ :} As mentioned in \citep{shafer2005probability}, for the previous step to work, a number of values of $\lambda$ have to be considered such that the indices $t$ where the martingale $M_t$ shows large values (especially those close to $\mathcal{O}(\sqrt{2t \log \log t})$) can be taken advantage of. However, $\lambda$ is unknown and this is solved by choosing a mixing distribution from which $\lambda$ may be sampled. In our case, we use a version of a distribution similar to the one proposed in Example 4 of \citep{robbins1970boundary}.\\
    \item \textbf{Lemma \ref{lemma:moment}:} A moment bound for $|M_t|$ corresponding to a fixed $\lambda_0$ is provided.\\
    \item \textbf{Lemma \ref{lemma:weaker}:} Time-uniform LIL is proved for  $\frac{|M_t|}{\sqrt{V_t \log \log V_t}}$ by first controlling the weaker condition $\frac{|M_t|}{\kappa V_t}$, where $V_t=\alpha(1-\alpha)t$.\\
    \item \textbf{Lemma \ref{lemma:lambda}:} Then, we prove the moment bound for the case when $\lambda$ is chosen stochastically.  The choice of density function is crucial here.\\
    \item \textbf{Lemma \ref{lemma:Bernstein}:}  Finally, a stopping time is defined and the Bernstein bound is established.
\end{enumerate}
\subsubsection{Statement of Lemma \ref{le:supermartingale} - Lemma \ref{lemma:Bernstein}}
\begin{lemma} \label{le:supermartingale} Let the global null hypothesis $\mathcal{H}_\infty$ hold and $\alpha\in(0,1/2]$.
If the $\{\indifunc{\test_i=1}\}_{i\in\mathbb{N}}$ are independent and $\{M_t\}_{t\in\mathbb{N}}$ is the martingale defined in \eqref{martingale_stat}, the process
\begin{align*}
X_t^\lambda \coloneqq \exp \{ \lambda M_t - \lambda^2 \kappa \alpha(1-\alpha)t \},
\end{align*}
is a super-martingale for any $\lambda \in [-e^{-2},e^{-2}]$, $\alpha\in(0,1/2]$ and any $\kappa\in\;[\kappa_0,\infty)$, where $$\kappa_0=\frac{\frac{1}{2}+\frac{1}{20e^{8}}-0.4\alpha+\max\{\frac{1}{6e^4}-0.1\alpha,0\}}{1-\alpha}.$$ 
\end{lemma}

\begin{proof}
 We need to show that  $\mathbb{E} [ X_t^\lambda | \mathbb{F}_{t-1}] \leq X_{t-1}^\lambda$ for any $t\in \N$. Since  the random variables $\indifunc{\test_t=1}$ are independent of $\mathbb{F}_{t-1}$ for any $t$ by assumption, we find
\begin{align*}
    \mathbb{E}[\exp(\lambda \xi_t) | \mathbb{F}_{t-1}] = \mathbb{E}[\exp(\lambda (\indifunc{\test_t=1}-\alpha) | \mathbb{F}_{t-1}]=\mathbb{E}[\exp(\lambda (\indifunc{\test_t=1}-\alpha) ].
\end{align*}
since $\indifunc{\test_t=1}$ is a Bernoulli random variable with success probability $\alpha$, we obtain
\begin{align*}
    \mathbb{E}[\exp(\lambda (\indifunc{\test_t=1}-\alpha) ]=(1-\alpha+\alpha e^{\lambda})e^{-\lambda\alpha}.
\end{align*}
This yields
\begin{align*}
    \mathbb{E} [ X_t^\lambda | \mathbb{F}_{t-1}]=X_{t-1}^\lambda\cdot\exp(-\lambda^2\kappa\alpha(1-\alpha)-\alpha\lambda)\cdot(1-\alpha+\alpha e^{\lambda}).
\end{align*}
Set $F(\alpha,\kappa,\lambda):=\exp(-\lambda^2\kappa\alpha(1-\alpha)-\alpha\lambda)\cdot(1-\alpha+\alpha e^{\lambda}).$
We now show that $\log(F(\alpha,\kappa,\lambda))\leq0$ for the parameters within the range specified in the formulation of this lemma. 
\begin{align*}
    \log(F(\alpha,\kappa,\lambda))=-\lambda^2\kappa\alpha(1-\alpha)-\alpha\lambda+\log\left(1-\alpha+\alpha e^{\lambda}\right)
\end{align*}
We now estimate the last term using a Taylor expansion:
\begin{align*}
  \log\left(1-\alpha+\alpha e^{\lambda}\right)&\leq\log(1)-\alpha(1-e^{\lambda})-\frac{\alpha^2}{2}(1-e^{\lambda})^2+\frac{2}{6}\alpha^3(1-e^{\lambda})^3,
\end{align*}
since $|\alpha(1-e^{\lambda})|<1$ for all $\alpha\in(0,1)$.
Using the Lagrangian form of the remainder in the Taylor expansion, the first non-zero term can be estimated as follows.
\begin{align*}
-\alpha(1-e^{\lambda})&\leq \alpha\lambda+\frac{\alpha\lambda^2}{2}+\frac{\alpha\lambda^3}{6}+\alpha\frac{\lambda^4}{24}e^{\lambda_{\max}}
\end{align*}
where $\lambda_{\max}=e^{-2}$.
This yields
\begin{align*}
-\alpha(1-e^{\lambda})&\leq \alpha\lambda+\frac{\alpha\lambda^2}{2}+\frac{\alpha\lambda^3}{6}+\alpha\frac{\lambda^4}{20}e^{\lambda_{\max}},
\end{align*}
and thus
\begin{align*}
   \log(F(\alpha,\kappa,\lambda))\leq -\lambda^2\kappa\alpha(1-\alpha)+\frac{\alpha\lambda^2}{2}+\frac{\alpha\lambda^3}{6}+\frac{\alpha\lambda^4}{20}-\frac{\alpha^2}{2}(1-e^{\lambda})^2+\frac{2}{6}\alpha^3(1-e^{\lambda})^3.
\end{align*}

If $\lambda>0$, the last two terms are negative and we can estimate
\begin{align*}
     \log(F(\alpha,\kappa,\lambda))\leq\lambda^2\alpha\left(-\kappa(1-\alpha)+\frac{1}{2}+\frac{1}{6e^4}+\frac{1}{20e^8}-\frac{\alpha}{2}\right),
\end{align*}
which gives that $\log(F(\alpha,\kappa,\lambda))\leq0$ if
\begin{align*}
   \kappa\geq \frac{\frac{1}{2}+\frac{1}{6e^4}+\frac{1}{20e^8}-\frac{\alpha}{2}}{1-\alpha}.
\end{align*}
If $\lambda\leq0,$ we find
\begin{align*}
    (1-e^\lambda)^2=1-2e^\lambda+e^{2\lambda}\geq \lambda^2+\lambda^3
\end{align*}
as well as
\begin{align*}
    (1-e^\lambda)^3\leq-\lambda^3,
\end{align*}
which implies
\begin{align*}
    \log(F(\alpha,\kappa,\lambda))&\leq-\lambda^2\kappa\alpha(1-\alpha)+\frac{\alpha\lambda^2}{2}+\frac{\alpha\lambda^3}{6}+\frac{\alpha\lambda^4}{23}-\frac{\alpha^2}{2}( \lambda^2+\lambda^3)-\frac{\lambda^3\alpha^3}{3}\\
    &\leq\lambda^2\alpha\left(-\kappa(1-\alpha)+\frac{1}{2}+\frac{\lambda^2}{23}-\frac{\alpha}{2}+\frac{2}{3}\alpha|\lambda|\right)\\
    &\leq\lambda^2\alpha\left(-\kappa(1-\alpha)+\frac{1}{2}+\frac{1}{20e^8}-0.4\alpha\right).
\end{align*}
From this we conclude that for $\lambda<0$
 $\log(F(\alpha,\kappa,\lambda))\leq0$ if
\begin{align*}
   \kappa\geq \frac{\frac{1}{2}+\frac{1}{20e^8}-0.4\alpha}{1-\alpha}.
\end{align*}
The statement of the lemma now follows.
\end{proof}

\begin{lemma}\label{lemma:moment}
For any stopping time $\tau$, $\kappa\geq\kappa_0),$ and a fixed $\lambda_0\in[-e^{-2},e^{-2}]$, we have that,
\begin{align*}
    \mathbb{E}  [ \exp (\lambda_0 M_\tau  - \lambda_0^2 \kappa V_\tau)] \leq 1.
\end{align*}
\end{lemma}

\begin{proof}
The statement of the lemma follows directly by the optional stopping theorem.
\end{proof}

\begin{lemma}\label{lemma:weaker}
Let $k\in(0,1)$,  $\kappa\geq\kappa_0)$, and $t_0=\left\lceil\frac{1}{\lambda_0^2\kappa\alpha(1-\alpha)}\log\left(\frac{1}{\delta}\right)\right\rceil$ with $|\lambda_0|\leq e^{-2}/(1+\sqrt{k}).$
Fix any $\delta > 0$. Then, the following holds
$$
 \mathbb{P} \bigg(\frac{M_t}{ \kappa \alpha(1-\alpha)t} \leq 2\lambda_0\quad \text{for all }t\geq t_0\bigg) \geq 1- \delta.
$$
\label{initBoundMarti}
\end{lemma}
Equivalently put, with the definition of the set
\begin{align*}
  A_{\delta}=\bigg\{\omega \in \Omega : \frac{M_t}{\kappa \alpha(1-\alpha)t} \leq 2\lambda_0, \forall t \geq t_0\bigg\},
\end{align*}
the statement of the above becomes
\begin{align*}
     \mathbb{P}(A_{\delta})\geq 1-\delta.
\end{align*}
\vspace{0.1cm}

\begin{proof}
Define the stopping time $\tau_1 = \min \{ t \geq t_0 : \frac{M_t}{\kappa \alpha(1-\alpha)t} > 2\lambda_0\}.$ Then it suffices to prove that $\mathbb{P} (\tau_1 < \infty) \leq \delta$ i.e., uniformly for all $t\geq\tau_0$, the rescaled martingale $\frac{M_t}{\kappa \alpha(1-\alpha)t} \leq 2\lambda_0$ with high probability. 
For this purpose, we use the previous lemma and get that,
\begin{align*}
    1&\geq \mathbb{E} [\exp(\lambda_0 M_t - \lambda_0^2 \kappa \alpha(1-\alpha)t)] \\
     &\geq \mathbb{E} [\exp(\lambda_0 M_t - \lambda_0^2 \kappa \alpha(1-\alpha)t) | \tau_1 < \infty ] \mathbb{P}(\tau_1 < \infty)\\
     &> \mathbb{E} [ \exp \{ \lambda_0^2 \kappa \alpha(1-\alpha)\tau_1\}| \tau_1 < \infty ] \mathbb{P}(\tau_1 < \infty).
\end{align*}
where the last estimate holds because $M_{\tau_1} > 2\lambda_0 \kappa\alpha(1-\alpha)\tau_1$ for $\tau_1 < \infty$. 
Since $\tau_1\geq t_0$ by construction, we further obtain
\begin{align*}
    1&
     \geq \mathbb{E} [ \exp \{ \lambda_0^2 \kappa \alpha(1-\alpha)t_0\}| \tau_1 < \infty ] \mathbb{P}(\tau_1 < \infty)\\
     & > \mathbb{E} [ \exp \{ -\log \big(\delta\big)\}| \tau_1 < \infty ] \mathbb{P}(\tau_1 < \infty)\geq \frac{1}{\delta} \mathbb{P}(\tau_1 < \infty),
\end{align*}
Therefore, $\mathbb{P}(\tau_1 < \infty) \leq \delta$, which concludes the proof of this lemma.
\end{proof}
\vspace{0.5cm}
We continue with Step 4 from our outline i.e., we prove the moment bound for the case when $\lambda$ is chosen stochastically. In the following, $\mathbb{E}[\cdot]$ denotes expectation with respect to the original probability space, $\mathbb{E}^{\lambda}[\cdot]$ denotes expectation with respect to the probability space $(\Omega_\lambda, \mathbb{F}_\lambda, \mathbb{P}_\lambda)$ and, for any event $A\in \mathbb{F}$, $\mathbb{E}_{A}[\cdot]$ denotes conditional expectation $\mathbb{E}[\cdot|A]$, i.e.,
\begin{align*}
    \mathbb{E}_A[X]=\frac{1}{\mathbb{P}(A)}\int_{A}X\,\mathrm{d}\mathbb{P}.
\end{align*}
With the latter notation, we can now state our next lemma.

\begin{lemma}\label{lemma:lambda} Given that $X_t^\lambda \coloneqq \exp \{ \lambda M_t - \lambda^2 \kappa \alpha(1-\alpha)t \}$ is a super-martingale for any $\lambda \in [-e^{-2}, e^{-2}]$ if  $\kappa\geq\kappa_0)$ and that $\lambda$ is sampled, independently of $\indifunc{\Phi_1=1},\indifunc{\Phi_2=1},\ldots$  from $(\Omega_\lambda, \mathbb{F}_\lambda, \mathbb{P}_\lambda)$, with density $f(\lambda) = 1/(|\lambda|\big(-\log |\lambda|\big)^2)$ on $\lambda \in [-e^{-2}, e^{-2}]\setminus \{0\}$, it holds that 
\begin{align*}
\mathbb{E}_{B_{\delta}} [\mathbb{E}^\lambda [X_t^{\lambda}]] \geq \mathbb{E}_{B_{\delta}} \bigg[ \frac{\exp  \big\{ \frac{M_t^2}{4 \kappa V_t} (1-k)\big\} }{\log^2 \big(\frac{2\kappa V_t}{|M_t| (1-\sqrt{k})}}\big) \bigg],
\end{align*}
where 
\begin{align*}
B_\delta \coloneqq \bigg\{\omega \in \Omega : \frac{|M_t|}{\kappa \alpha(1-\alpha)t} \leq 2\lambda_0, \forall t \geq t_0\bigg\},\quad \lambda_0\in\left(0,e^{-2}/\left(1+\sqrt{k}\right)\right).
\end{align*}

\label{momentBound}
\end{lemma}

\begin{proof}
Let $\lambda$ be chosen stochastically from $(\Omega_\lambda, \mathbb{F}_\lambda, P_\lambda)$  with probability density $f$ such that $f(\lambda) = \frac{1}{|\lambda|\big(\log \frac{1}{|\lambda|}\big)^2}$ on $\lambda \in [-e^{-2}, e^{-2}]\setminus \{0\}$, a mixing distribution suggested in \cite{robbins1970boundary}. Then,  with $V_t=\alpha(1-\alpha)t$
\begin{align*}
    \mathbb{E}^\lambda[X_t^\lambda] &= \int_{-1/e^2}^{1/e^2} \exp \big( \lambda M_t - \lambda^2 \kappa V_t\big) f(\lambda) d\lambda\\
    &= \int_{-1/e^2}^{1/e^2} \exp\bigg\{ \frac{M_t^2}{4\kappa V_t}- \lambda^2 \kappa V_t -\frac{M_t^2}{ 4\kappa V_t} + \lambda M_t \bigg \}f(\lambda) d\lambda\\
    &= \exp \bigg \{ \frac{M_t^2}{4\kappa V_t} \bigg \} \int_{-1/e^2}^{1/e^2} \exp \bigg\{ -\kappa V_t \bigg( \lambda^2 + \frac{M_t^2}{4 \kappa^2 V_t^2} - \frac{\lambda M_t}{\kappa V_t}\bigg) \bigg\}  f(\lambda) d\lambda.
\end{align*}
This yields
\begin{align*}
    \mathbb{E}^\lambda[X_t^\lambda]
    & =\exp \bigg \{ \frac{M_t^2}{4\kappa V_t} \bigg \} \int_{-1/e^2}^{1/e^2} \exp \bigg\{ -\kappa V_t \bigg( \lambda - \frac{M_t}{2 \kappa V_t} \bigg)^2 \bigg\}  f(\lambda) d\lambda\\
    &=\exp \bigg \{ \frac{M_t^2}{4\kappa V_t} \bigg \} \int_{-1/e^2}^{0} \exp \bigg\{ -\kappa V_t \bigg( \lambda - \frac{M_t}{2 \kappa V_t} \bigg)^2 \bigg\}  f(\lambda) d\lambda\\
    &~+\exp \bigg \{ \frac{M_t^2}{4\kappa V_t} \bigg \} \int_{0}^{1/e^2} \exp \bigg\{ -\kappa V_t \bigg( \lambda - \frac{M_t}{2 \kappa V_t} \bigg)^2 \bigg\}  f(\lambda) d\lambda.
\end{align*}
By a change of variables, using that $f(\lambda)=f(-\lambda)$, we find
\begin{align*}
    \mathbb{E}^\lambda[X_t^\lambda]
    &=\exp \bigg \{ \frac{M_t^2}{4\kappa V_t} \bigg \} \int_{0}^{1/e^2} \exp \bigg\{ -\kappa V_t \bigg( -\lambda - \frac{M_t}{2 \kappa V_t} \bigg)^2 \bigg\}  f(\lambda) d\lambda\\
    &~+\exp \bigg \{ \frac{M_t^2}{4\kappa V_t} \bigg \} \int_{0}^{1/e^2} \exp \bigg\{ -\kappa V_t \bigg( \lambda - \frac{M_t}{2 \kappa V_t} \bigg)^2 \bigg\}  f(\lambda) d\lambda.
\end{align*}
Since both integrals are non-negative, we obtain
\begin{align*}
    \mathbb{E}^\lambda[X_t^\lambda]
    &\geq\exp \bigg \{ \frac{M_t^2}{4\kappa V_t} \bigg \} \int_{0}^{1/e^2} \exp \bigg\{ -\kappa V_t \bigg( \lambda - \frac{|M_t|}{2 \kappa V_t} \bigg)^2 \bigg\}  f(\lambda) d\lambda.
\end{align*}
Notice that the exponential in the integral is maximal, i.e., $$\exp \bigg\{ -\kappa V_t \bigg( \lambda - \frac{|M_t|}{2 \kappa V_t} \bigg)^2 \bigg\}=1,$$ if $\lambda = |M_t|/(2 \kappa V_t)$. To further estimate the integral, we therefore restrict the integral over a region around  $\lambda = |M_t|/(2 \kappa V_t)$ and find for $k\in(0,1)$
\begin{align*}
    \mathbb{E}^\lambda[X_t^\lambda]
    &\geq\exp \bigg \{ \frac{M_t^2}{4\kappa V_t} \bigg \} \int_{\frac{|M_t|}{2\kappa V_t}(1-\sqrt{k})}^{{\frac{|M_t|}{2\kappa V_t}(1+\sqrt{k})}} \exp \bigg\{ -\kappa V_t \bigg( \lambda - \frac{|M_t|}{2 \kappa V_t} \bigg)^2 \bigg\}  f(\lambda) d\lambda.
\end{align*}
Notice that on the set $B_{\delta}$, we have for $k \in (0,1)$ that
\begin{align*}
   \left(\frac{|M_t|}{2\kappa V_t}(1-\sqrt{k}),\frac{|M_t|}{2\kappa V_t}(1+\sqrt{k})\right)\subset(0,e^{-2}), 
\end{align*}
i.e., the domain of integration is included in the support of $\mathbb{P}_{\lambda}$ if we remain on the set $B_{\delta}$, which we assume in the following.
Now, the exponential in the integral is lower bounded by $\exp \bigg \{-k \frac{M_t^2}{4\kappa V_t} \bigg \}$, which yields
\begin{align*}
    \mathbb{E}^\lambda[X_t^\lambda]
    &\geq\exp \bigg \{ \frac{M_t^2}{4\kappa V_t} -k \frac{M_t^2}{4\kappa V_t} \bigg \} \int_{\frac{|M_t|}{2\kappa V_t}(1-\sqrt{k})}^{{\frac{|M_t|}{2\kappa V_t}(1+\sqrt{k})}}   f(\lambda) d\lambda\\
    &=\exp \bigg \{(1-k) \frac{M_t^2}{4\kappa V_t} \bigg \}\left[\frac{1}{\log(\frac{1}{\lambda})}\right]_{\frac{|M_t|}{2\kappa V_t}(1-\sqrt{k})}^{{\frac{|M_t|}{2\kappa V_t}(1+\sqrt{k})}}.
\end{align*}
Finally,
\begin{align*}
    \mathbb{E}^\lambda[X_t^\lambda]
    &\geq
   \exp  \bigg\{ \frac{M_t^2}{4\kappa V_t} (1-k)\bigg\}\frac{\log \big( \frac{1+\sqrt{k}}{1-\sqrt{k}}\big)}{\log^2 \left(\frac{2\kappa V_t}{|M_t|(1-\sqrt{k})}\right)}.
\end{align*}
Taking $\tau$ to be any stopping time and using the definition of $B_\delta$ for $t\geq \tau$, we get that,
\begin{align*}
\mathbb{E}_{B_\delta}[\mathbb{E}^\lambda [X_\tau^\lambda]]&\geq \mathbb{E}_{B_\delta}\bigg[\exp  \bigg\{ \frac{M_t^2}{4\kappa V_t} (1-k)\bigg\}\frac{\log \big( \frac{1+\sqrt{k}}{1-\sqrt{k}}\big)}{\log^2 \frac{2\kappa V_t}{|M_t|(1-\sqrt{k})}} \bigg]\\
& = \mathbb{E}_{B_\delta} \bigg[ \frac{\mathcal{K} \exp  \big\{ \frac{M_t^2}{4\kappa V_t} (1-k)\big\}}{\log^2 \left(\frac{2\kappa V_t}{|M_t| (1-\sqrt{k})}\right)}\bigg].
\end{align*}
where $\mathcal{K} \coloneqq \log \big( \frac{1+\sqrt{k}}{1-\sqrt{k}}\big)$.
\end{proof}

\begin{lemma}\label{lemma:Bernstein}
Let $k\in(0,1)$,  $\kappa\geq\kappa_0$, $\mathcal{K}= \log \big( \frac{1+\sqrt{k}}{1-\sqrt{k}}\big)$, $\lambda_0\in\big(0,e^{-2}/(1+\sqrt{k})\big)$ and $V_t=\alpha(1-\alpha)t$.
Define a stopping time $\tau$ as 
\begin{align*}
    \tau &= \min \bigg\{  t \geq t_0: M_t> 2\lambda_0 \kappa V_t \, \, \vee \\
    & \bigg(M_t \leq 2\lambda_0 \kappa V_t \wedge M_t  >\sqrt{\frac{1}{1-k}\kappa V_t \bigg(2 \log\log \frac{\kappa V_t}{ |M_t| (1-\sqrt{k})} + \log \frac{2}{\delta \mathcal{K}}}\bigg)\bigg) \bigg\}.
\end{align*}
Then, it holds that,
\begin{align*}
    P(\tau = \infty) \geq 1- \delta.
\end{align*}
\end{lemma}

\begin{proof}
At stopping time $\tau$, we can bound the martingale $M_t$ from below by
\begin{align*}
M_t  &>\sqrt{\frac{4}{1-k}\kappa V_t \bigg(2 \log\log \frac{2\kappa V_t}{ |M_t| (1-\sqrt{k})} + \log \frac{2}{\delta \mathcal{K}}}\bigg).
\end{align*}
This implies
\begin{align*}
\frac{M_t^2 (1-k)}{ 4\kappa V_t} & > 2\log\log\bigg(\frac{2\kappa V_t}{|M_t| (1-\sqrt{k})}\bigg) + \log \frac{2}{\delta \mathcal{K}}\\
\Leftrightarrow\exp \bigg\{\frac{M_t^2 (1-k)}{4 \kappa V_t}  \bigg \}& >   \frac{2}{\delta \mathcal{K}}\log^2 \bigg(\frac{ 2\kappa V_t}{|M_t| (1-\sqrt{k})}\bigg),
\end{align*}
and finally
\begin{align}\label{eq:Lemma10}
\frac{\mathcal{K}\exp \bigg\{\frac{M_t^2 (1-k)}{ 4\kappa V_t}  \bigg \} }{\log^2 \bigg(\frac{ 2\kappa V_t}{|M_t| (1-\sqrt{k})}\bigg)}& >   \frac{2}{\delta}.
\end{align}

Using the previous lemma, on the event $B_{\delta/2}$, we have that, 
\begin{align*}
    \frac{4}{3}&\geq \frac{1}{1-\frac{\delta}{2}}= \frac{\mathbb{E}^\lambda [\mathbb{E}[X_0^\lambda]]}{1-\frac{\delta}{2}}\stackrel{(i)}{\geq}  \frac{\mathbb{E}^\lambda [\mathbb{E}[X_\tau^\lambda]]}{1-\frac{\delta}{2}} \geq \mathbb{E}^\lambda [\mathbb{E}_{B_{\delta/2}} [X_\tau^\lambda]]
    \stackrel{(ii)}{=}  \mathbb{E}_{B_{\delta/2}} [\mathbb{E}^\lambda [X_\tau^\lambda]] \\
    &\stackrel{(iii)}{\geq} \mathbb{E}_{B_{\delta/2}} \bigg[ \tfrac{\mathcal{K}\exp \big\{\frac{|M_{\tau}|^2 (1-k)}{ 4\kappa V_{\tau}}  \big \} }{\log^2 \big(\frac{ 2\kappa V_{\tau}}{|M_{\tau}| (1-\sqrt{k})}\big)}\bigg] \geq \mathbb{E}_{B_{\delta/2}} \bigg[ \tfrac{\mathcal{K}\exp \big\{\frac{|M_{\tau}|^2 (1-k)}{ 4\kappa V_{\tau}}  \big\} }{\log^2 \big(\frac{ 2\kappa V_{\tau}}{|M_{\tau}| (1-\sqrt{k})}\big)} \bigg | \tau < \infty \bigg] P_{B_{\delta/2}} (\tau <\infty)\\
    & \stackrel{(iv)}{>} \frac{2}{\delta} P_{B_{\delta/2}} (\tau < \infty),
\end{align*}
where $(i)$ is due to the optional stopping theorem, $(ii)$ follows  from Tonelli's theorem, $(iii)$ is due to the \cref{momentBound} for fixed $\lambda \in (e^{-2},e^{-2})$ and (iv) follows from \eqref{eq:Lemma10}. We deduce that
\begin{align*}
    P_{B_{\delta/2}} (\tau < \infty)<\frac{2}{3}\delta.
\end{align*}
Next, we need to relate $P_{B_{\delta/2}} (\tau < \infty)$ to $P_{A_{\delta/2}} (\tau < \infty)$.
Since $B_{\delta}\subset A_{\delta}$, clearly
\begin{align*}
    \frac{1}{P(A_\delta)}\leq\frac{1}{P(B_\delta)}.
\end{align*}
Furthermore,
\begin{align*}
    P\left(\{\tau<\infty\}\cap A_{\delta}\right)&=P\left(\left(\{\tau<\infty\}\cap B_{\delta}\right)\cup\left( \{\tau<\infty\}\cap\{M_t<-2\lambda_0\kappa V_t\}\right)\right)\\
    &=P\left(\{\tau<\infty\}\cap B_{\delta}\right)+P\left( \{\tau<\infty\}\cap\{M_t<-2\lambda_0\kappa V_t\}\right).
\end{align*}
Since $\{\tau<\infty\}\cap\{M_t<-2\lambda_0\kappa V_t\}=\emptyset$, we conclude
\begin{align*}
    P\left(\{\tau<\infty\}\cap A_{\delta}\right)
 =P\left(\{\tau<\infty\}\cap B_{\delta}\right).
\end{align*}
Therefore, we have, 
\begin{align*}
   \frac{2}{3}\delta\geq P_{B_{\frac{\delta}{2}}}(\tau<\infty)\geq P_{A_{\frac{\delta}{2}}}(\tau<\infty).
\end{align*}
Next,
\begin{align*}
    P_{A_{\delta/2}} (\tau <\infty) <\frac{2}{3} \delta \implies P_{A_{\delta/2}} (\tau = \infty) \geq 1-\frac{2}{3}\delta.
\end{align*}
Further, since $\delta<\frac{1}{2}$ by assumption, we have that, 
\begin{align*}
    P(\tau = \infty) &\geq P({\tau = \infty} \cap A_{\delta/2}) \stackrel{(i)}{=} P_{A_{\delta/2}}(\tau = \infty) P(A_{\delta/2})\\
    & \stackrel{(ii)}{\geq} \left(1-\frac{2}{3}\delta\right) \left(1-\frac{\delta}{2}\right) \geq 1-\delta,
\end{align*}
where $(i)$ follows from the definition of $A_\delta$ and $(ii)$ follows from \Cref{initBoundMarti}. 
\end{proof}

\subsubsection{Proof of the theorem}
The claim of Theorem \ref{thm:BoundLarget}  is an immediate consequence of Lemma \ref{lemma:Bernstein}, where the bound on $\kappa$ follows from the requirement that $X_t^{\lambda}$ is a super-martingale and $s_{0,\text{LIL}}$ results from choosing the largest possible value for $\lambda_0$ that is in agreement with Lemma \ref{le:supermartingale} - Lemma \ref{lemma:Bernstein}, i.e., $\lambda_0=e^{-2}/(1+\sqrt{k})$ for the initial time $t_0$ defined  in Lemma \ref{lemma:weaker}.\\

\hfill$\Box$

\end{document}